\documentclass[useAMS,usenatbib]{mn2e}

\usepackage{graphicx}
\usepackage{times}
\usepackage{natbib}
\usepackage{amsmath}

\renewcommand{\d}{\mathrm{d}}

\newcommand{\gtrsim}{\ga}
\newcommand{\lesssim}{\la}
\def\Zsun{{\rm Z_\odot}}

\def\msun{{\rm M_\odot}}

\def\Mpch{{\rm Mpc/{\it h}}}

\def\Omegat{{\Omega_{0,\rm tot}}}
\def\Omegab{{\Omega_{0,\rm b}}}
\def\Omegam{{\Omega_{0,\rm m}}}
\def\Omegal{{\Omega_{0,\rm \Lambda}}}

\title[Radiative sources]{Radiative feedback and cosmic molecular gas: the role of different radiative sources}
\author[U.~Maio et al.]{
Umberto~Maio$^{1,2}$\thanks{E-mail: umaio@aip.de, maio@oats.inaf.it},
Margarita~Petkova$^{3,4}$,
Gabriella De~Lucia$^{2}$
and
Stefano~Borgani$^{2,5,6}$
\\
${}^1$Leibniz Institute for Astrophysics, An der Sternwarte 16, 14482 Potsdam, Germany\\
${}^2$INAF-Osservatorio Astronomico di Trieste, via G. Tiepolo 11, 34143 Trieste, Italy\\
${}^3$Faculty of Physics, LMU Munich, Scheinerstr. 1, 81679 Munich, Germany\\
${}^4$Excellence Cluster Universe, Boltzmannstr. 2, 85748 Garching, Germany\\
${}^5$Department of Physics, University of Trieste, Piazzale Europa 1, 34128 Trieste, Italy\\
${}^6$INFN-Sezione di Trieste, Via Valerio 2, 34127 Trieste, Italy\\
}

\begin{document}

\date{(draft)}
\pagerange{\pageref{firstpage}--\pageref{lastpage}}\pubyear{0}
\maketitle
\label{firstpage}

\begin{abstract}
We present results from multifrequency radiative hydrodynamical chemistry simulations addressing primordial star formation and related stellar feedback from various populations of stars, stellar energy distributions (SEDs) and initial mass functions.
Spectra for massive stars, intermediate-mass stars and regular solar-like stars are adopted over a grid of 150 frequency bins and consistently coupled with hydrodynamics, heavy-element pollution and non-equilibrium species calculations.
Powerful massive population~III stars are found to be able to largely ionize H and, subsequently, He and He$^+$, causing an inversion of the equation of state and a boost of the Jeans masses in the early intergalactic medium.
Radiative effects on star formation rates are between a factor of a few and 1~dex, depending on the SED.
Radiative processes are responsible for gas heating and photoevaporation, although emission from soft SEDs has minor impacts.
These findings have implications for cosmic gas preheating, primordial direct-collapse black holes, the build-up of ``cosmic fossils'' such as low-mass dwarf galaxies, the role of AGNi during reionization, the early formation of extended disks and angular-momentum catastrophe.
\end{abstract}


%************************************************************************

\section{Introduction}\label{Sect:introduction}

%************************************************************************

With recent experimental advancements \cite[][]{PlanckXXI2015arXiv} it is becoming more and more feasible to define the general framework within which the cosmological structures form and grow from the formation of the first mini-haloes at very high redshift, $z$, up to the following gas collapse and star forming events with consequent feedback mechanisms 
\cite[][]{GunnGott1972,PressSchechter1974,WhiteRees1978,ShethTormen1999,Peacock1999,ColesLucchin2002,PR2003, BarkanaLoeb2001,CiardiFerrara2005,BrommYoshida2011}.
\\
Structure formation depends strongly on the initial density fluctuations imprinted on the primordial matter distribution.
Cold-dark-matter overdensities are expected to be originated in the early inflationary epoch\footnote{
Seminal works are by e.g. \cite{Starobinsky1980,Guth1981,Linde1990}.
}
and to evolve into commonly observed structures, via molecular cooling and star formation processes\cite[see e.g.][and references therein]{Schmidt1959,Kennicutt1996,Kennicutt1998,GP1998,Abel2002,McKeeOstriker2007}.
\\
These events sign the beginning of the cosmic dawn and the passage from the primordial dark ages to the subsequent brighter ages enlightened by the first generations of stars and (proto-)galaxies.
Many speculations are still open on the physical properties of such primordial stars \cite[][]{Abel2002, Yoshida2006, Yoshida_et_al_2007,CampbellLattanzio2008,SudaFujimoto2010, Stacy2014}.
\\
In the latest years, it has been fairly well established that the primordial regime should have existed for a relatively short period of, at most, $\sim 10^7-10^8\,\rm yr$
 \cite[see][]{Tornatore2007, Maio2010, Maio2011, Wise2012, Muratov2013}.
Afterwards, standard solar-like stars, similar to the ones populating the local Universe, took place dominating the star formation on cosmological scales.
This scenario should hold independently of the assumed primordial initial mass function (IMF), supernova ranges and metal yields \cite[][]{Maio2010, MaioTescari2015},
as well as of the effects of alternative dark-energy cosmologies
\cite[][]{Maio2006},
early non-Gaussianities
\cite[][]{MaioIannuzzi2011,Maio2011cqg,MaioKhochfar2012}
or dark-matter nature
\cite[][]{MaioViel2014arXiv}.
\\
Stellar radiation emitted by the first hot massive stars is generally thought to be responsible for dissociation, ionization and heating of molecular gas
\cite[e.g.][]{Draine1996, Ricotti2001, Ahn2007, Whalen2008,Wise2009, Wolcott2011, Yajima2011, Petkova2011, PetkovaMaio2012, Greif2014, Hartwig2014arXiv}.
\\
It should also contribute to cosmic hydrogen reionization
\cite[e.g.][]{Shapiro1986, Gnedin1997, Gnedin2000, Ciardi2003, OnkenMiraldaEscude2004, Kohler2007, Lee2008, Aubert2015, Bauer2015, Pawlik2015, Pawlik2016}.
\\
A few self-consistent attempts, via numerical hydrodynamical simulations including multifrequency radiative transfer, have been performed to address the intrinsic physical role of radiative feedback on cosmic structure formation.
\cite{Haworth2015} pointed out that detailed modeling of radiative processes is generally needed to properly describe star forming sites and HII regions.
Moreover, they also showed that stellar metallicity and radiation pressure are usually minor actors for radiative effects.
\cite{Wise2014} explored the effects of momentum transfer from ionizing radiation, i.e. radiation pressure, by assuming very massive hot stars with typical surface temperatures of $\sim 10^5\,\rm K$.
These sources can emit radiation up to the UV range and contribute to the establishment of a UV background, responsible for heating the low-density gas after first structure formation episodes \cite[][]{Salvaterra2013,Dayal2013, Yajima2014arXiv}.
\\
When comparing radiation pressure to photoionization from young stars, the former appears to be relatively weak, hence a proper treatment of stellar feedback on the interstellar medium should primarily account for the ionization power of stellar radiation \cite[e.g.][]{Hopkins2014, Lopez2014, Sales2014, Kannan2014, Moody2014, Ceverino2014, Murante2015}.
This is particularly true for cosmic star formation in primordial epochs when mini-haloes are very sensitive to radiative feedback from the first stars.
\\
Inspired by such considerations, in the present work we investigate the role of different radiative sources and their impacts on the ambient medium, by performing cosmological multifrequency hydrodynamical chemistry simulations of structure formation \cite[according to the implementation presented in][]{PetkovaMaio2012} and by adopting different assumptions for the central engines of radiative transfer.
We study the evolution of the thermodynamical properties of primordial gas and the changes in the gaseous and stellar content during the growth of cosmic structures.
\\
The paper is structured as follows: after presenting in Sect.~\ref{Sect:simulations} the method and the simulations, in Sect.~\ref{Sect:results} we show the main results, in Sect.~\ref{Sect:discussion} we discuss our findings and in Sect.~\ref{Sect:conclusion} we summarise our main conclusions.

%*****************************************************************************

\section{Simulations}\label{Sect:simulations}

%*****************************************************************************

The simulations considered here were run by using the parallel N-body smoothed-particle hydrodynamic (SPH) \cite[][]{Monaghan1992,Price2012} code Gadget-3, an updated version of the publicly available code Gadget-2 \cite[][]{Springel2005}, modified by \cite{PetkovaMaio2012} to include self-consistently the relevant features of gas cooling, chemistry and multifrequency radiative treatment from both pristine (population III, popIII) and metal-enriched (population II-I, popII-I) stellar populations.
\\
Here we give the basic information of the implementation and refer the interested reader to our previous technical work \cite[][]{PetkovaMaio2012} for further details.

\subsection{Implementation}\label{Sect:implementation}

In the following we summarize the basic approaches behind the various implementations of gravity, cooling, chemistry (including molecule evolution from a reaction network for e$^-$, H, H$^+$, H$^-$, He, He$^+$, He$^{++}$, H$_2$, H$^+$, D, D$^+$, HD and HeH$^+$), radiative transfer (RT), gas self-shielding to Lyman-Werner (LW) radiation, stellar evolution and metal spreading of a number of different elements (readily, He, C, Ca, O, N, Ne, Mg, S, Si, Fe) for popIII and popII-I regimes.

\begin{figure}
  \includegraphics[width=0.475\textwidth, height=0.3\textheight]{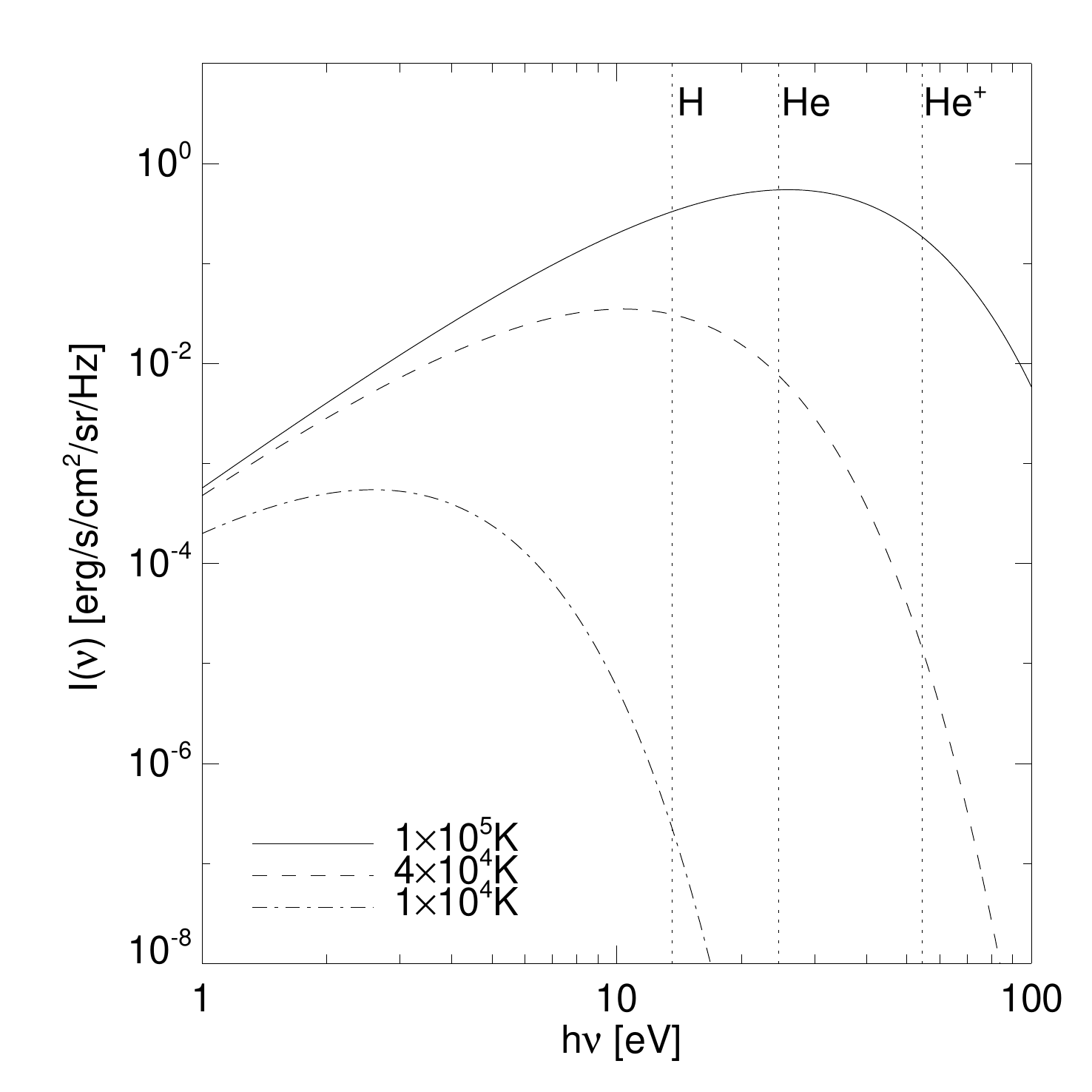}\\
  \vspace{-0.3cm}
  \caption[]{\small
    Emission spectra for a $10^5\,\rm K$ black body (solid line), $4\times 10^4\,\rm K$ black body (dashed line) and $10^4\,\rm K$ black body (dot-dashed line). The vertical lines are drawn in correspondence of the ionization energy of H ($13.6~\rm eV$), He ($24.6~\rm eV$) and He$^+$ ($54.4~\rm eV$), as indicated by the labels.
  }
  \label{fig:cooling_spectra}
\end{figure}

\begin{itemize}

\item
  Resonant and fine-structure transition gas cooling \cite[][]{SD1993,Maio2007} and non-equilibrium chemistry for primordial species -- e$^-$, H, H$^+$, H$^-$, He, He$^+$, He$^{++}$, H$_2$, H$^+$, D, D$^+$, HD and HeH$^+$ \cite[][]{Yoshida2003, Maio2007, Maio2010} -- are coupled with the thermal and hydrodynamical evolution of the gas by solving for the abundance fractions each 1/10th the hydrodynamical time-step \cite[][]{Anninos1997}. 
More exactly, for each species, $i$, the time variation of its number density $n_i$ is computed from collisional, photoionization and photodissociation events via
\begin{equation}
\label{eq:noneq_eq}
\frac{\d n_i}{\d t}= \sum_p\sum_q k_{pq,i} n_p n_q -  \sum_l k_{li} n_l n_i - k_{\gamma i}n_i,
\end{equation}
where $k_{pq,i}$ is the rate of creation of the species $i$ from species $p$ and $q$, $k_{li}$ is the destruction rate of the species $i$ due to collisions with species $l$ and $k_{\gamma i}$ is the photoionization or photodissociation rate of species $i$ interacting with radiation.
Once the abundances have been computed, the corresponding cooling at various temperatures and metallicities can be evaluated.
\\

\item
  The radiative treatment relies on \cite{Petkova2009, Petkova2011}'s implementation, based on the variable Eddington formalism \cite[][]{GnedinAbel2001} and extended to a multi-frequency regime \cite[][]{PetkovaMaio2012} within the star formation and winds module \cite[][]{SH2003}.
Radiative transfer equations are solved for $\sim 150$ frequencies and, at each time-step and for each radiative source, the evolution of the photon number density in each frequency bin, $n_\gamma(\nu)$, is computed according to:
\begin{equation}
\label{eq:rt}
\frac{\partial n_\gamma(\nu)}{\partial t}= c \frac{\partial}{\partial
x_j}\left( \frac{1}{\kappa(\nu)}\frac{\partial n_\gamma(\nu) h^{ij}}{\partial
x_i} \right) - c\,\kappa(\nu)\, n_\gamma(\nu) + s_\gamma(\nu) ,
\end{equation}
where
$t$ is time,
$x_i$ and $x_j$ are the coordinate components,
$c$ is the speed of light, 
$\kappa(\nu)$ is the absorption coefficient, 
$h^{ij}$ are the components of the Eddington tensor,
$s_\gamma(\nu)$ is the source function and
the Einstein summation convention is adopted for the indices $i$ and $j$ of the first term on the right-hand side.
For sake of compactness, the cosmological term, $3 H/c$ with $H$ expansion parameter, has been absorbed into $\kappa$, such that  $\kappa \rightarrow \kappa + 3 H/c$ \cite[see eq.~17 and additional details in][]{Petkova2009}.
We follow radiative emission from individual sources with no additional assumptions on the cosmological background.
The spectral energy distribution (SED) of central sources, $s(\nu)$, can be chosen independently of the numerical scheme. 
Here we focus on stellar-like black-body emission: 
in Fig.~\ref{fig:cooling_spectra} the black-body spectra adopted throughout this work are displayed.
We consider typical popII-I OB stellar sources, whose temperatures\footnote{
  The hottest early-type O star in the Milky Way \cite[][]{Underhill1979} is HD~93129A, with an effective temperature of $5.2\times 10^4 \, \rm K$ \cite[][]{BenagliaKoribalski2004}, while the closest O star to Earth is $\theta^1$~Orionis~C, with an effective temperature of $4.5\times 10^4\,\rm K$ \cite[]{Gagne1997}.
}
are between $10^4\,\rm K$ and $4\times 10^4\,\rm K$, while massive primordial popIII stars might have temperatures up to $\sim 10^5\,\rm K$ \cite[][]{Baraffe2001,Woosley2002,HegerWoosley2010}.
As clear from the SEDs in Fig.~\ref{fig:cooling_spectra}, the differences between the possible cases are found mostly in the high-frequency tail, where the hotter sources give larger contributions.
For sake of comparison, we also mark the frequencies corresponding to H, He and He$^+$ ionization thresholds.
\\

\item
  The SEDs shown contribute to the LW band ([$11.2$, $13.6$]~eV), with larger and larger emission in the case of hotter stars. Thus, self-shielding of dense material \cite[e.g.][]{Draine1996, Wolcott2011,GnedinDraine2014} is taken into account when solving for equations~(\ref{eq:noneq_eq}) and (\ref{eq:rt}).
 We follow the prescriptions by \cite{Draine1996} who suggest a simple analytic approximation for treating self-shielding in different density regimes through corrections of the H$_2$ photodissociation rate.
By inspecting Fig.~\ref{fig:cooling_spectra}, it is easy to see that typical spectra of OB stars are in principle able to ionize H, while hot massive popIII stars produce a larger amount of UV ($h\nu > 13.6\,\rm eV$) photons (widely distributed in the spectral peak of their Planckian law) which can impact both He and He$^+$ chemical balance.
\\

\item
  Finally, we included stellar evolution\footnote{
	We stress that a proper accounting of time delays due to stellar evolution, like the one we perform here \cite[see also e.g.][]{DeLucia2014}, has crucial implications on the cosmic enrichment history, because it determines the actual fraction of SNe at different epochs and, indirectly, the available metal content of different baryonic components and the resulting abundance ratios as a function of cosmic time \cite[][]{MaioTescari2015}.
In fact, simplified assumptions, such as the instantaneous recycling approximation, might severely bias theoretical predictions.
}
and metal spreading for several heavy elements (He, C, Ca, O, N, Ne, Mg, S, Si, Fe, etc.) by employing tabulated input yields for different stellar masses \cite[][]{WW1995, vdHoek1997, Woosley2002, Thielemann2003}, lifetimes \cite[][]{PM1993} and metallicities, as in \cite{Tornatore2007,Tornatore2010}, and coupled to the chemistry and hydro parts \cite[][]{Maio2010, Maio2011b} with a transition threshold between popIII and popII regimes of $Z_{crit} = 10^{-4}\Zsun$ \cite[][]{Bromm_Loeb_2003, Schneider_et_al_2002, Schneider_et_al_2006}.

\end{itemize}

%*****************************************************************************

\begin{table*}
\centering
\caption[Simulation set-up]{Initial parameters and comoving dimensions for the different runs.}
\begin{tabular}{lccccccccc}
\hline
Runs & Box side & Particle mass [$\msun/h$]  & Softening & PopIII IMF   & PopII IMF       & metal     & RT    & PopIII BB      & PopII BB \\
        & [\Mpch]  & gas (dark matter)   & [kpc/{\it h}] & range [M$_\odot$] & range [M$_\odot$] & spreading &  & T$_{\rm eff}$[K]& T$_{\rm eff}$[K]\\
\hline
Run128.noRT   & 0.5  & $ 6.6\times 10^2\quad$ ($4.3\times 10^3 $)& 0.2  & [100, 500]& [0.1, 100]& on & off & - & -\\
%Run064.noRT   & 0.5  & $ 5.3\times 10^3\quad$ ($3.4\times 10^4 $)& 0.4  & [100, 500]& [0.1, 100]& on & off & - & -\\
%Run032.noRT   & 0.5  & $ 4.2\times 10^4\quad$ ($2.7\times 10^5 $)& 0.8  & [100, 500]& [0.1, 100]& on & off & - & -\\
%\hline
Run128   & 0.5  & $ 6.6\times 10^2\quad$ ($4.3\times 10^3 $)& 0.2  & [100, 500]& [0.1, 100]& on & on & $10^5$ & $10^4$ \\
%Run064   & 0.5  & $ 5.3\times 10^3\quad$ ($3.4\times 10^4 $)& 0.4  & [100, 500]& [0.1, 100]& on & on & $10^5$ & $10^4$\\
%Run032   & 0.5  & $ 4.2\times 10^4\quad$ ($2.7\times 10^5 $)& 0.8  & [100, 500]& [0.1, 100]& on & on & $10^5$ & $10^4$\\
%\hline
Run128.SL.noRT   & 0.5  & $ 6.6\times 10^2\quad$ ($4.3\times 10^3 $)& 0.2  & [0.1, 100]& [0.1, 100]& on & off & - & -\\
%Run064.SL.noRT   & 0.5  & $ 5.3\times 10^3\quad$ ($3.4\times 10^4 $)& 0.4  & [0.1, 100]& [0.1, 100]& on & off & - & -\\
%Run032.SL.noRT   & 0.5  & $ 4.2\times 10^4\quad$ ($2.7\times 10^5 $)& 0.8  & [0.1, 100]& [0.1, 100]& on & off & - & -\\
%\hline
 Run128.SL.1e4 & 0.5  & $ 6.6\times 10^2\quad$ ($4.3\times 10^3 $)& 0.2  & [0.1, 100]& [0.1, 100]& on & on & $10^4$ & $10^4$ \\
%Run064.SL.1e4  & 0.5  & $ 5.3\times 10^3\quad$ ($3.4\times 10^4 $)& 0.4  & [0.1, 100]& [0.1, 100]& on & on & $10^4$ & $10^4$\\
%Run032.SL.1e4  & 0.5  & $ 4.2\times 10^4\quad$ ($2.7\times 10^5 $)& 0.8  & [0.1, 100]& [0.1, 100]& on & on & $10^4$ & $10^4$\\
%\hline
Run128.SL.4e4 & 0.5  & $ 6.6\times 10^2\quad$ ($4.3\times 10^3 $)& 0.2  & [0.1, 100]& [0.1, 100]& on & on & $4\times10^4$ & $4\times 10^4$ \\
%Run064.SL.4e4  & 0.5  & $ 5.3\times 10^3\quad$ ($3.4\times 10^4 $)& 0.4  & [0.1, 100]& [0.1, 100]& on & on & $4\times10^4$ & $4\times 10^4$\\
%Run032.SL.4e4 & 0.5  & $ 4.2\times 10^4\quad$ ($2.7\times 10^5 $)& 0.8  & [0.1, 100]& [0.1, 100]& on & on & $4\times 10^4$ & $4\times 10^4$\\
\hline
\label{tab:runs}
\end{tabular}
\begin{flushleft}
\vspace{-0.5cm}
{\small
}
\end{flushleft}
\end{table*}

\subsection{Runs}\label{Sect:runs}

In order to investigate how different radiative sources impact cosmic gas, star formation episodes and the state of the inter-galactic medium (IGM), we perform a number of N-body hydrodynamical chemistry and multifrequency radiative simulations with different assumptions.
For each of them, the softening is $\sim 1/20$ the mean inter-particle separation \cite[gravitational softening and SPH softening are taken to be equal, e.g.][]{Springel2005} and the number of particles used is  $2\times128^3$, for gas and dark-matter species, respectively.
Hydrodynamical densities are estimated by (B-spline) kernel smoothing over 32 neighbours.
\\
We generate initial conditions at redshift $z=100$ and evolve them down to the reionization epoch, $z\sim 6$.
We sample the cosmic medium according to standard density parameters
$\Omegat = 1.0$,
$\Omegam = 0.3$,
$\Omegal = 0.7$,
$\Omegab = 0.04$,
for total, cold dark matter, $\Lambda$ and baryon components following an initial Gaussian density distribution\footnote{
  For non-Gaussian effects see e.g. \cite{MaioIannuzzi2011,Maio2011cqg,MaioKhochfar2012}.
},
expansion parameter normalized to $100 \, \rm km/s/Mpc$  $h = 0.7$ and spectral parameters 
$\sigma_8=0.8$ and $n=1$.
\\
The adopted box side of the simulated primordial volume is $\rm 0.5\, Mpc/{\it h} $ which, although small to be a complete representation of the whole Universe, is well suited to investigate local star formation and radiative feedback in the early interstellar medium.
\\
Stars are generated stochastically above a density threshold of $1\, \rm cm^{-3}$ and a number of generations of 4 is assumed when spawning stars from gas particles.
For more discussion we refer the reader to our previous works \cite[][]{Maio2009,Maio2010}. 
As an example, Fig.~8 in \cite{Maio2010} shows the effects of varying the threshold density within a few orders of magnitude and clearly demonstrates that values of the order of $1\, \rm cm^{-3}$ or higher converge without introducing significant discrepancies.
In the no-RT case, SN feedback determines energy deposition in the surrounding gas particles according to the SPH kernel.
\\
Concerning the stellar populations included, different assumptions for masses and effective temperatures have been made, as described in the following.
For popII regimes, we adopted typical OB effective temperatures of $~\rm T_{eff}=10^4\,\rm K$  and  $4\times 10^4\,\rm K$ while the popII IMF has been chosen to be Salpeter over the range [0.1, 100]~$\rm M_\odot$\footnote{
  It is worth mentioning that alternative IMF shapes can be chosen. In general, different IMFs determine different ratios between low-mass and high-mass stars and consequantly they can impact e.g. gas duty cicle and star formation rates. As an example, using a more sophisticated Chabrier IMF would alter the star formation rates and corresponding UV luminosities by a factor of up to 2. Given the lack of precise information at high redshift we neglect these details and, for sake of simplicity, we adopt a simple Salpeter IMF.
}.
For popIII regimes, we tested both the massive case with a top-heavy IMF over [100, 500]~$\rm M_\odot$ with slope $-2.35$ and the more regular case with a Salpeter IMF over the range [0.1, 100]~$\rm M_\odot$.
In the former scenario, pair-instability SNe (PISNe) are the main responsible for metal enrichment and radiative feedback having surface effective temperatures of $\rm T_{eff}\sim 10^5\,\rm K$;
in the latter scenario, we bracket the effects of different lower-mass popIII progenitors \cite[e.g.][]{Abel2002,Hosokawa2011,Stacy2012,Stacy2014}, by choosing both a relatively cold spectrum with $\rm T_{eff}=10^4\,\rm K$ and a hotter one with $\rm T_{eff}=4\times 10^4\,\rm K$.
\\
The numerical outputs are processed by a friends-of-friends algorithm (FoF) with a linking length of $20\%$ the mean inter-particles separation and a minimum number of 32 constituting particles to define formed structures and their basic properties.
\\
The different assumptions and characteristics of the simulations are listed in Tab.~\ref{tab:runs}.
In the table, the various runs are named by tags indicating their resolution and properties.
The tag 'noRT' stands for 'no radiative transfer', while 'SL' stands for 'Salpeter-like' popIII IMF and the values '1e4' and '4e4' indicate the effective temperature of emitting sources.
If not specified in their name, the runs assume a top-heavy popIII IMF with effective temperature of $10^5\,\rm K$, a Salpeter popII-I IMF with effective temperature of $10^4\,\rm K$ and include multifrequency radiative treatment over $150$ frequency bins.

%*****************************************************************************

\section{Results}\label{Sect:results}

We perform several runs at different resolutions, but in the following we will consider only the higher-resolution runs that can catch early structure formation in a more detailed way and allow us to discuss more precisely the effects of radiative feedback from different stellar sources.
\\
In the next, we show results for the five cases corresponding to:

\begin{itemize}
\item
	no RT with top-heavy popIII IMF and Salpeter popII-I IMF; 
\item
	RT with top-heavy popIII IMF, Salpeter popII-I IMF, popIII $~\rm T_{eff}=10^5\, K$ and popII-I $~\rm T_{eff}=10^4\,K$;
\item
  	no RT with SL popIII and popII-I IMFs; 
\item
  	RT with SL popIII and popII-I IMFs, $~\rm T_{eff}=10^{4}\, K$;
\item
  	RT with SL popIII and popII-I IMFs, $~\rm T_{eff}=4\times 10^{4}\,K$.
\end{itemize}

%*****************************************************************************

\subsection{Thermal evolution}\label{Sect:maps}

\begin{figure*}
\centering
{\underline{\bf No RT} \hspace{5cm}  \underline{\bf with RT} $\quad \rm (T_{\rm eff}=10^5\,K)$}\\
\vspace{-1cm}
 \includegraphics[width=0.35\textwidth]{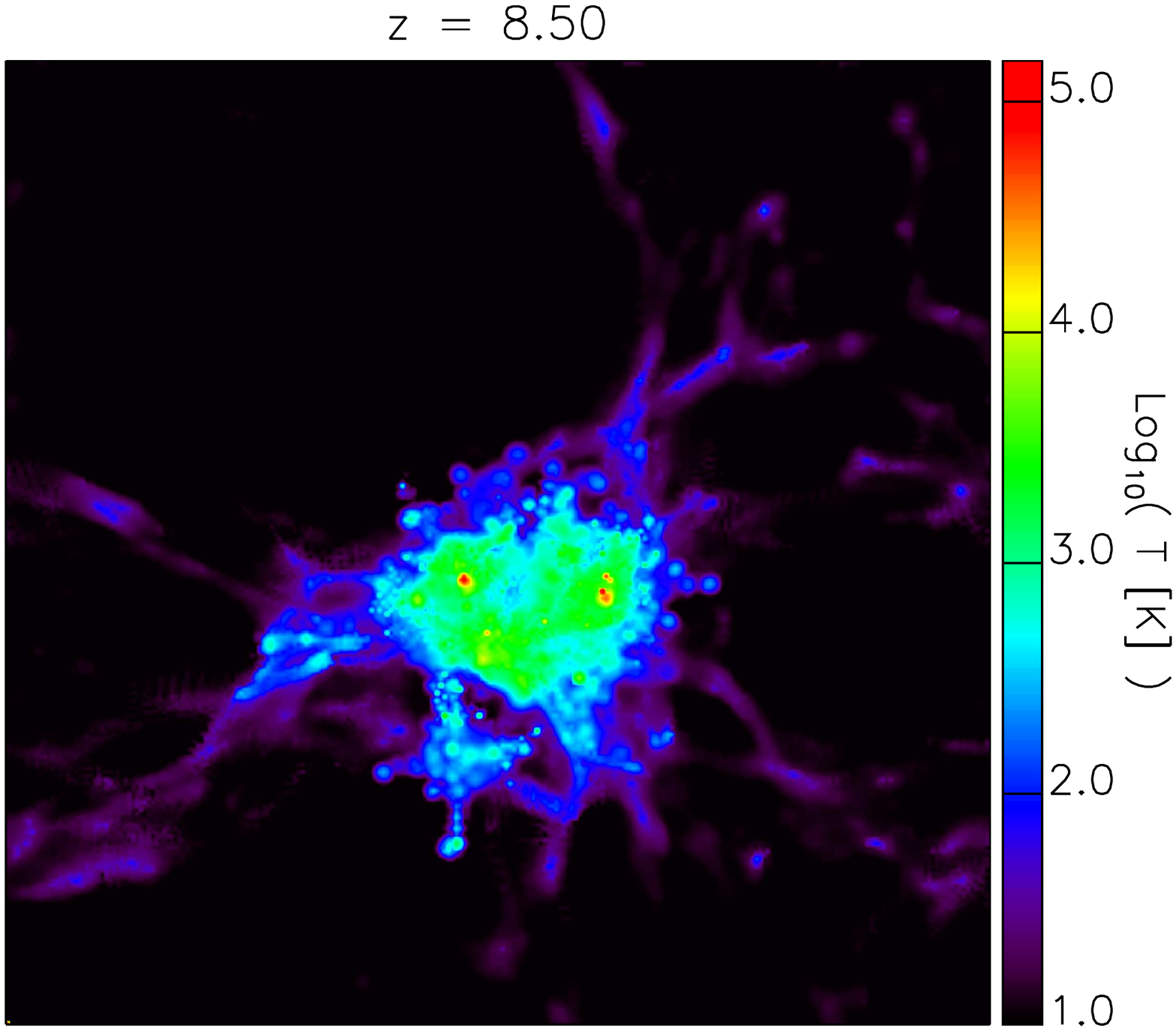}
 \includegraphics[width=0.35\textwidth]{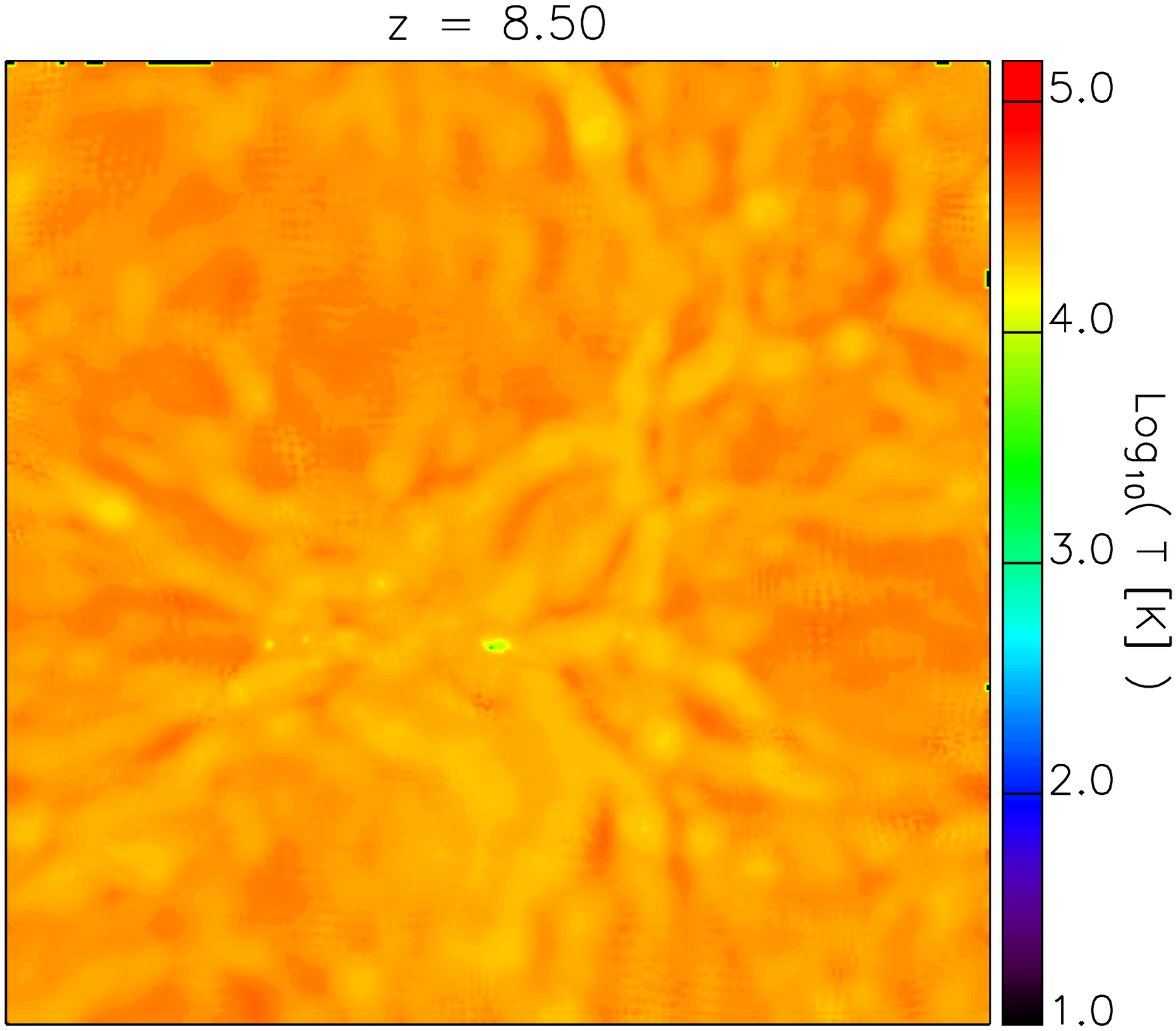}\\
 \vspace{-2.5cm}
 \includegraphics[width=0.35\textwidth]{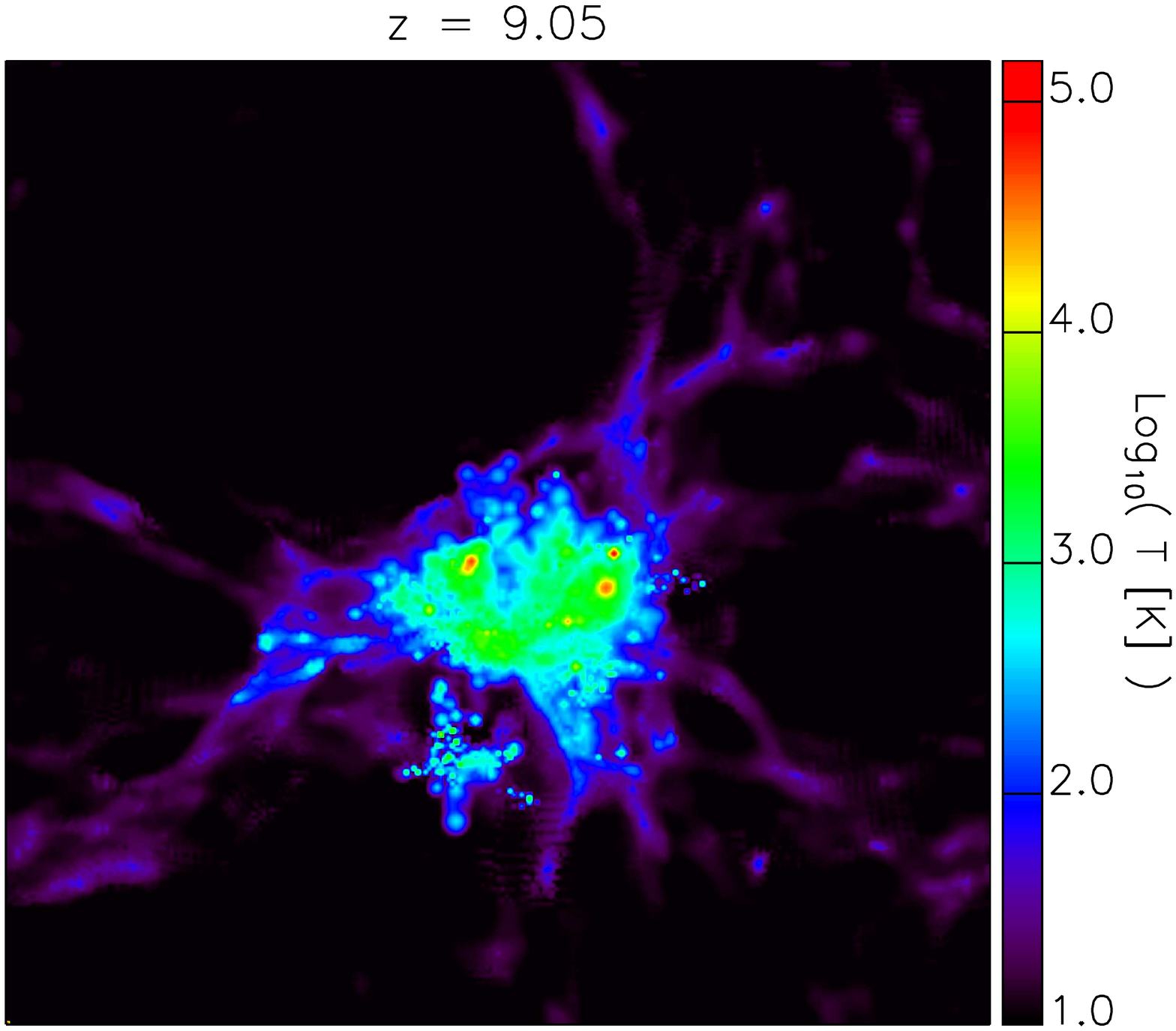}
 \includegraphics[width=0.35\textwidth]{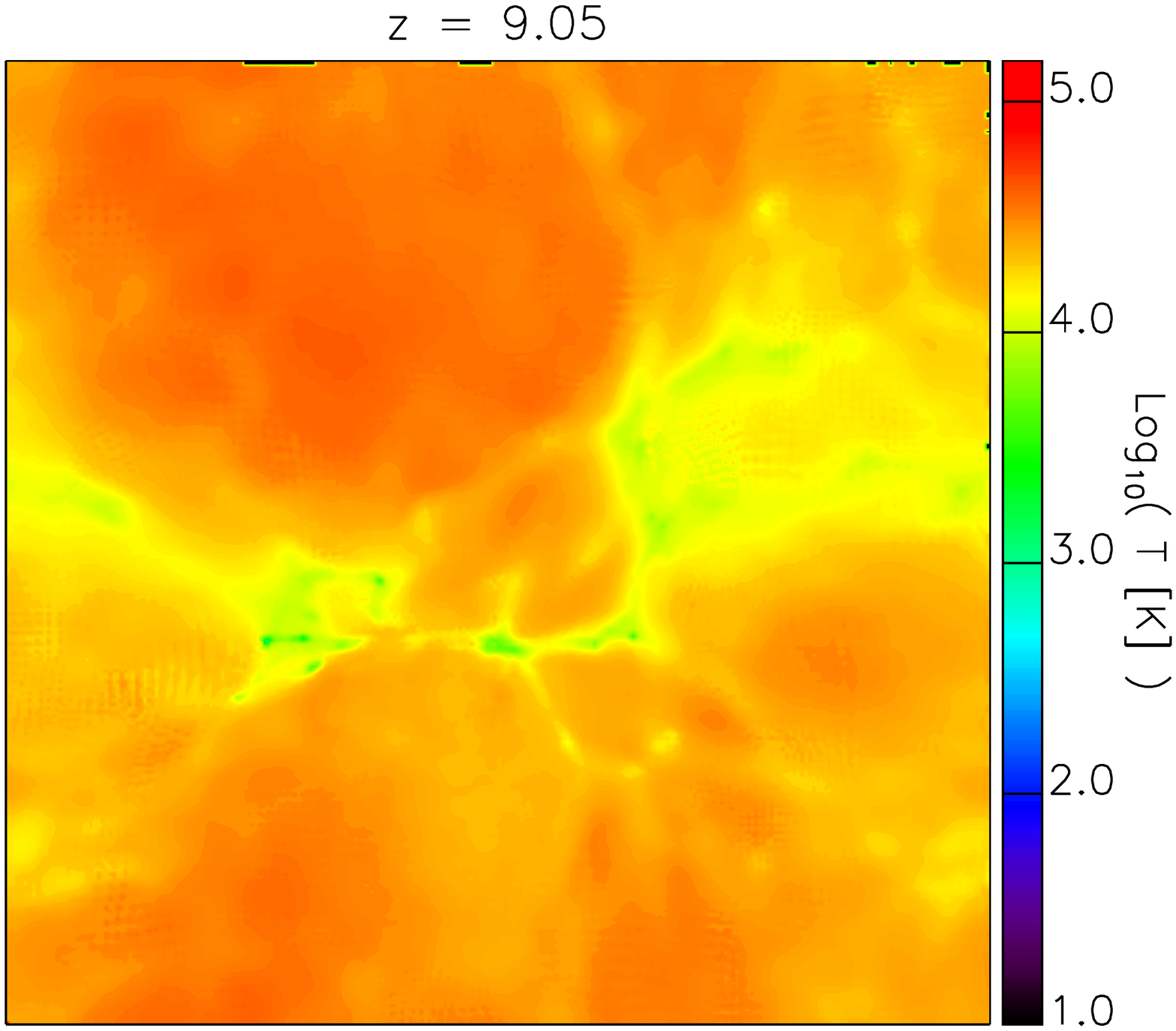}\\
 \vspace{-2.5cm} 
 \includegraphics[width=0.35\textwidth]{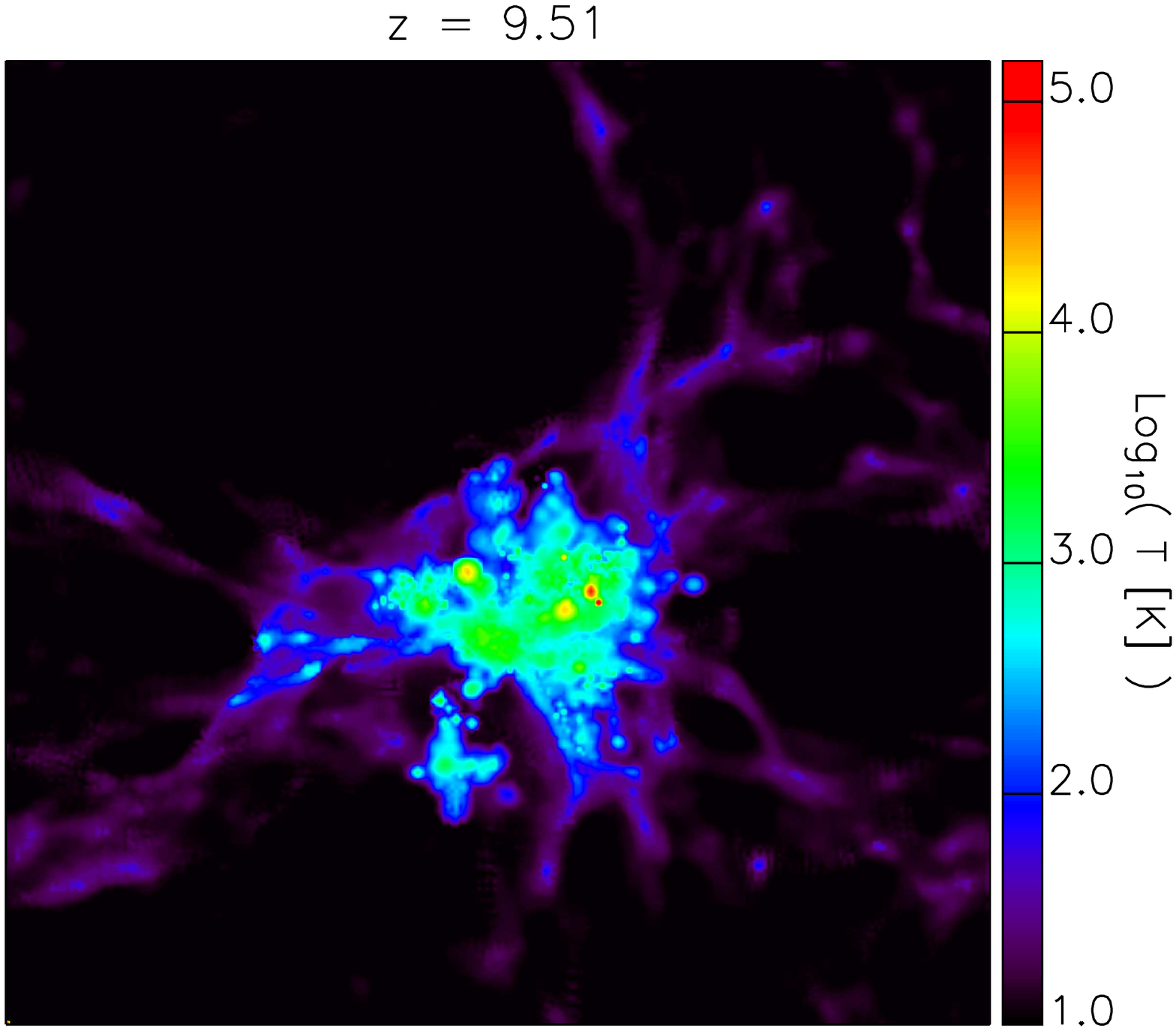}
 \includegraphics[width=0.35\textwidth]{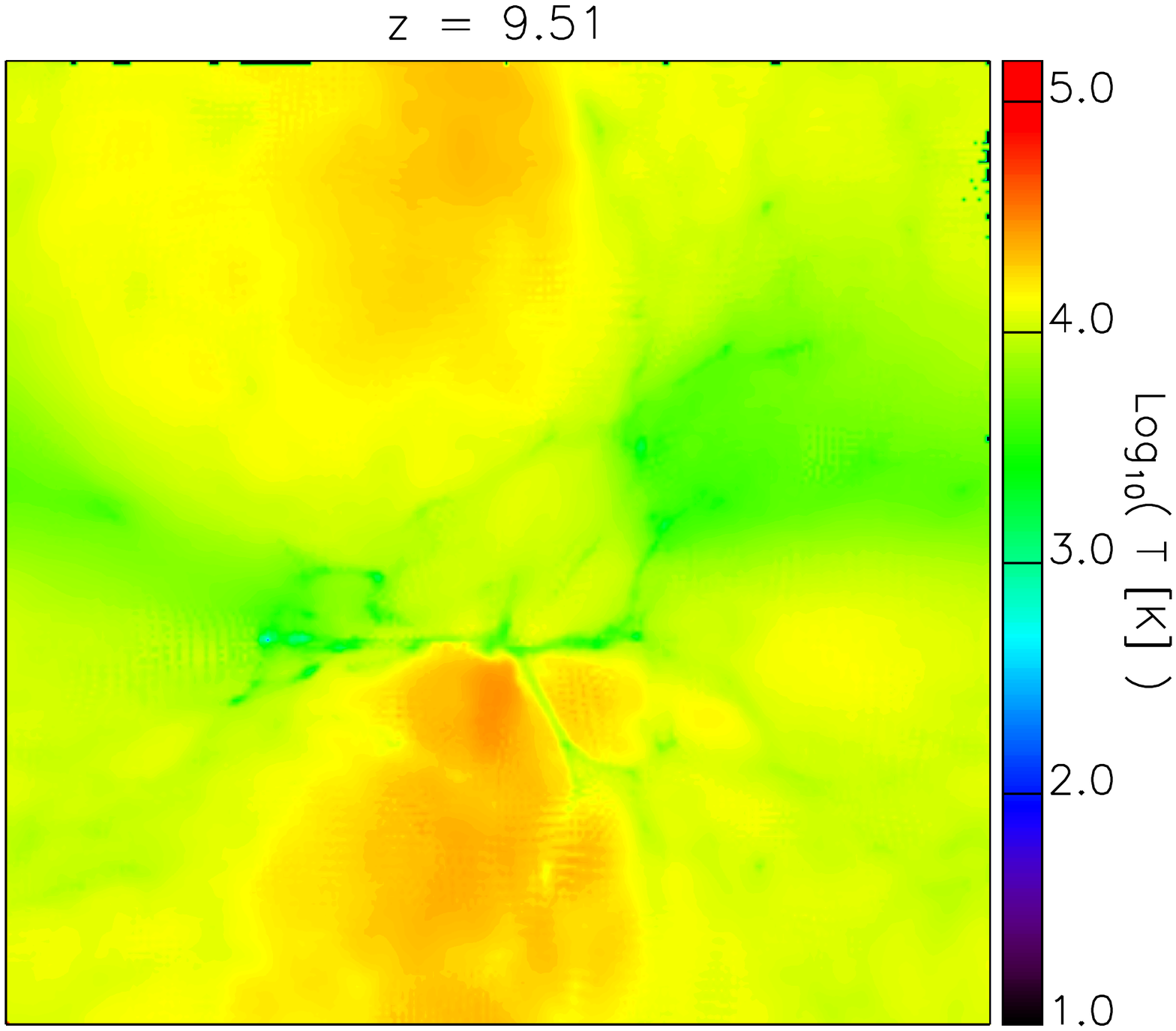}\\
  \vspace{-2.5cm}
 \includegraphics[width=0.35\textwidth]{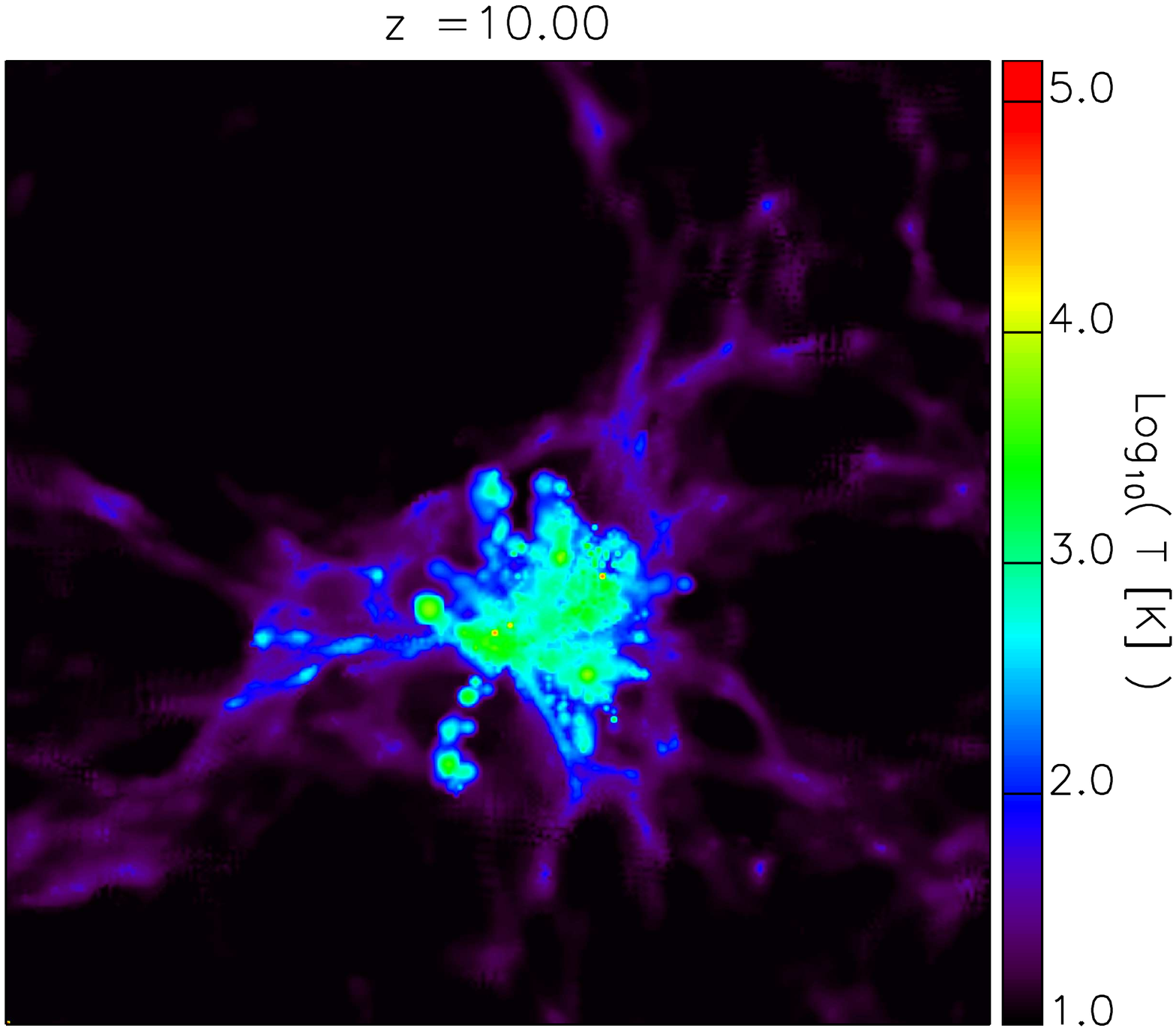}
 \includegraphics[width=0.35\textwidth]{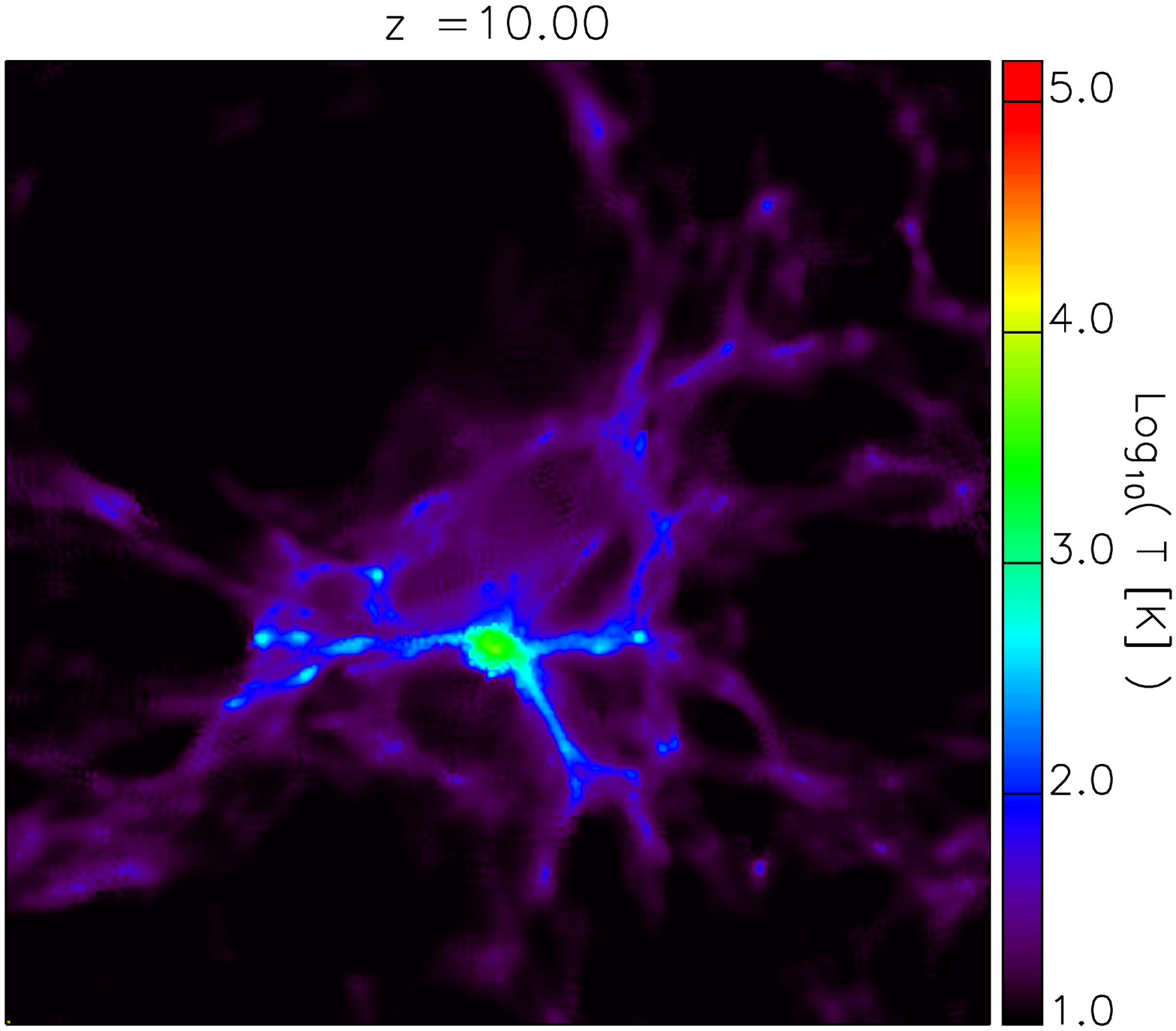}\\
  \vspace{-1cm}
\caption[]{\small
Temperature maps at different redshifts for the simulations having a top-heavy popIII IMF and a Salpeter popII-I IMF. Results refer to the no-RT run (left) and to the run with RT (right). In this latter case the assumed SED follows a black-body spectrum with effective temperature $\rm T_{eff} = 10^5\,K$ for popIII sources and $\rm T_{eff} = 10^4\,K$  for popII-I sources.
}
\label{fig:maps}
\end{figure*}

\begin{figure*}
\centering
{
  \underline{\bf No RT} \hspace{3.1cm} 
  \underline{\bf with RT} $\quad\rm (T_{\rm eff}=4\times 10^4\,K)$ \hspace{1.1cm} \underline{\bf with RT} $\quad\rm (T_{\rm eff}=10^4\,K)$
}
\vspace{-1cm}\\
\hspace{-1.2cm}
\includegraphics[width=0.35\textwidth]{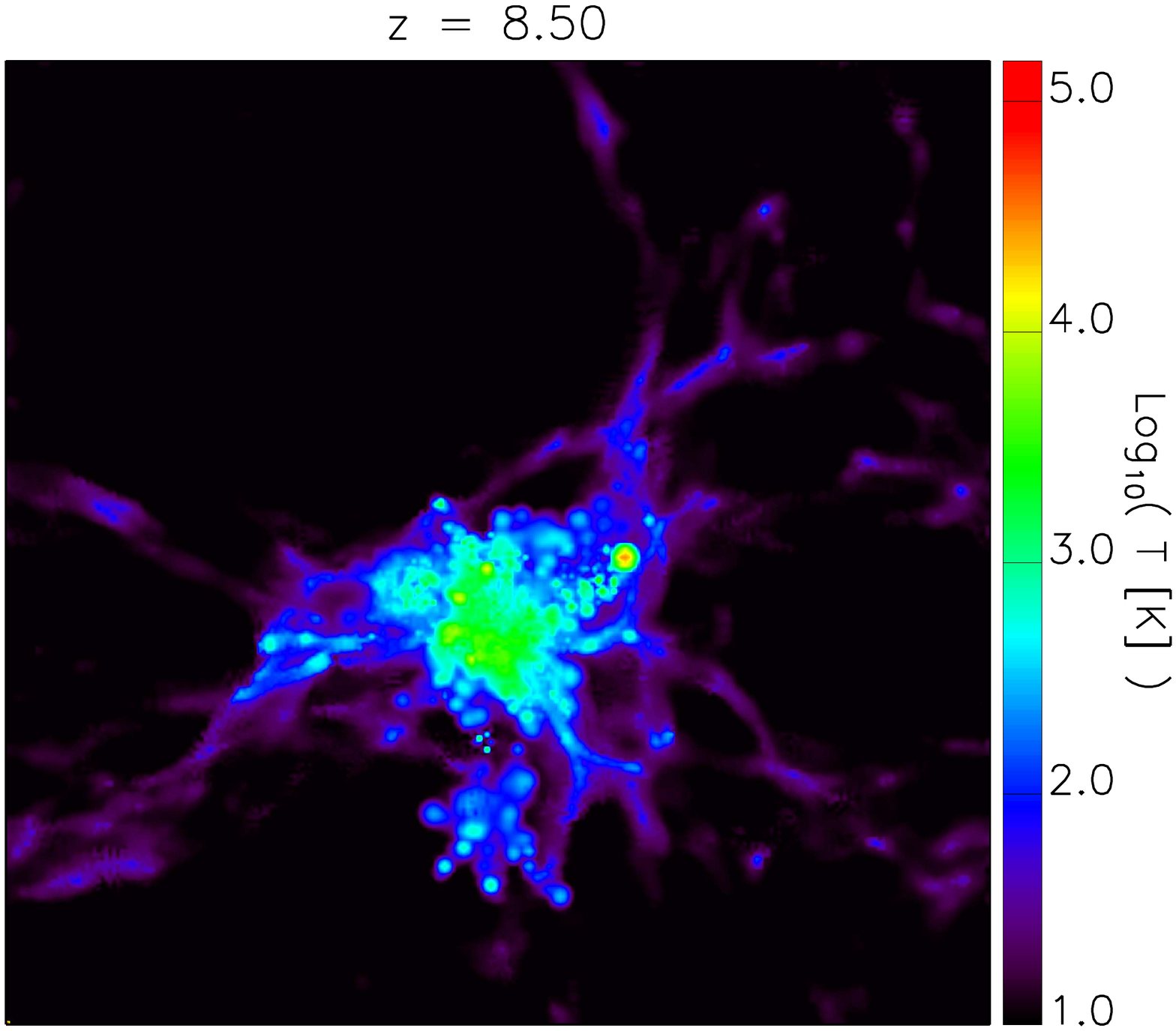}\hspace{-1.45cm}
\includegraphics[width=0.35\textwidth]{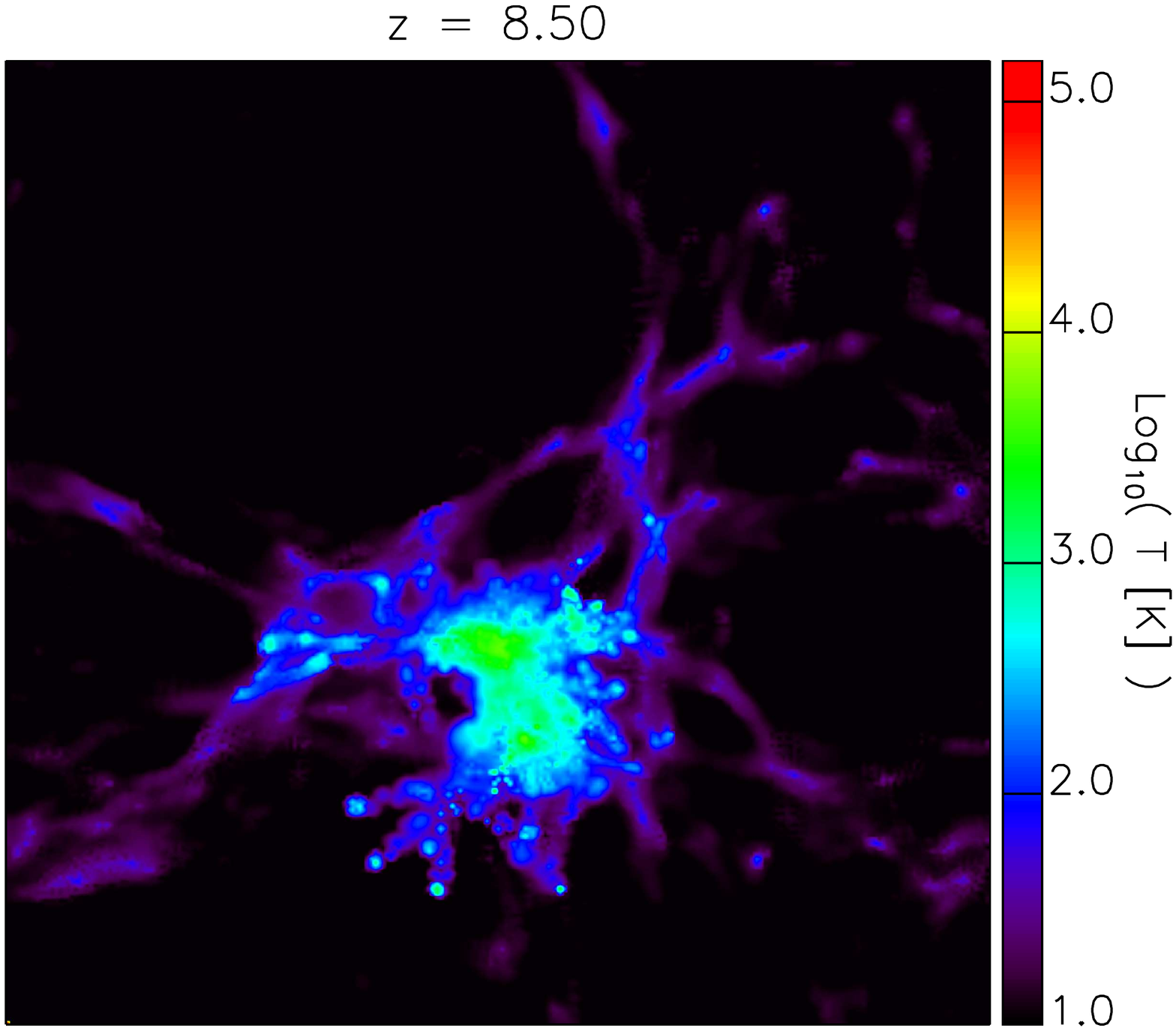}\hspace{-1.45cm}
\includegraphics[width=0.35\textwidth]{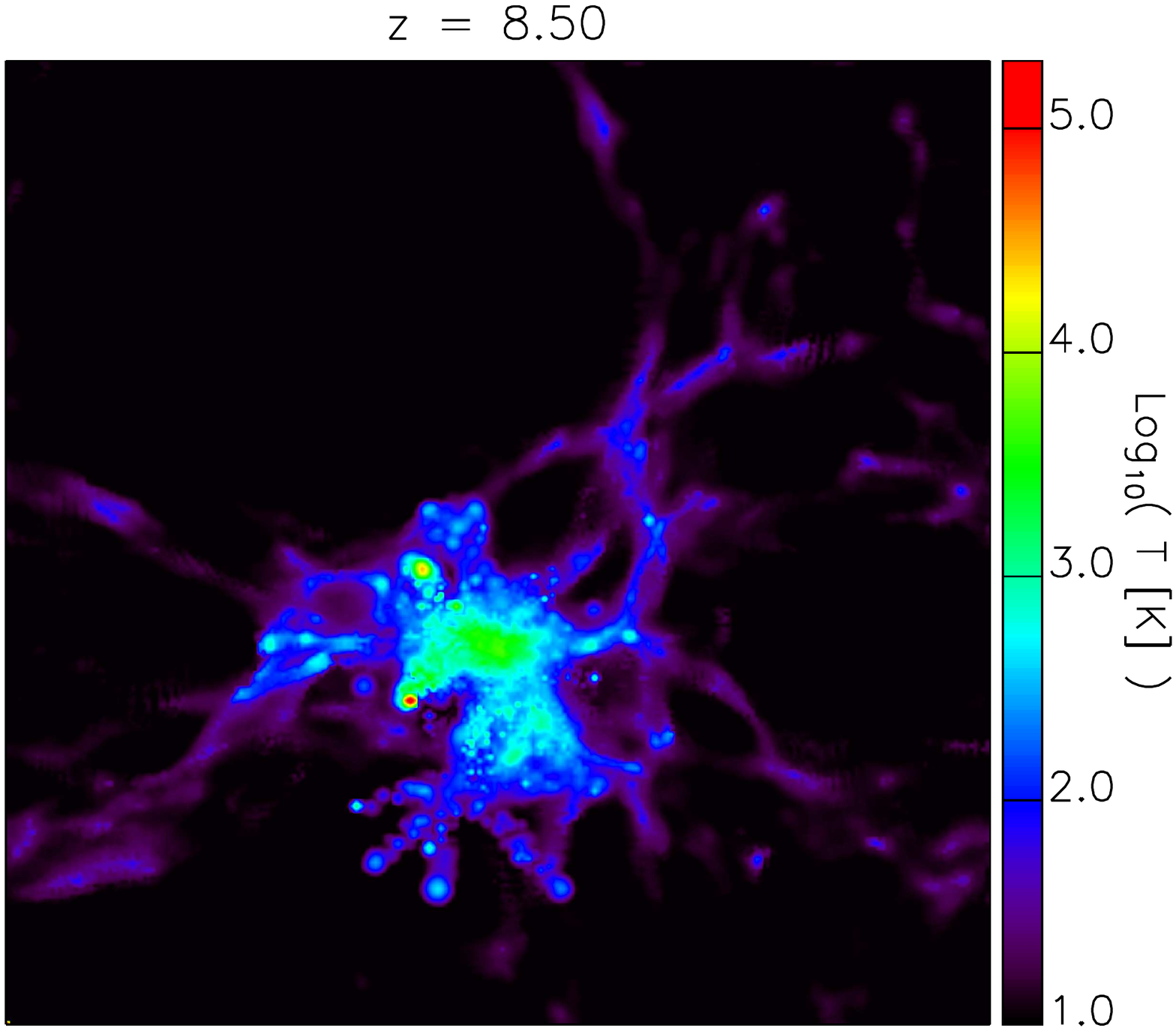}\\
\vspace{-2.5cm}
\hspace{-1.2cm}
\includegraphics[width=0.35\textwidth]{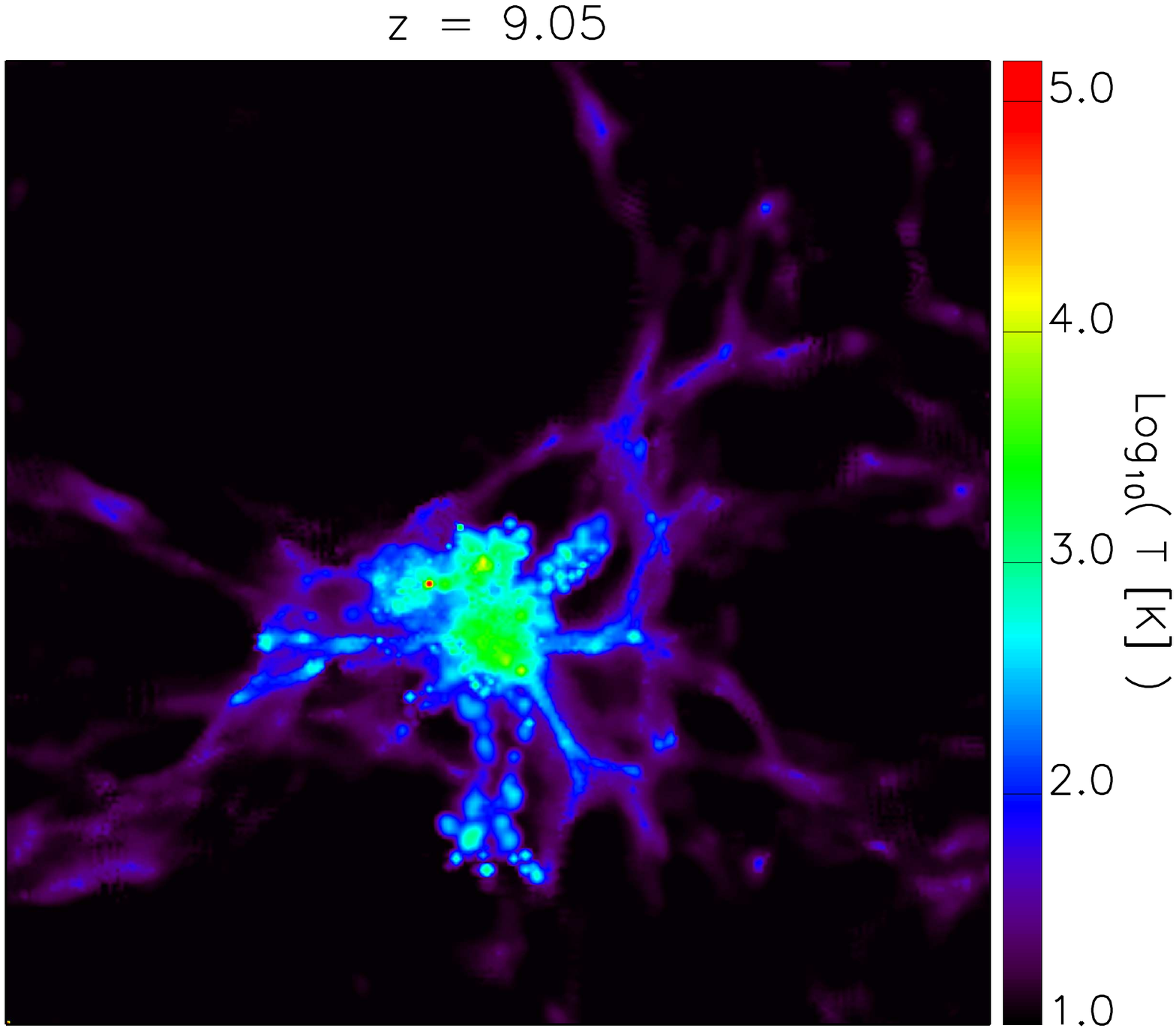}\hspace{-1.45cm}
\includegraphics[width=0.35\textwidth]{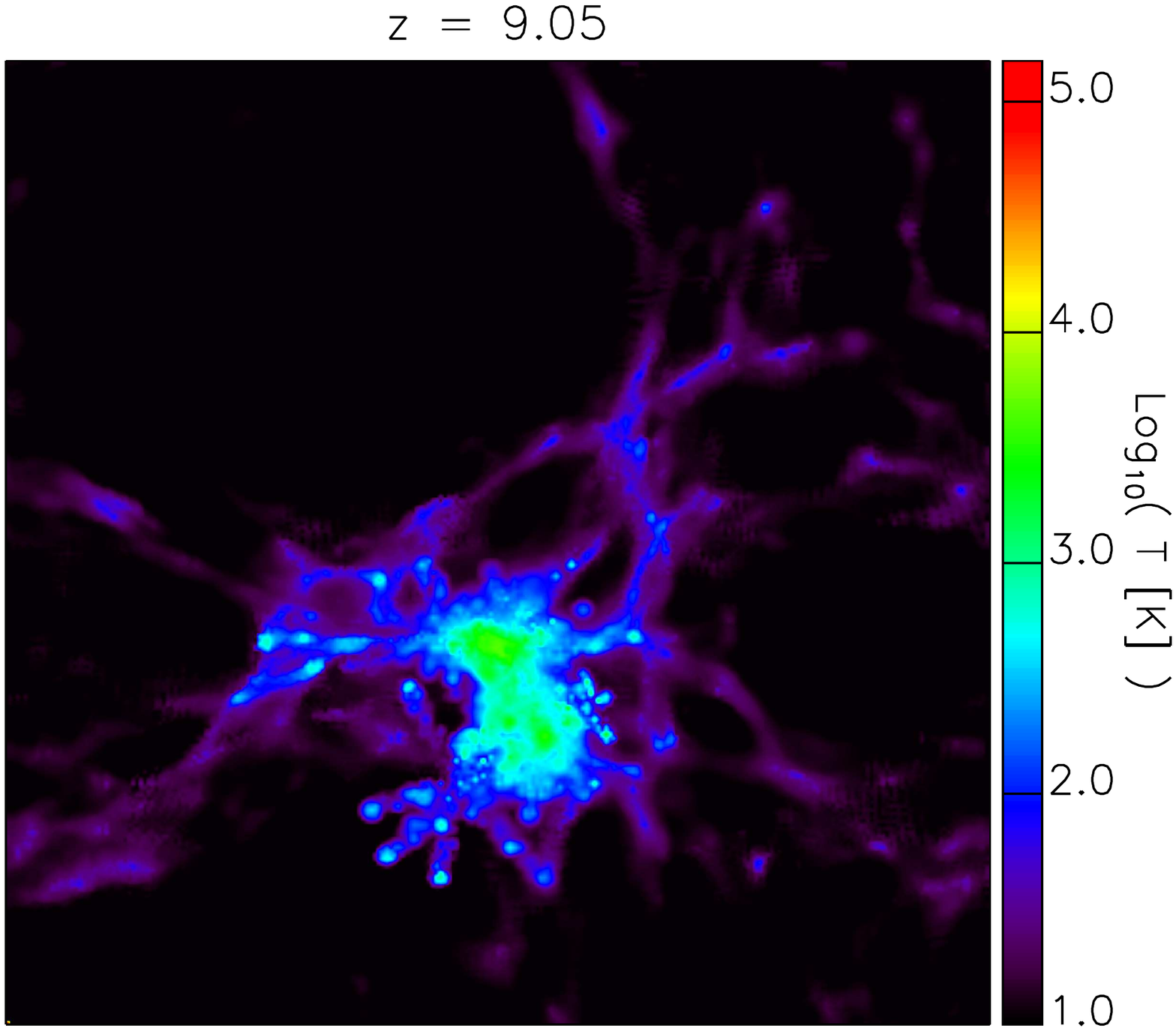}\hspace{-1.45cm}
\includegraphics[width=0.35\textwidth]{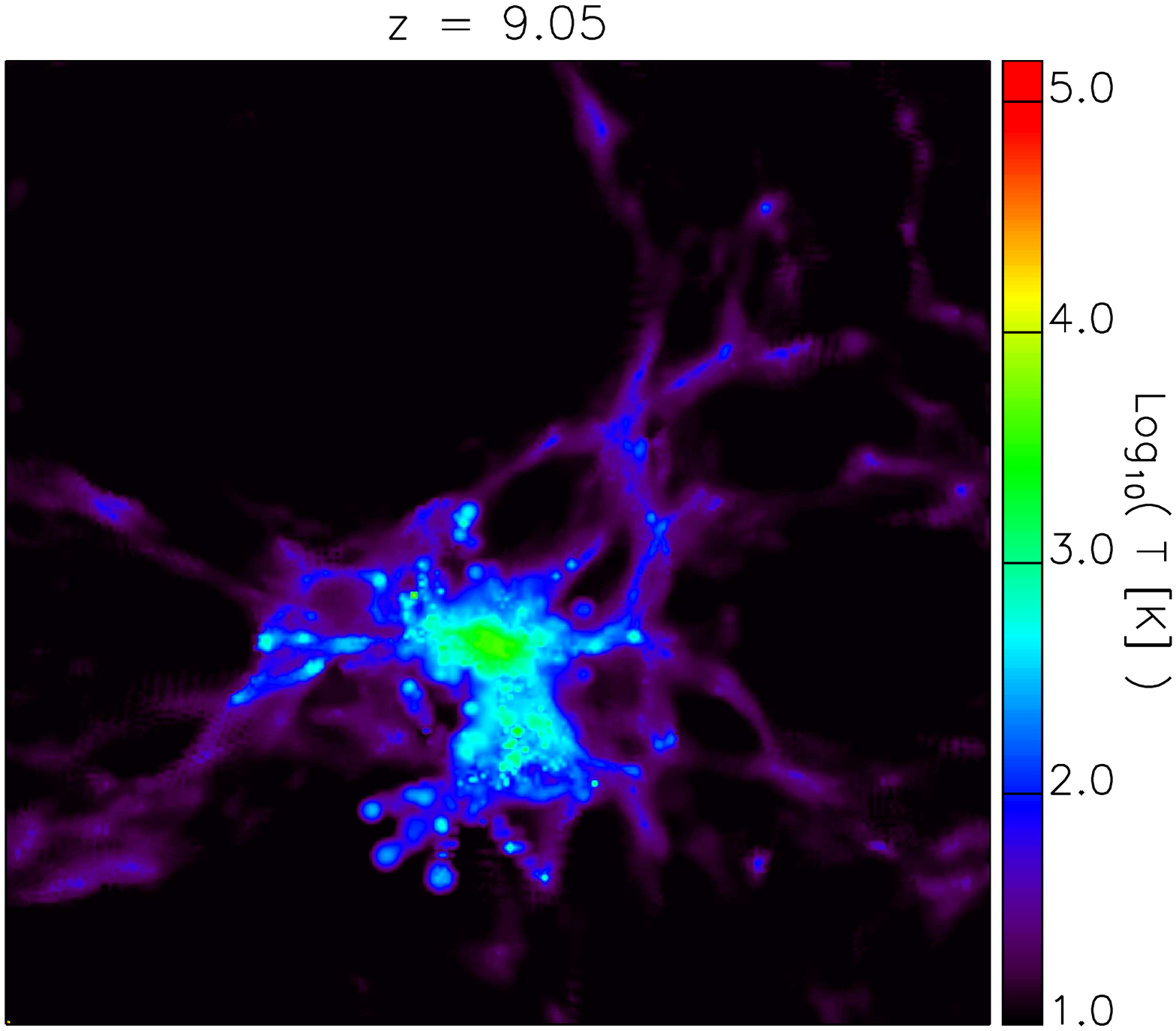}\\
\vspace{-2.5cm}
\hspace{-1.2cm}
\includegraphics[width=0.35\textwidth]{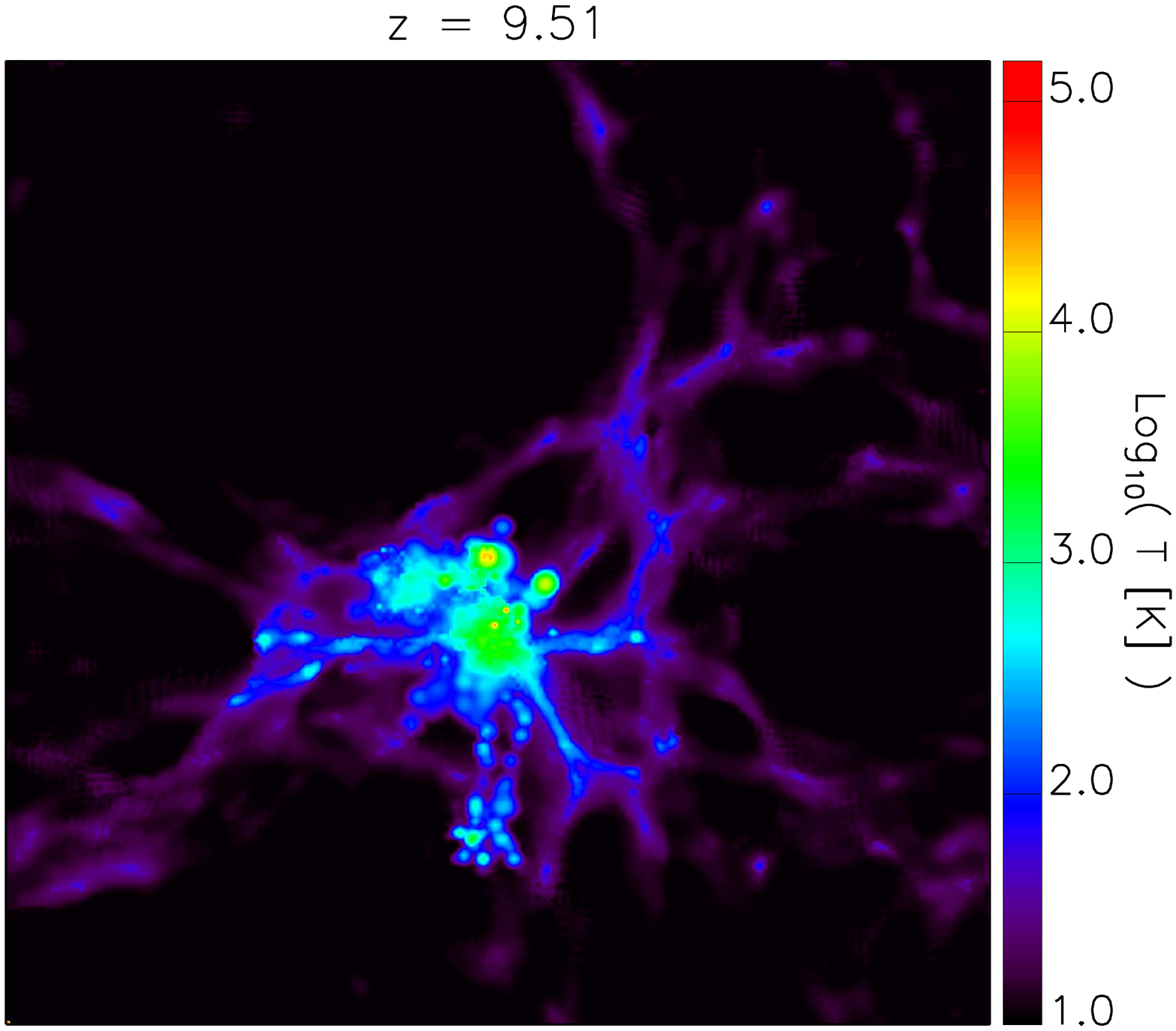}\hspace{-1.45cm}
\includegraphics[width=0.35\textwidth]{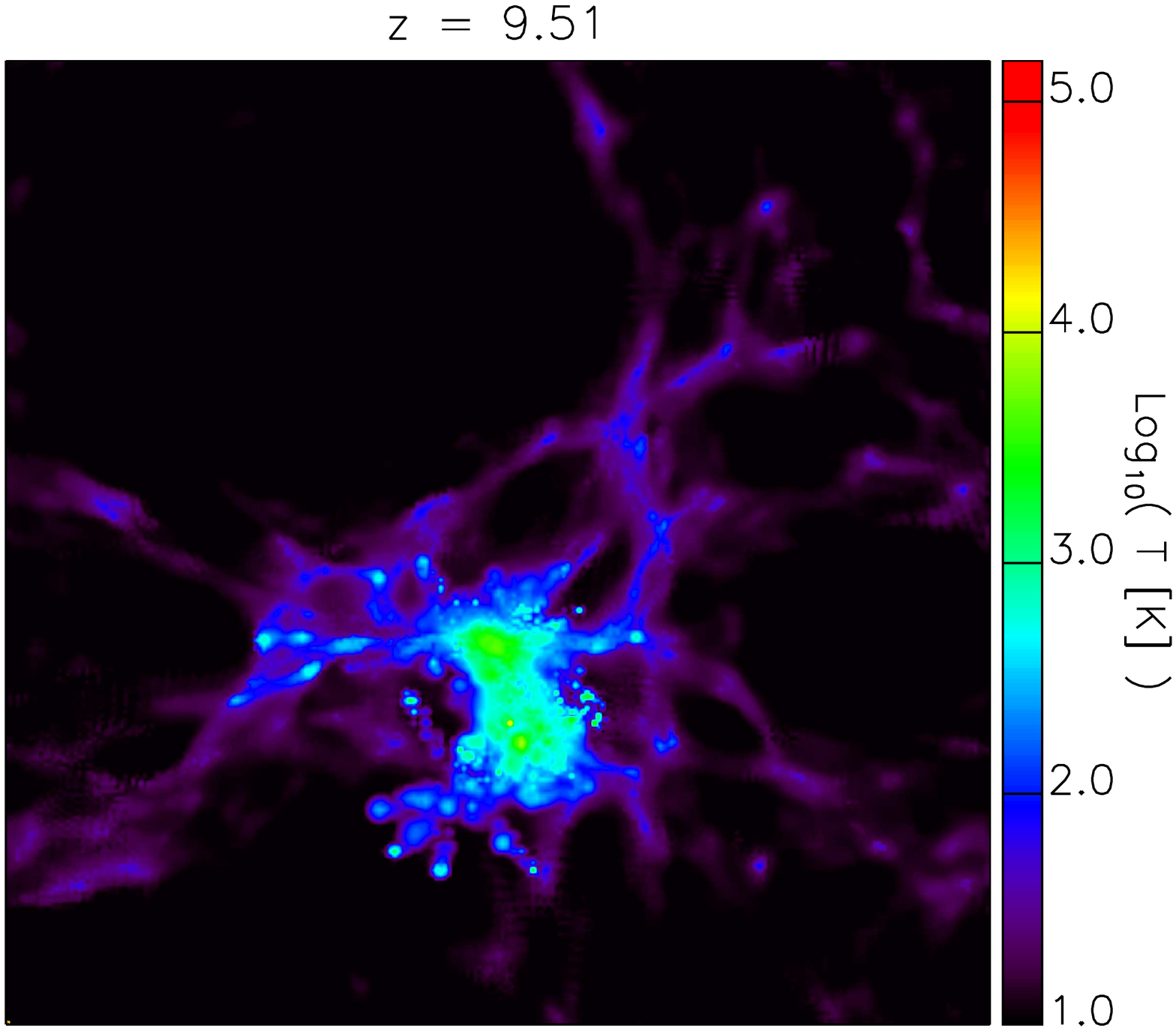}\hspace{-1.45cm}
\includegraphics[width=0.35\textwidth]{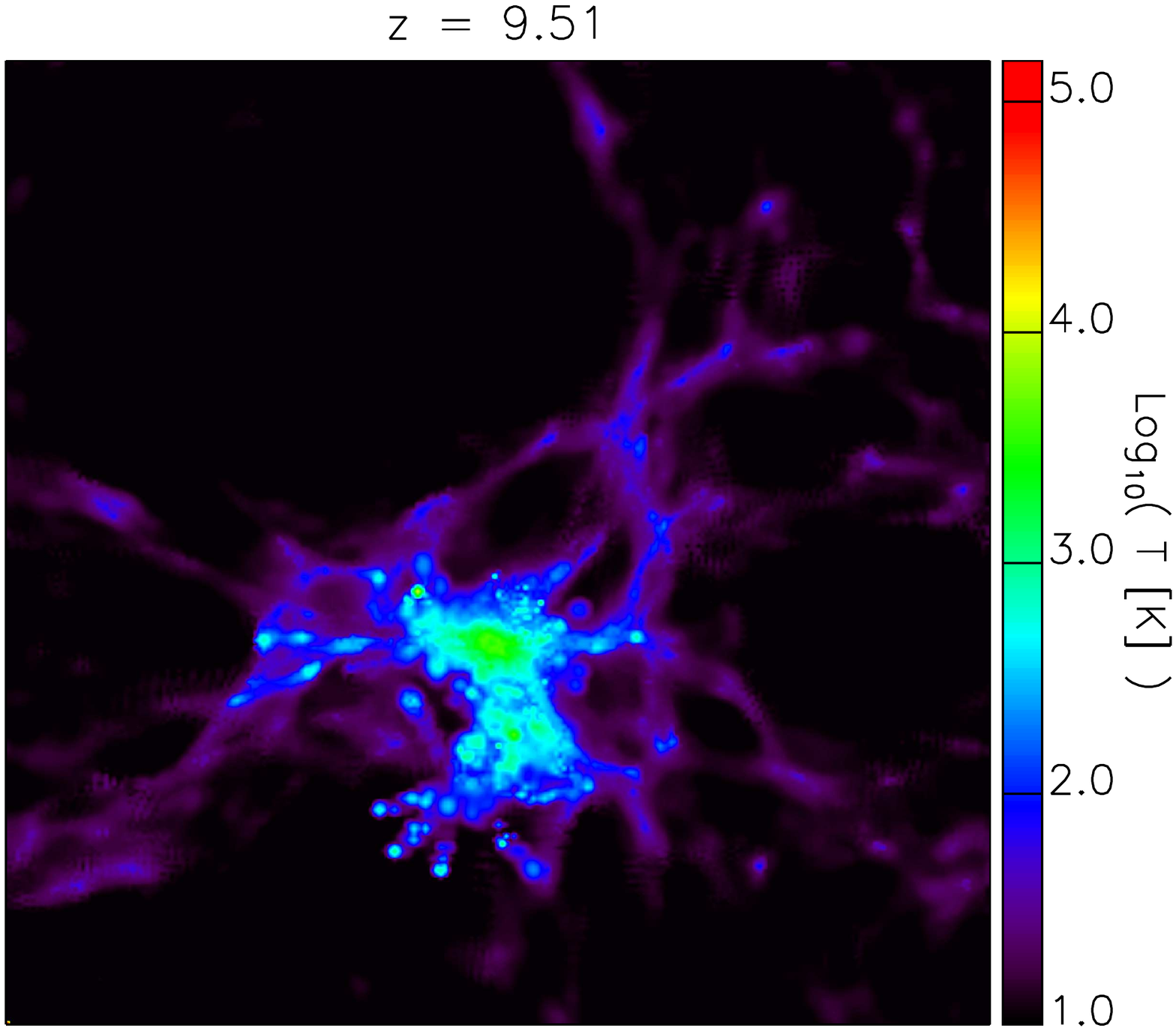}\\
\vspace{-2.5cm}
\hspace{-1.2cm}
\includegraphics[width=0.35\textwidth]{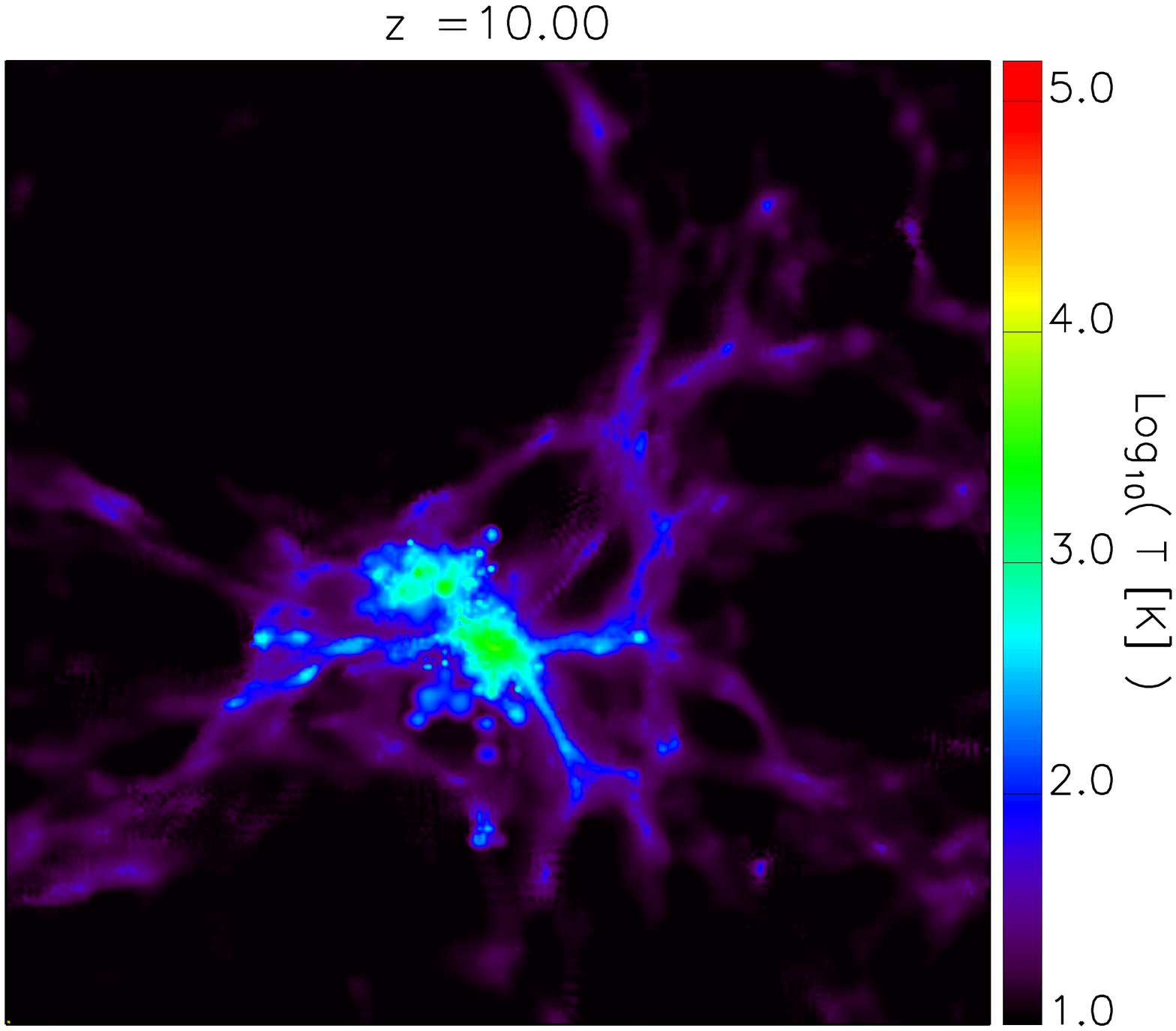}\hspace{-1.45cm}
\includegraphics[width=0.35\textwidth]{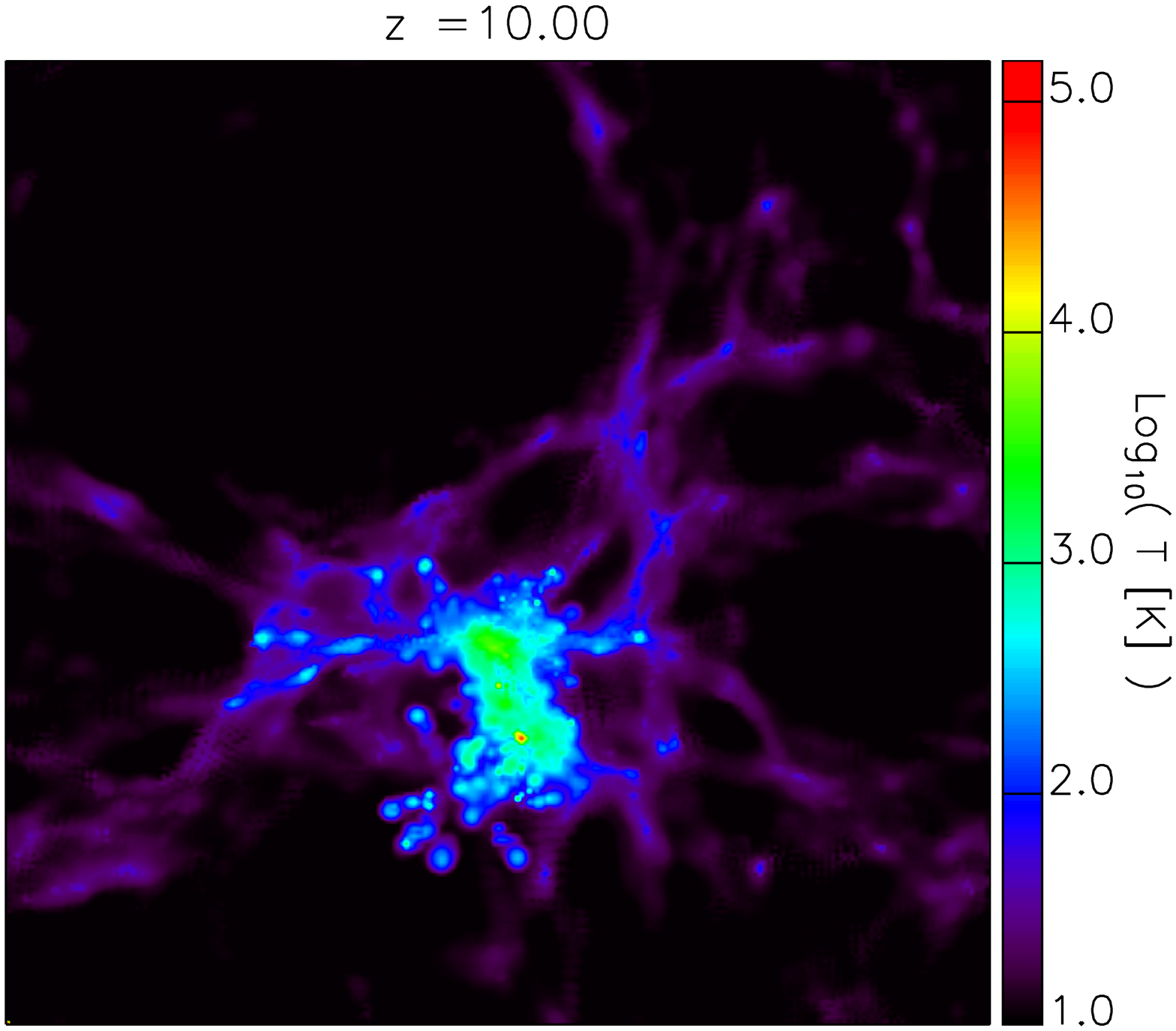}\hspace{-1.45cm}
\includegraphics[width=0.35\textwidth]{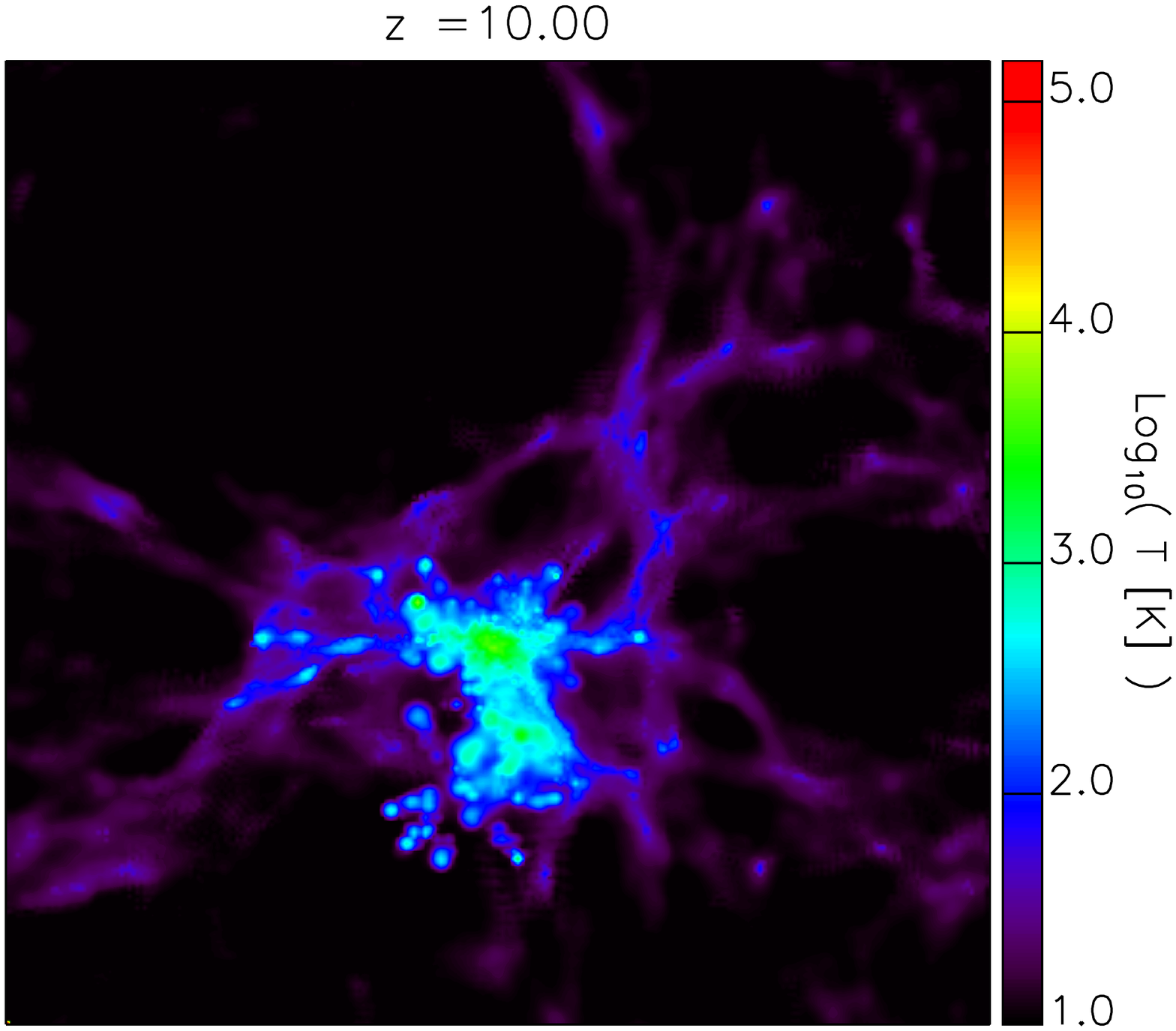}\\
\vspace{-1cm}
\caption[]{\small
  Temperature maps at different redshifts for the simulations having Salpeter-like  popIII and popII-I IMFs. Results refer to the no-RT run (left) and to the runs including RT according to black-body spectra with effective temperature $\rm T_{eff} = 4\times10^4\,K$ (center) and $\rm T_{eff} = 10^4\,K$ (right) for both popIII and popII-I sources.
}
\label{fig:maps.lowmass}
\end{figure*}

Visual representations of the simulated boxes are displayed through temperature maps in Fig.~\ref{fig:maps} (for massive and hot stellar sources) and Fig.~\ref{fig:maps.lowmass} (for more standard sources).
We start by investigating the former case in the next Sect.~\ref{Sect:Tmaps} and the latter case in the following Sect.~\ref{Sect:Tmaps.lowmass} with a discussion of the consequent implications for both entropy, in Sect.~\ref{Sect:entropy}, and IGM, in Sect.~\ref{Sect:IGMimplications}.

\subsubsection{Massive hot stellar sources}\label{Sect:Tmaps}

In Fig.~\ref{fig:maps}, we focus on the redshift evolution of the two runs (with and without RT) with top-heavy popIII IMF over the range [100, 500]~$\msun$.
PopII-I IMF is assumed to follow a Salpeter shape for star forming gas parcels at metallicities $Z > 10^{-4} Z_\odot$.
On the left panels, results for the case without RT calculations is presented, while on the right ones we display the results for the run including RT.
The typical IGM temperatures are significantly different in the two cases and this is essentially due to the different treatment of radiative sources.
\\
In the case without RT, the cosmic gas keeps relatively low temperatures (below $\sim 10^2\,\rm K$) and only in the dense star forming regions thermal feedback mechanisms locally heat up the medium.
Since in this run RT is not considered, star formation has no direct effect on the thermodynamical state of the gas in lower density environments and in the IGM.
Impacts are found only in the neighbouring ambient medium that gets metal enriched.
On the other hand, the RT run (right) shows a very strong evolution with first bursts that shut down following star formation episodes and significantly heat the primordial IGM up to temperatures of at least $\sim 10^3-10^4\, \rm K$.
Due to the extremely short lifetimes of massive $\sim 10^2\,\rm \msun$-stars this process takes place already from the early star formation episodes that, at $z\sim 10$, result delayed in comparison to the no-RT simulation.
Denser shielded filaments are recognizable in the maps, as they feature lower temperatures and partially halt photon propagation.
\\
In the right-hand panels of Fig.~\ref{fig:maps} this is easily inferred from the greenish and yellowish areas at around $\sim 10^3-10^4\, \rm K$, in contrast to the orange and red colours of the lower-density hotter medium, whose maximum temperatures reach a few $10^4\,\rm K$.
Low-entropy cores detected at temperatures of $\lesssim 10^3\,\rm K$ and surrounded by warmer material are visible, as well.
Already at higher redshift ($z=10$) first star forming events in the RT and no-RT runs show some differences, since the regions affected by stellar feedback in the no-RT case (left) are more extended than in the RT case (right), where radiation from the very first short-lived stars inhibits subsequent star formation (see also discussion in Sect.~\ref{Sect:sfr}).
\\
In general, the thermodynamic behaviour of the cosmic gas sampled in the two volumes in presence of massive hot radiative sources is significantly different in the two cases. This suggests that simple feedback mechanisms (employed e.g. in the no-RT run) are not able to mimic or even partly reproduce the overall trends of the IGM structure.

\subsubsection{Standard stellar sources}\label{Sect:Tmaps.lowmass}

In Fig.~\ref{fig:maps.lowmass}, we show the expected evolution of IGM temperature in the case of a standard Salpeter IMF for both popIII and popII-I stellar populations.
This assumption implicitly states that there is no actual stellar mass transition when passing from a regime to another.
We exploit such scenario as an extreme counterpart of the previous case, because of the many uncertainties on the determinations of primordial stellar masses \cite[][]{Abel2002, Yoshida2006, Yoshida_et_al_2007,CampbellLattanzio2008,SudaFujimoto2010, Hosokawa2011, Stacy2012, Stacy2014}.
On the left panels, results for the no-RT case are shown, while on the central and right panels the corresponding evolutions for the RT cases are displayed, both for the assumption of $\rm T_{eff} =4\times 10^4\,\rm K$ (center) black-body emission and for the assumption of $\rm T_{eff} =10^4\,\rm K$ (right).
\\
The global large-scale structure is evident in all the maps and star forming regions show only little difference.
Indeed, the panels on the left referring to the no-RT case show a colder evolution with respect to the corresponding no-RT case of the previous Fig.~\ref{fig:maps}, as a consequence of the different energies emitted by the two types of popIII sources.
In the low-mass IMF case (Fig.~\ref{fig:maps.lowmass}) typical energies of $\sim 10^{51}\,\rm erg$ are mechanically injected into the surrounding medium, while in the top-heavy case (Fig.~\ref{fig:maps}) much larger energies (up to $\sim 10^{53}\,\rm erg$), can heat the environment and make the temperature drastically increase.
\\
When comparing the no-RT run of Fig.~\ref{fig:maps.lowmass} with the two runs employing radiation propagation, similar trends are found.
Despite the propagating fronts of ionizing radiation shaping the surrounding gas, the distribution of warm and hot material only slightly differs from the no-RT case (left), while cold low-density medium is basically unaffected. 
 \\
On the central panels, instead, radiation propagation determines a more irregular shape of the warm/hot material at any redshift since gas temperatures at $\sim 10^3-10^4\,\rm K$ are smeared out due to radiative feedback.
This is already readily seen at $z\sim 10$ when the thin (blueish/greenish) filaments of the no-RT run (left panels) are replaced by ``fatter'' and more extended blobs, led by RT fronts (center).
At later times, the different {\it loci} where star formation takes place are quite distinguishable from gas temperatures in the no-RT case (left), while they are more hidden by the radiative feedback in the corresponding RT runs.
\\
Similar conclusions hold for the right panels, as well.
There, early gas is also heated up by radiative feedback and the underlying thin filaments where star formation takes place are again covered by RT effects.
Despite the weaker sources, gas temperatures are quite similar to the ones on the central panels, although slightly colder (consistently with the source spectrum).
\\
We stress that, while the formation (see previous Sect.~\ref{Sect:Tmaps}) of massive hot stars with powerful emissions can heat up the whole box within a very short time ($\ll 10^8\,\rm yr$), the ongoing formation of standard low-mass stars is able to heat up only material near to the star forming regions.
Therefore, inclusion of detailed radiative transfer calculations from central sources appears to be extremely crucial in the case of powerful emissions, but less dramatic in case of more standard sources.

\subsubsection{Entropy}\label{Sect:entropy}

\begin{figure}
\centering
{\underline{\bf No RT} }\\
\vspace{-2cm}
\includegraphics[width=0.4\textwidth]{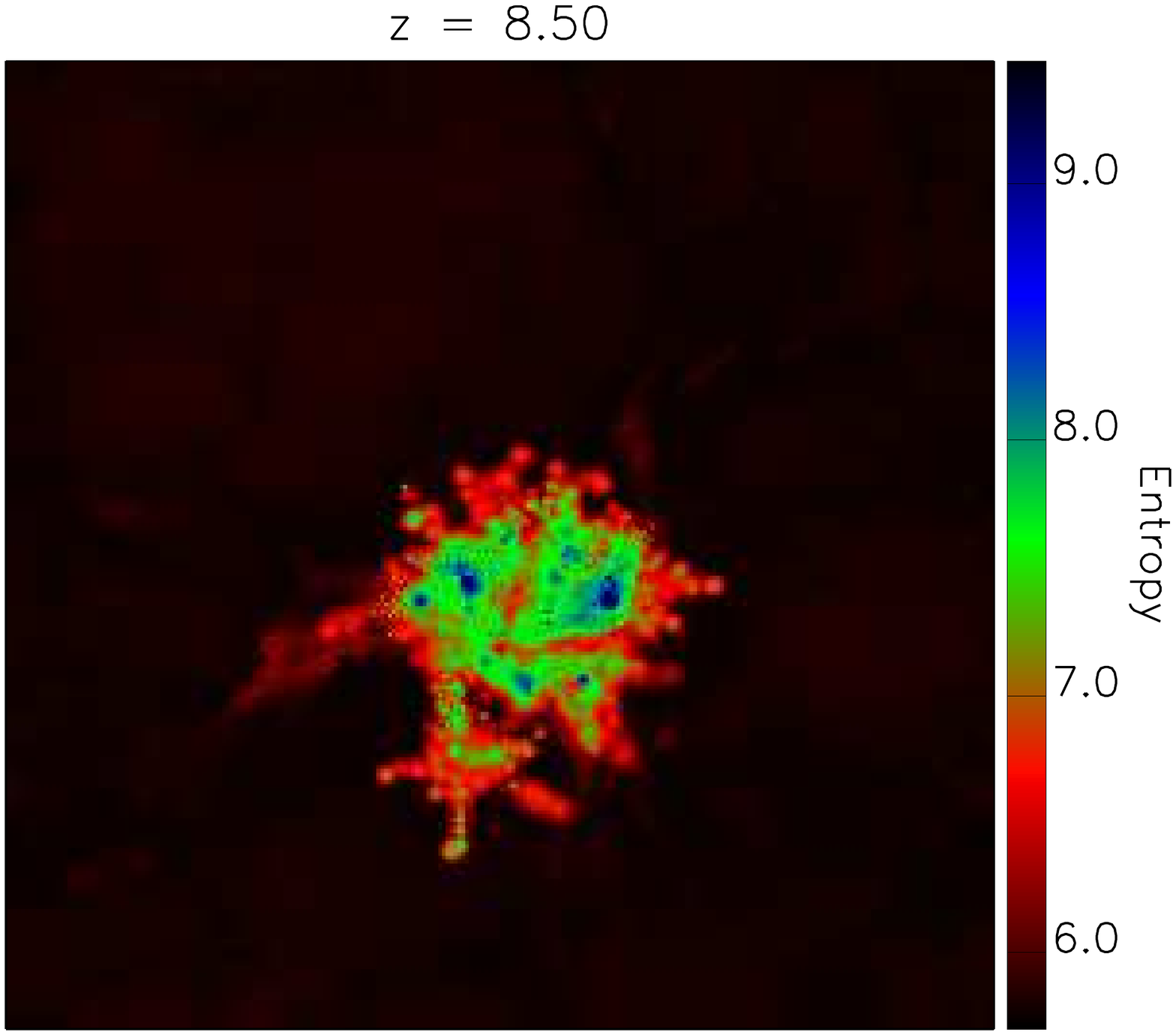}\\
\vspace{-1cm}
{\underline{\bf With RT} $\qquad$   ($\rm T_{\rm eff} = 10^5\,\rm K$)}\\
\vspace{-2cm}
\includegraphics[width=0.4\textwidth]{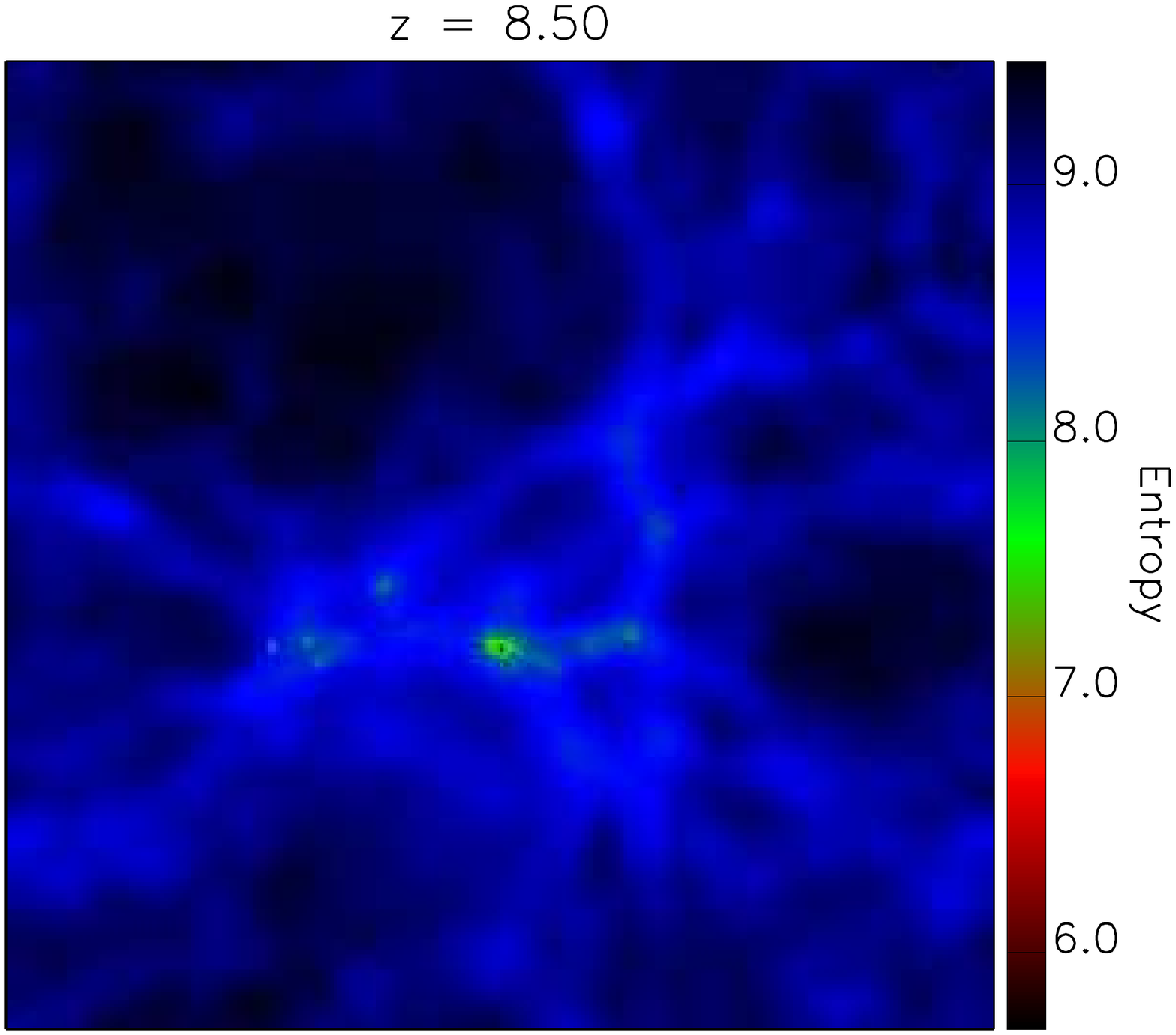}
\vspace{-1cm}
\caption[]{\small
Entropy (comoving) of the cosmic gas at redshift $z=8.50$ for the case with no RT implementation (upper) and the case including RT implementation (lower) with massive popIII sources. Units are in cgs and colour scale is logarithmic. Both panels have the same color range for a more direct comparison.
}
\label{fig:entropy}
\end{figure}

In Fig.~\ref{fig:entropy}, we check the entropic state of the gas for the no-RT and RT cases including powerful popIII sources at $z=8.50$.
We map the quantity $S=P/\rho^{\gamma}$, with $P$ pressure, $\rho$ gas density and $\gamma\simeq 5/3$ adiabatic index.
The differences are striking, as a result of the different temperature patterns in the two different models (compare to Fig.~\ref{fig:maps}).
\\
In the no-RT model diffuse cosmic gas features low entropy values due to the ongoing gas cooling during cosmic expansion.
Effects from clumped dense star forming regions do not reach the sites far away from the radiative sources,  because of the lack of photon propagation.
Nevertheless, gas parcels near star forming regions are locally heated by feedback effects and this drives larger entropy values.
In such environments the entropy contrast is up to 2 orders of magnitude and this allows us to recognise the underlying colder material with low entropy (red) against the rarefied hot gas shock-heated by star formation (green and blue).
\\
In the RT case, IGM gas gets heated through the more accurate treatment of photons and hence its entropy values are more-than-one-dex larger than the previous case.
Gas recombination and cooling in denser regions bring down the temperatures and also in this case it is possible to recover the underlying filamentary patterns (blue) and the cold collapsing regions (green) traced by the lowest entropy values.
The trends found here are consistent with Fig.~\ref{fig:maps}.
\\
As before, the corresponding maps for weaker sources are similar to the no-RT case, thus we do not show them  for simplicity.

\subsubsection{Implications on early IGM structure}\label{Sect:IGMimplications}

\begin{figure*}
\centering
{\underline{\bf No RT}}\\
\includegraphics[width=0.25\textwidth]{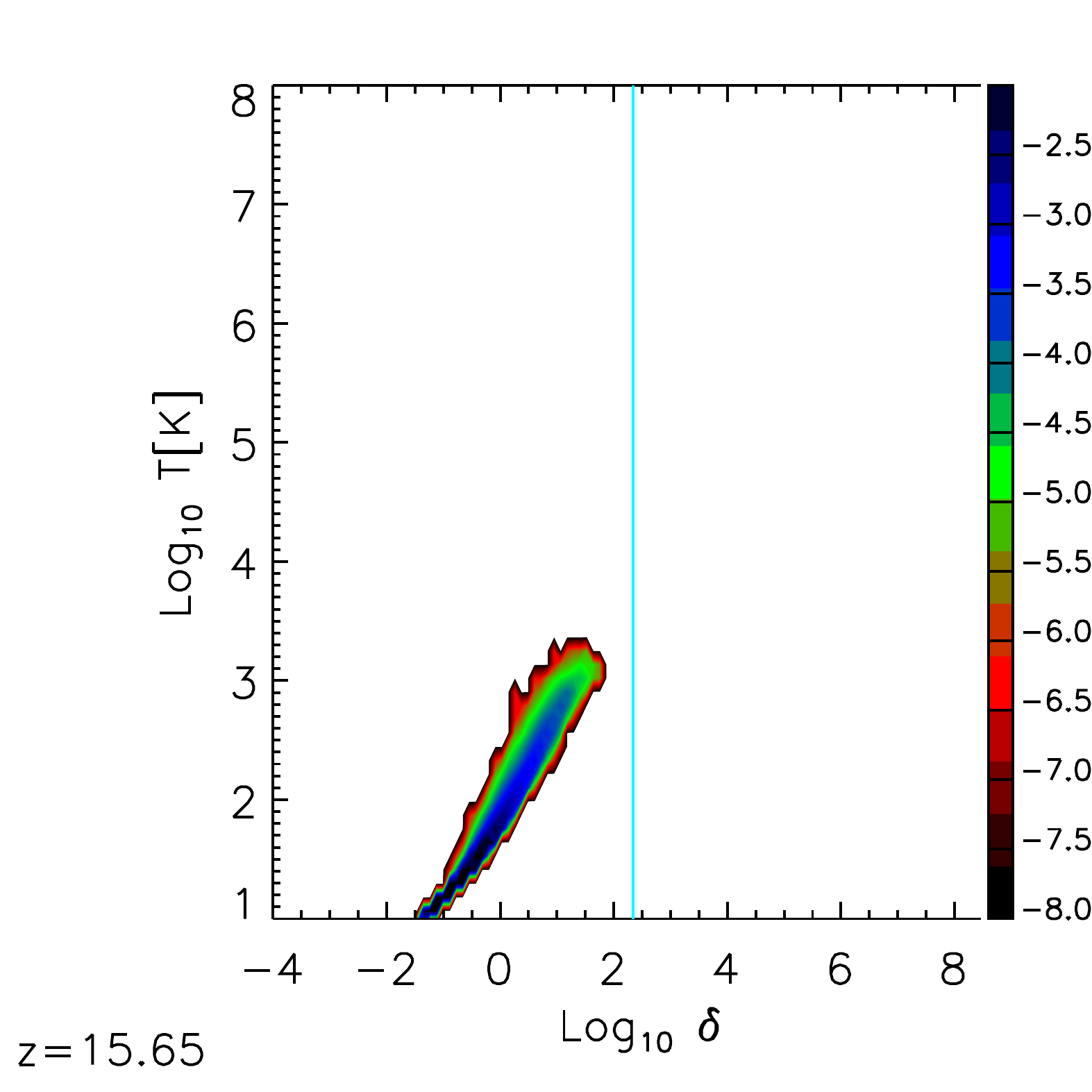}
\includegraphics[width=0.25\textwidth]{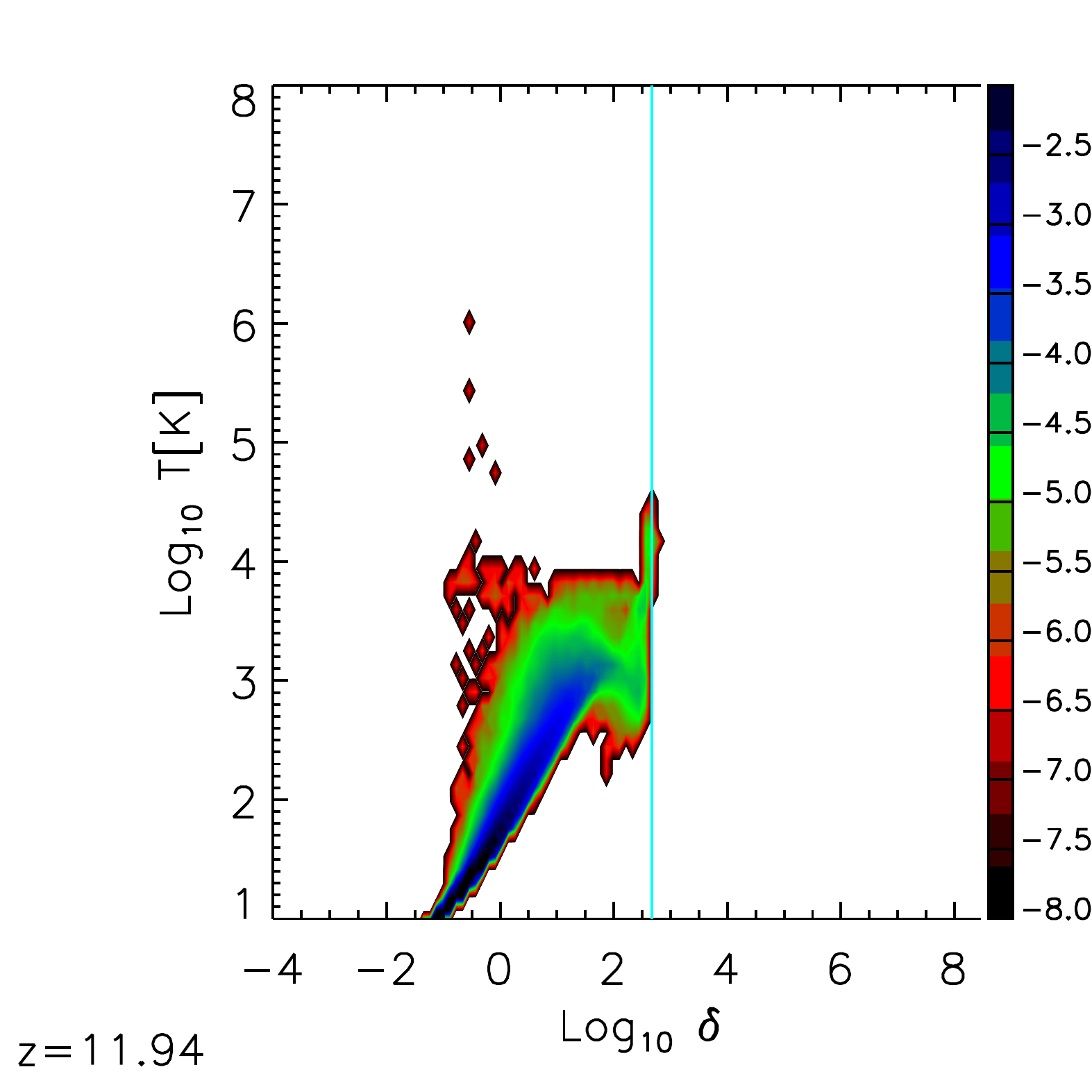}
\includegraphics[width=0.25\textwidth]{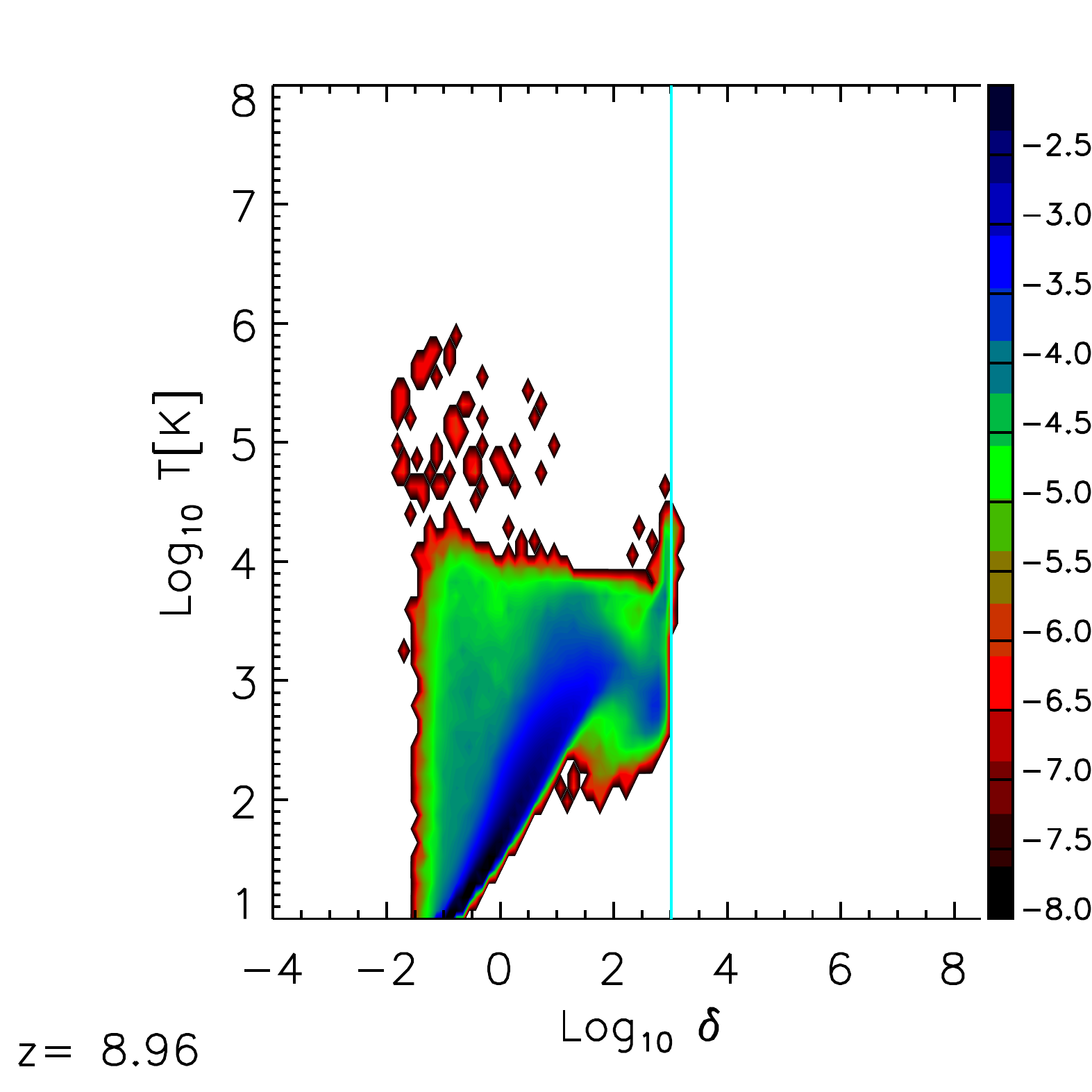}
\\
{\underline{\bf With RT}}\\
\includegraphics[width=0.25\textwidth]{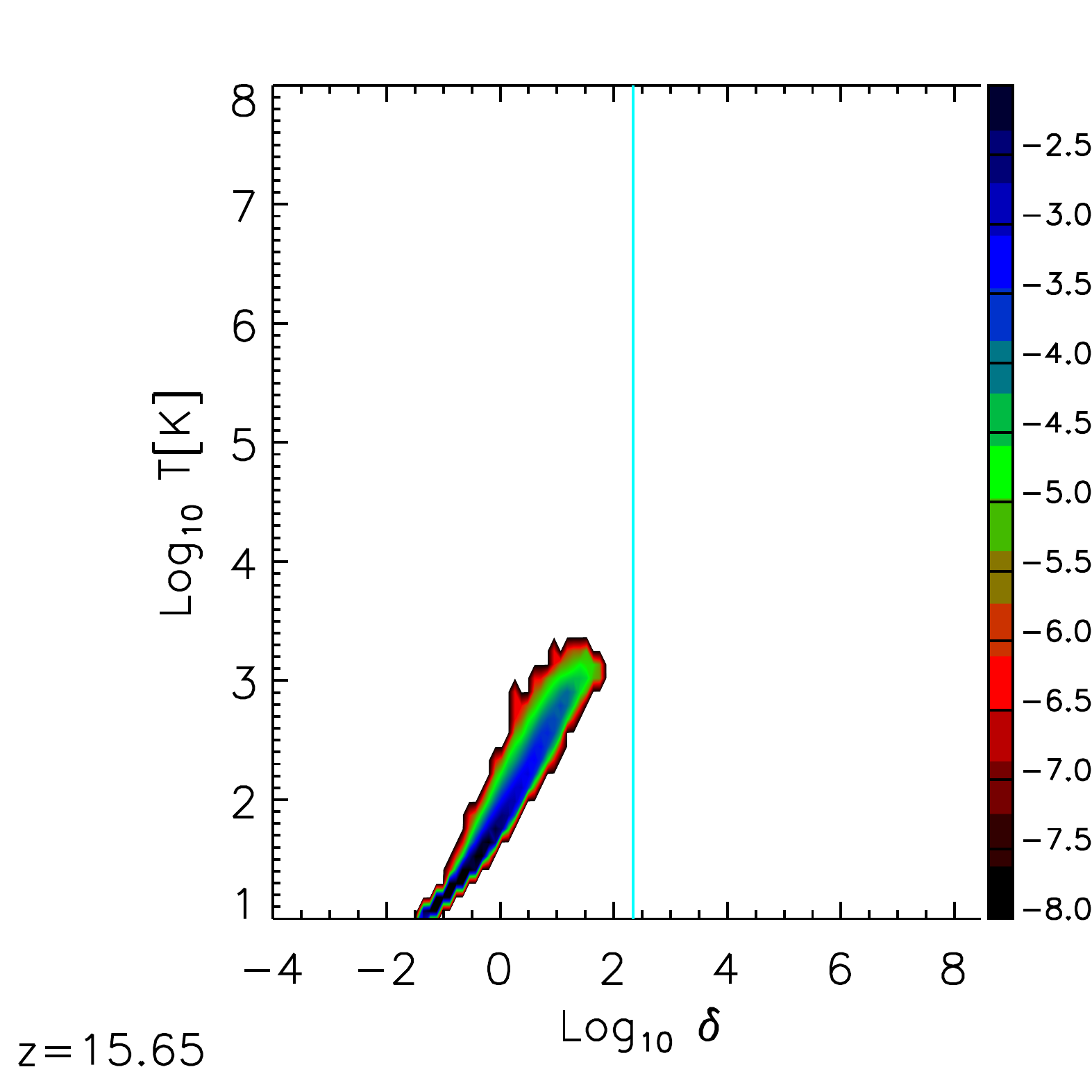}
\includegraphics[width=0.25\textwidth]{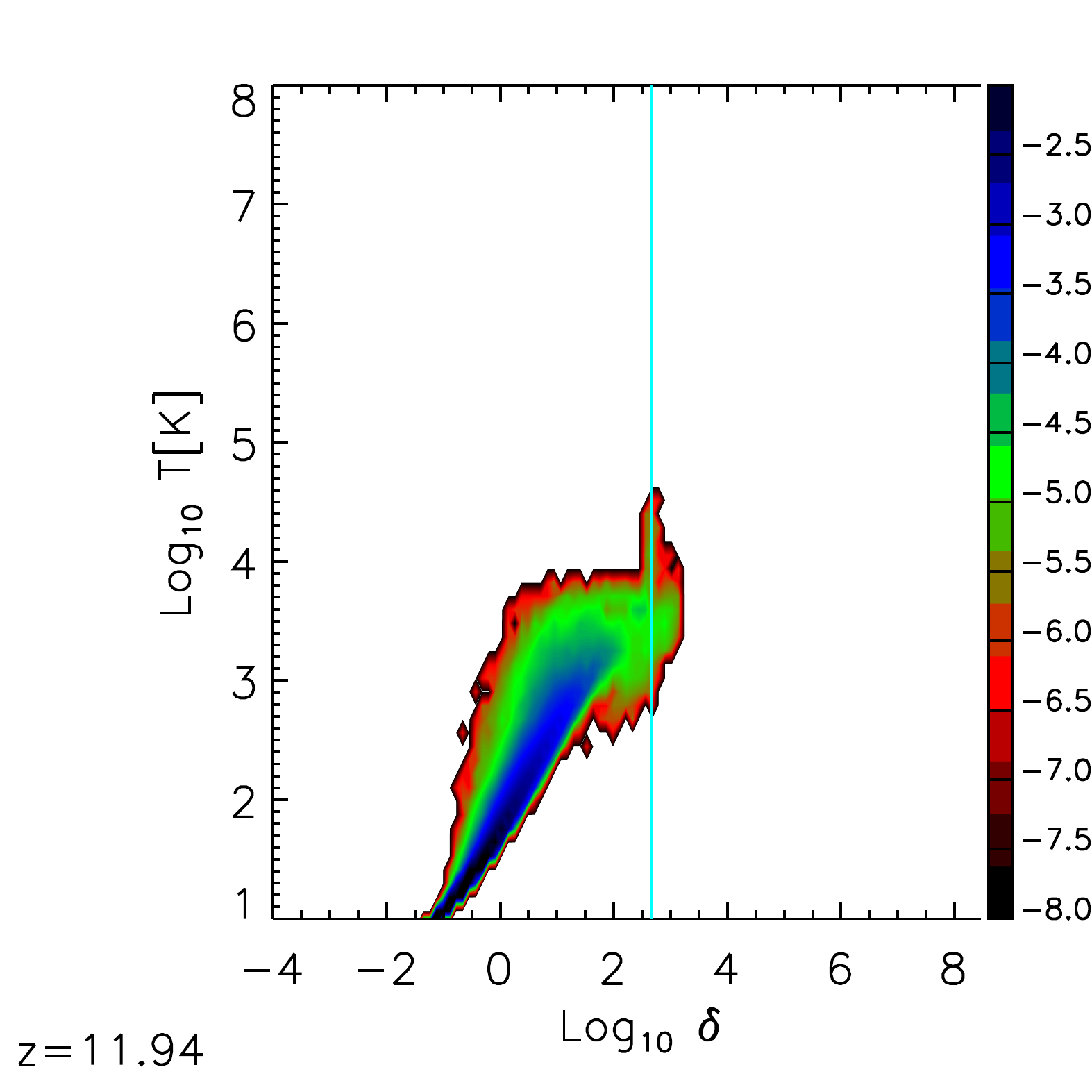}
\includegraphics[width=0.25\textwidth]{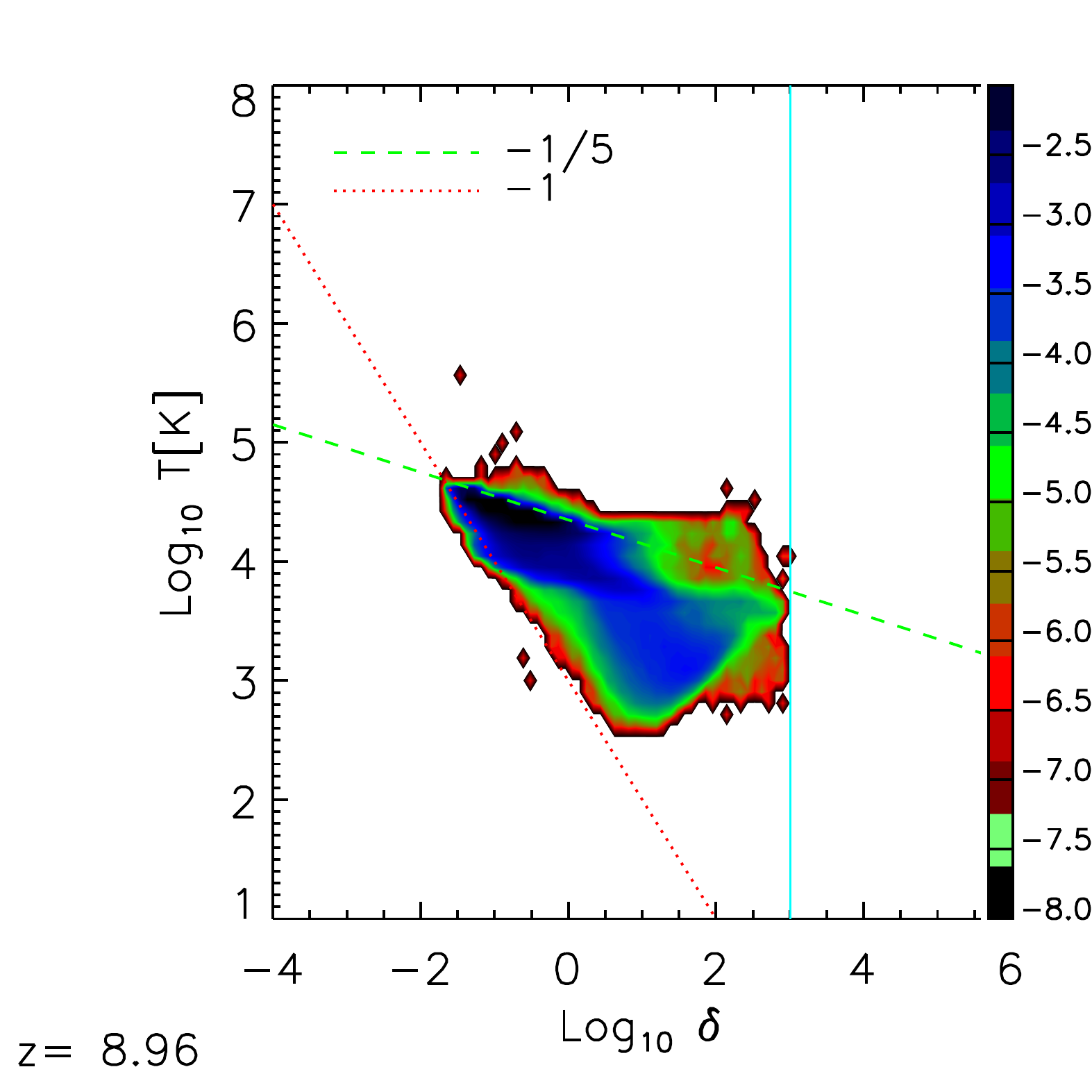}
\caption[]{\small
Phase diagrams (over-density versus temperature) at three different redshifts for the cases with no RT implementation (upper) and the cases including RT implementation (lower) with massive popIII sources. 
The vertical cyan solid line represents the star formation over-density threshold (see text).
In the bottom right panel, the red dotted line marks a slope of $-1$, while the green dashed line marks a slope of $-1/5$.
The colour coding refers to the probability density function in each panel.
}
\label{fig:phase}
\end{figure*}
Radiative feedback has immediate effects on the thermodynamical state of the gas, because it influences both its density and temperature, with obvious consequences on derived quantities such as the resulting IGM equation of state and Jeans mass.
\\
In order to explore and understand more in depth the status of primordial medium and its global behaviour, as emerging from the complex interplay of thermodynamics and feedback effects, we address the trends of gas density and temperature via phase diagrams at different cosmological epochs for the different cases considered.
In Fig.~\ref{fig:phase} we plot temperature, $T$, versus overdensity, $\delta$, at $z =~$15.6, 11.94 and 8.96, for the top-heavy popIII IMF no-RT and RT runs.
Data are gridded and colour-coded according to the resulting probability density function.
\\
At early times, the infall into dark-matter halos determines gas shock-heating up to $\sim 10^4\rm~K$, with a slope of $\sim 2/3$ and with a following loitering phase due to the counterbalance of gas atomic cooling. Then, molecules form and lead a run-away collapse, with consequent star formation (taking place above the density threshold of $\rm 1~cm^{-3}$ highlighted by the vertical line in the plots) and feedback effects.
These latter are mainly responsible for the shape of the phase diagrams at later times, when the RT and the no-RT runs show significant deviations.
Indeed, in the no-RT case of Fig.~\ref{fig:phase} the hotter particles, at $\sim 10^5-10^6\rm~K$, have been heated and pushed into low-density environments by the sub-grid SN feedback.
In fact, collapsing gas parcels are stochastically converted into stars which explode and inject entropy in the surrounding regions by heating the gas and kicking it away from its original site.
These processes act only on the dense, star forming particles, though, and have no effect on the quiescent inter-galactic medium, with $\rm T < 10^4\rm~K$ and $\delta \ll 10^2$.
On the other hand, in the RT run radiation from stellar sources propagates and acts also on the lower-density gas heating it up to $\sim 10^3-10^4\rm~K$.
Since kinetic and thermal feedback effects from SNe are included in the same way in the two runs, the heating of low-density gas is entirely due to this additional radiative feedback.
This is sufficient to explain the inverted trend in the latter case.
Higher-density gas is instead less heated, due to more shielding along filaments and in clumped collapsing regions.
Such inversion in the effective equation of state of the IGM was also suggested from studies of the flux distribution in the Ly{$\alpha$} forest \cite[e.g.][]{Bolton2008, Bolton2009} with a typical IGM slope of $\sim -1$ \cite[see also][]{McQuinn2009}.
Therefore, according to our results, radiative feedback from powerful radiative sources represents a viable mechanism to invert the IGM equation of state.
The slope of the temperature-density relation changes from roughly $2/3$ at early times and in the no-RT case, to values bracketed between $\sim -1$ and nearly $-1/5$ (see straight lines in the panel) in the RT case.\\
We note that in the cases of milder standard sources the phase diagram experience little modifications with respect to the no-RT run and no inversion takes place.
\\
\begin{figure*}
\centering
{\underline{\bf No RT}}\\
\includegraphics[width=0.25\textwidth]{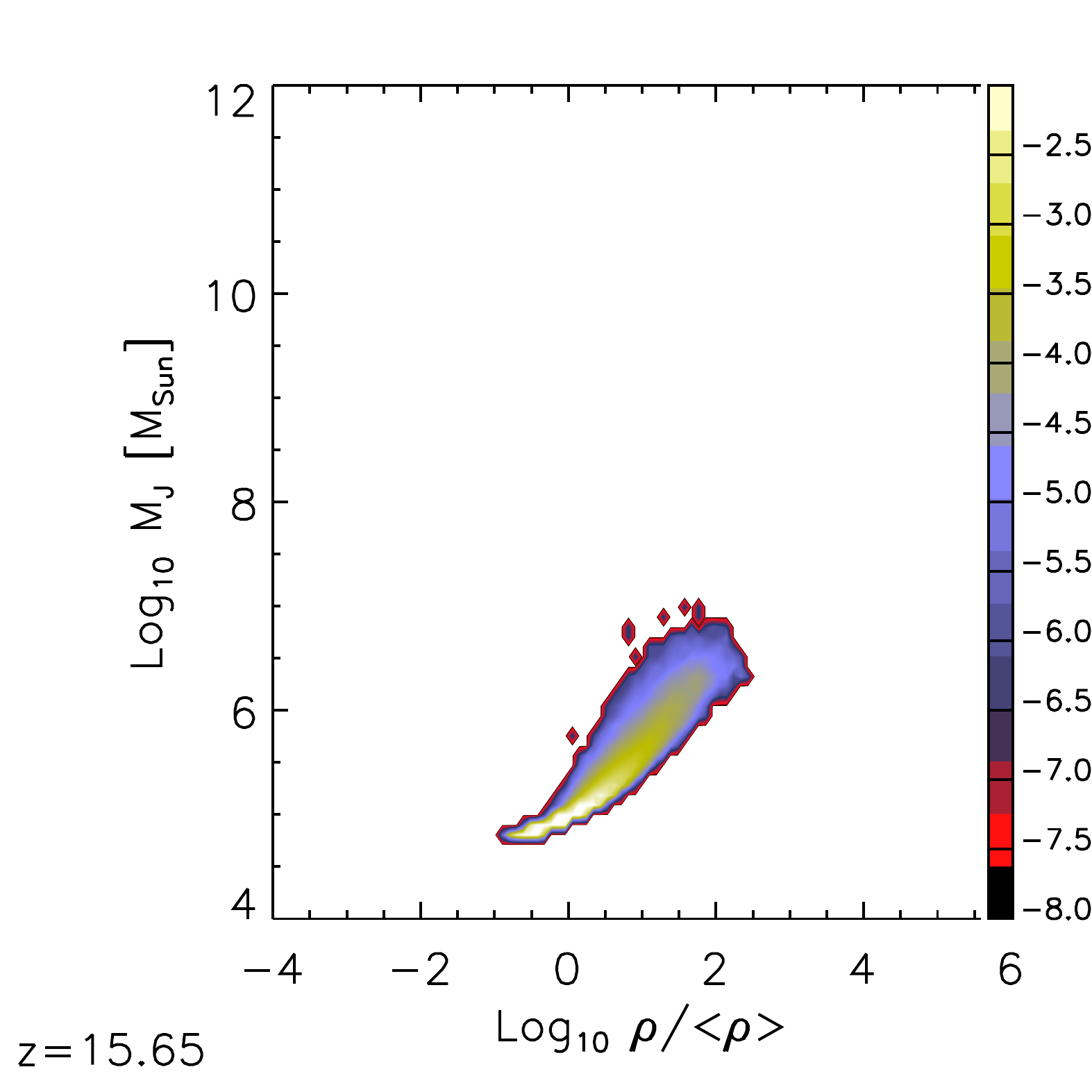}
\includegraphics[width=0.25\textwidth]{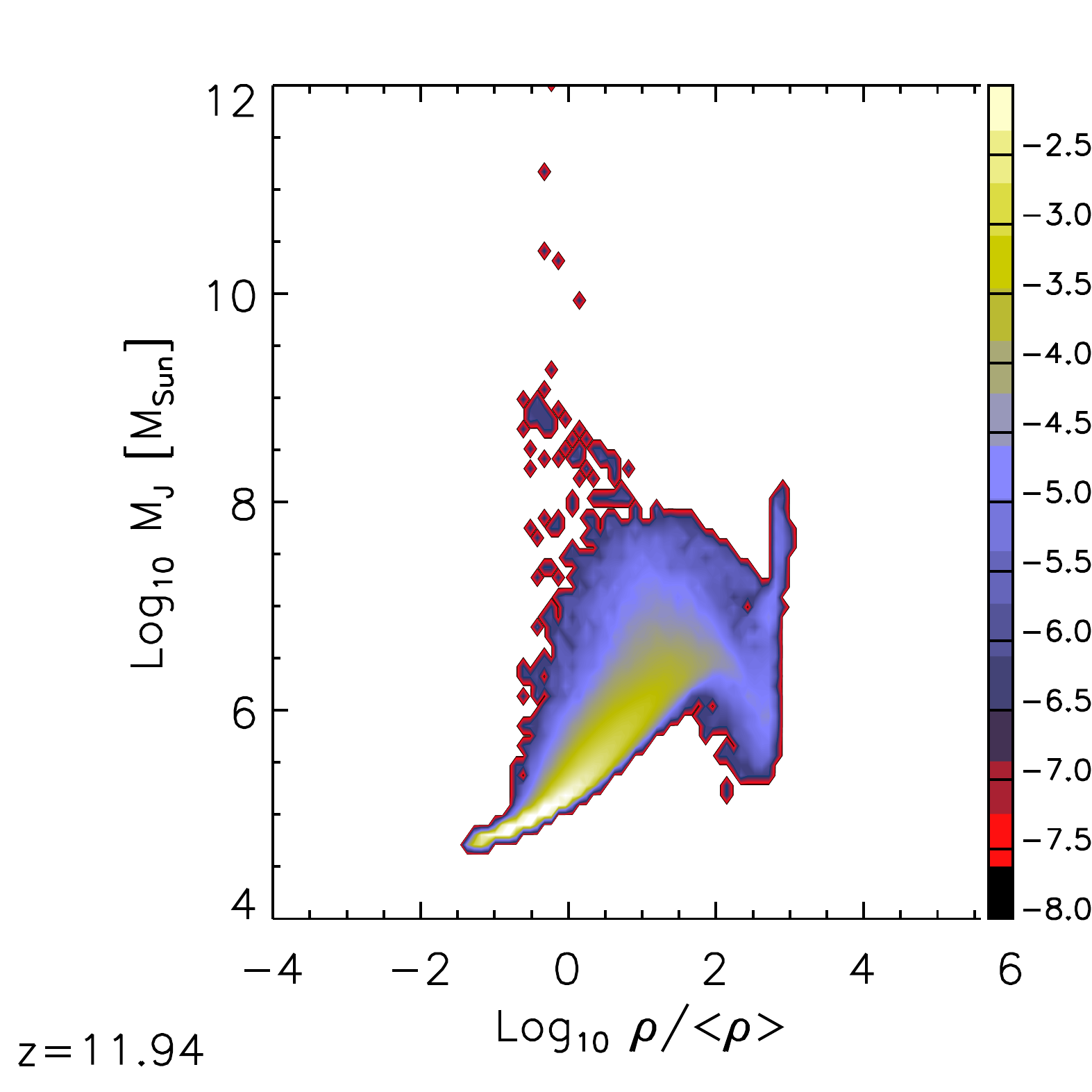}
\includegraphics[width=0.25\textwidth]{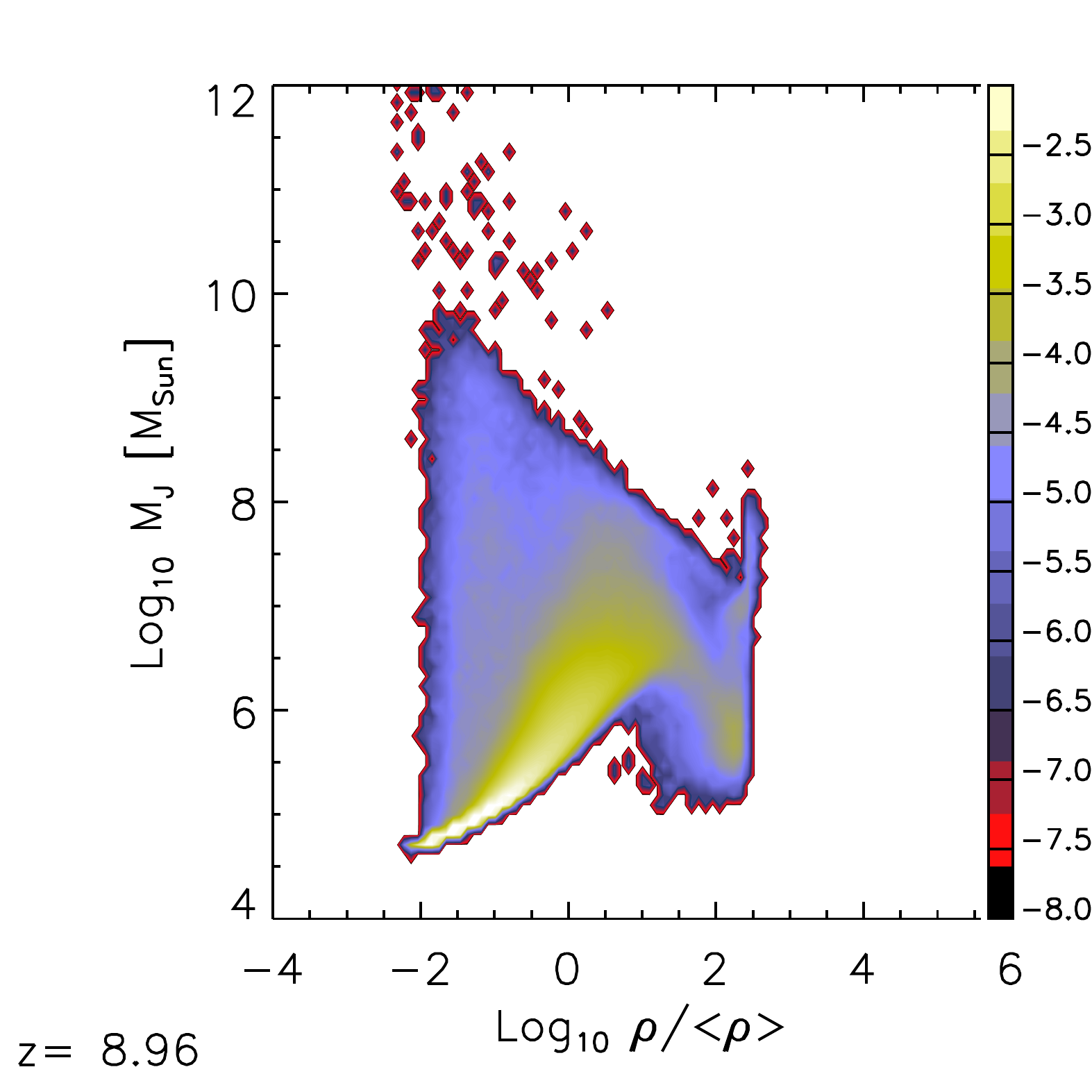}
\\
{\underline{\bf With RT}}\\
\includegraphics[width=0.25\textwidth]{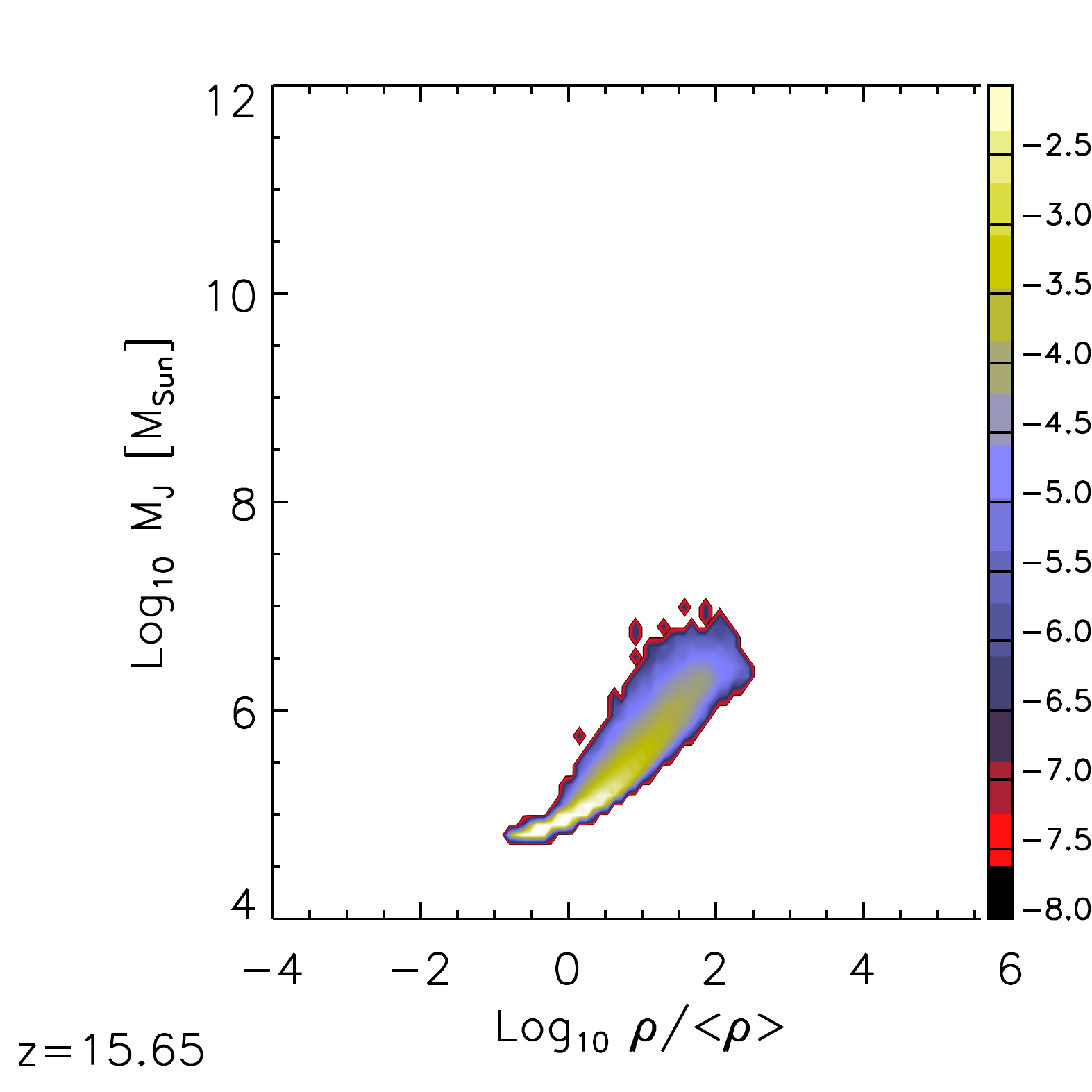}
\includegraphics[width=0.25\textwidth]{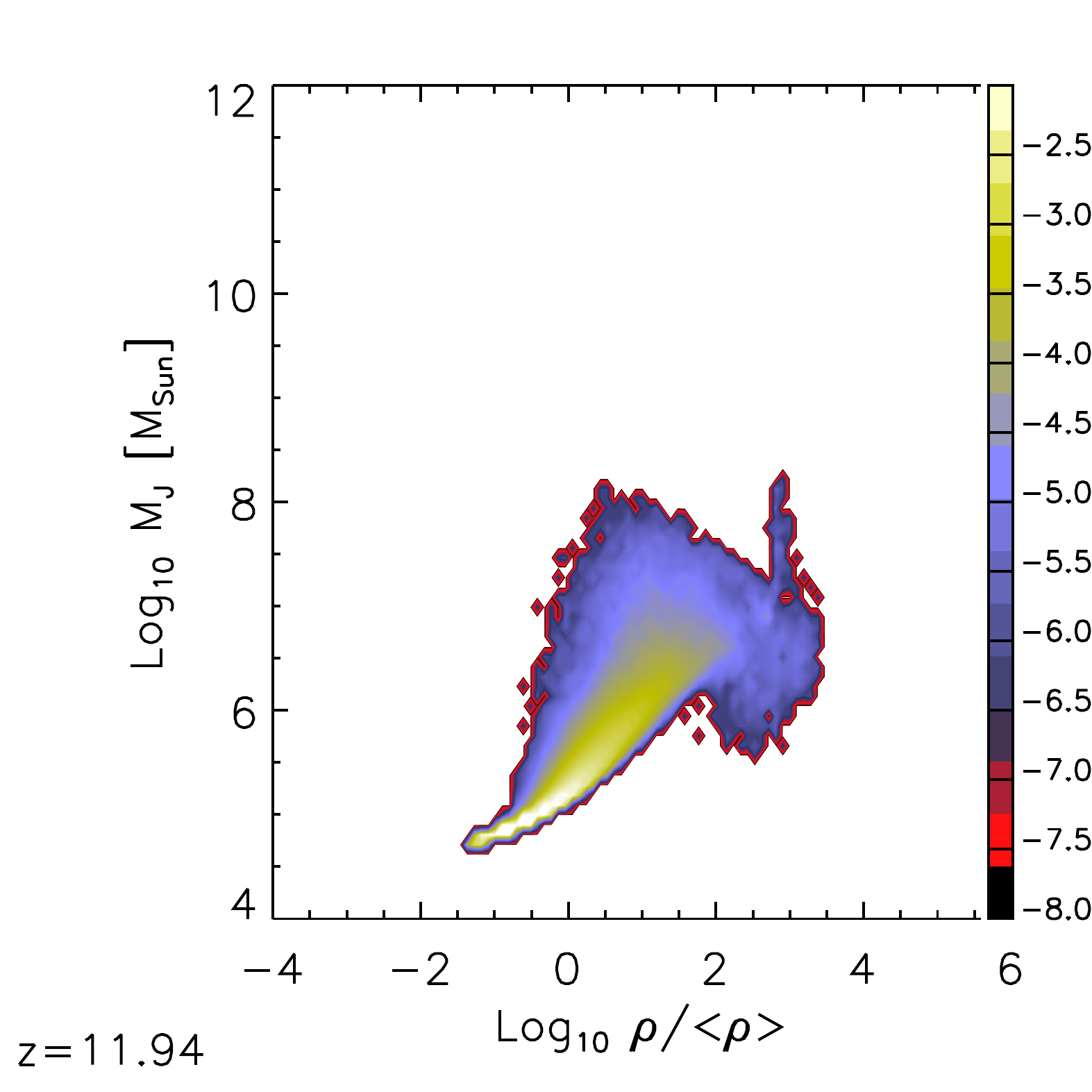}
\includegraphics[width=0.25\textwidth]{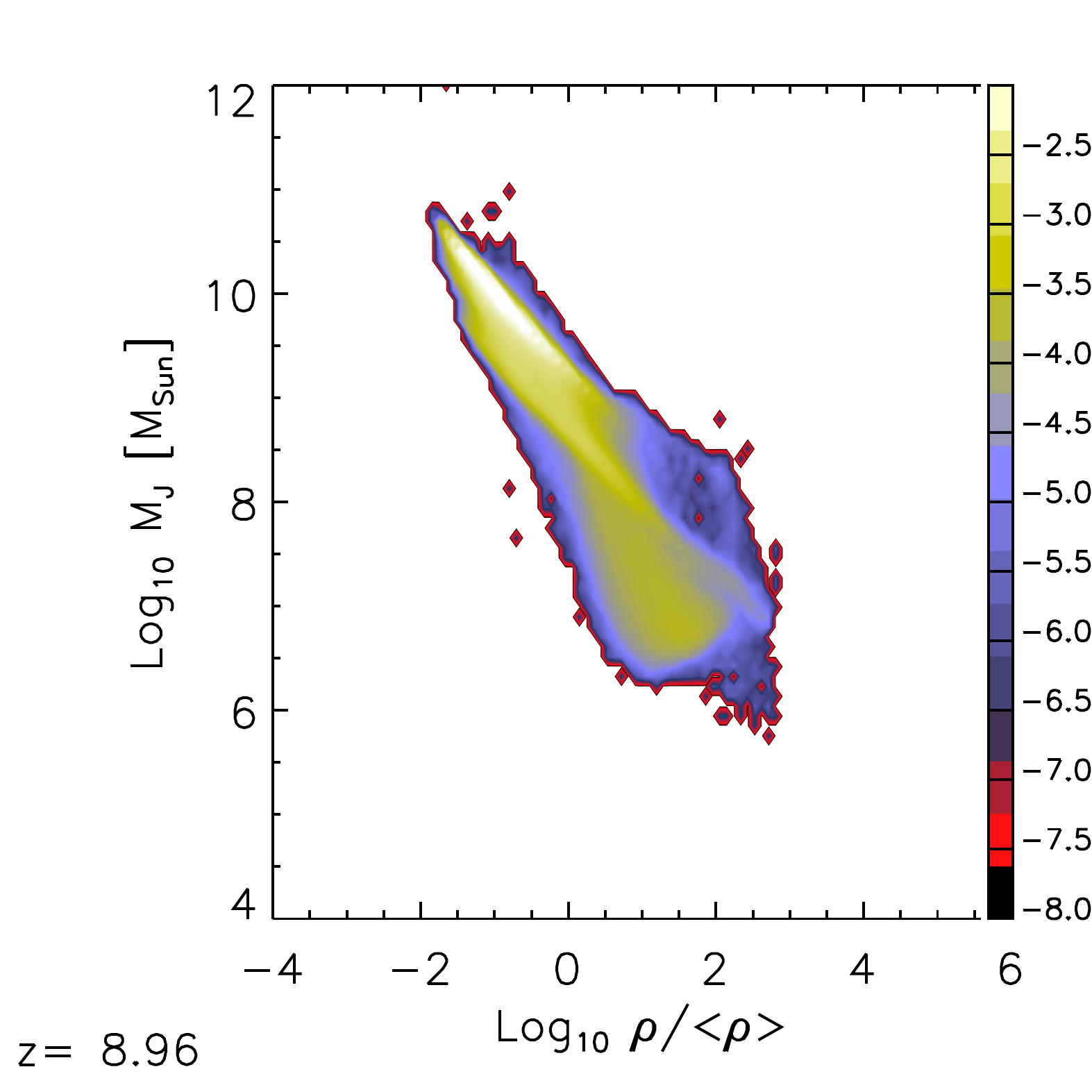}
\caption[]{\small
Jeans mass versus density at three different redshifts for the cases with no RT implementation (upper) and the cases including RT implementation (lower) with massive popIII sources.
The colour coding refers to the probability density function in each panel.
}
\label{fig:MjeansRho}
\end{figure*}
Fig.~\ref{fig:MjeansRho} shows the expected values for the Jeans mass, $M_J$, at the same redshifts as Fig.~\ref{fig:phase}, colour-coded according to the corresponding probability distribution.
The Jeans mass is an interesting quantity because it tells us the typical gas masses able to collapse at fixed density and temperature.
Also in Fig.~\ref{fig:MjeansRho} we refer to the no-RT run and the RT run including massive popIII sources, while we neglect the RT runs with low-mass popIII sources, because their differences from the no-RT run are minor.
The trends in the figure feature early $M_J$ values around $\sim 10^5-10^7\,\rm M_\odot$, which are the usual values expected in primordial mini-haloes \cite[][]{BiffiMaio2013, Wise2014}.
By redshift $z\sim 12$, $M_J$ increases above $\sim 10^8\,\rm M_\odot$ as a result of the increased temperatures of shock-heated gas.
Some parcels of gas in the no-RT run reach values as high as $\gtrsim 10^{10}\,\rm M_\odot$ due to the previously discussed feedback effects and the more advanced stages of star formation that is not slowed down by radiative feedback (photoionization and photodissociation of the surrounding gas).
The changes are even more evident from the panels at redshift $z=8.96$ where the RT case demonstrates $M_J$ values mostly around $\sim 10^6-10^{11}\,\rm M_\odot$ for low-density gas ($\delta <1$).
This means that smaller gaseous structures at these densities are unlikely to start collapsing processes.
The situation is completely different in the no-RT scenario, where $M_J$ values span a wide range of $\sim 7$ orders of magnitude and fragmentation processes are in principle allowed down to $\sim 10^4\,\rm M_\odot$ scales.
Therefore, radiative feedback from massive sources plays a remarkable role for the IGM structure and seriously affects gas fragmentation capabilities.
\\
Radiative feedback from regular sources (both with $T_{\rm eff } = 10^4\,\rm K$ and $T_{\rm eff } = 4\times 10^4\,\rm K$ black-body spectra) is less efficient and the photoionization effects remain confined in a very local volume, so that no distinct large-scale features appear with respect to the no-RT case. 
Without powerful sources it becomes rather difficult to explain reionization in these epochs.
\\
An intriguing conclusion (see more in the discussion section) can also be drawn about low-mass dwarf galaxies.
They have masses comparable to the limits on the Jeans mass derived for the RT scenario and, hence, could be explained as fossils of reionization (in this case led by very powerful massive popIII stars).
Dwarf galaxies could have been assembled from gas collapse before reionization and their growth would have been eventually stopped by cosmic gas (re)heating due to the explosion of early massive SNe \cite[][]{Simpson2013}.
In this respect, the inversion of the IGM equation of state might be a hint of the formation of these fossil structures.

\subsection{Chemistry evolution}\label{Sect:Hmaps}
We now discuss the main impacts of radiative feedback on the principal species that are relevant for early gas cooling, collapse and IGM ionization state, such as H, He, H$_2$, HD.
Radiative impacts are interesting, because chemical abundances can be severely altered and hence the whole baryonic structure formation history could result strongly affected.
cd
\begin{figure*}
  \centering
  \underline{\bf No RT} (top-heavy popIII IMF)\\
  \vspace{-1.5cm}
  \includegraphics[width=0.33\textwidth]{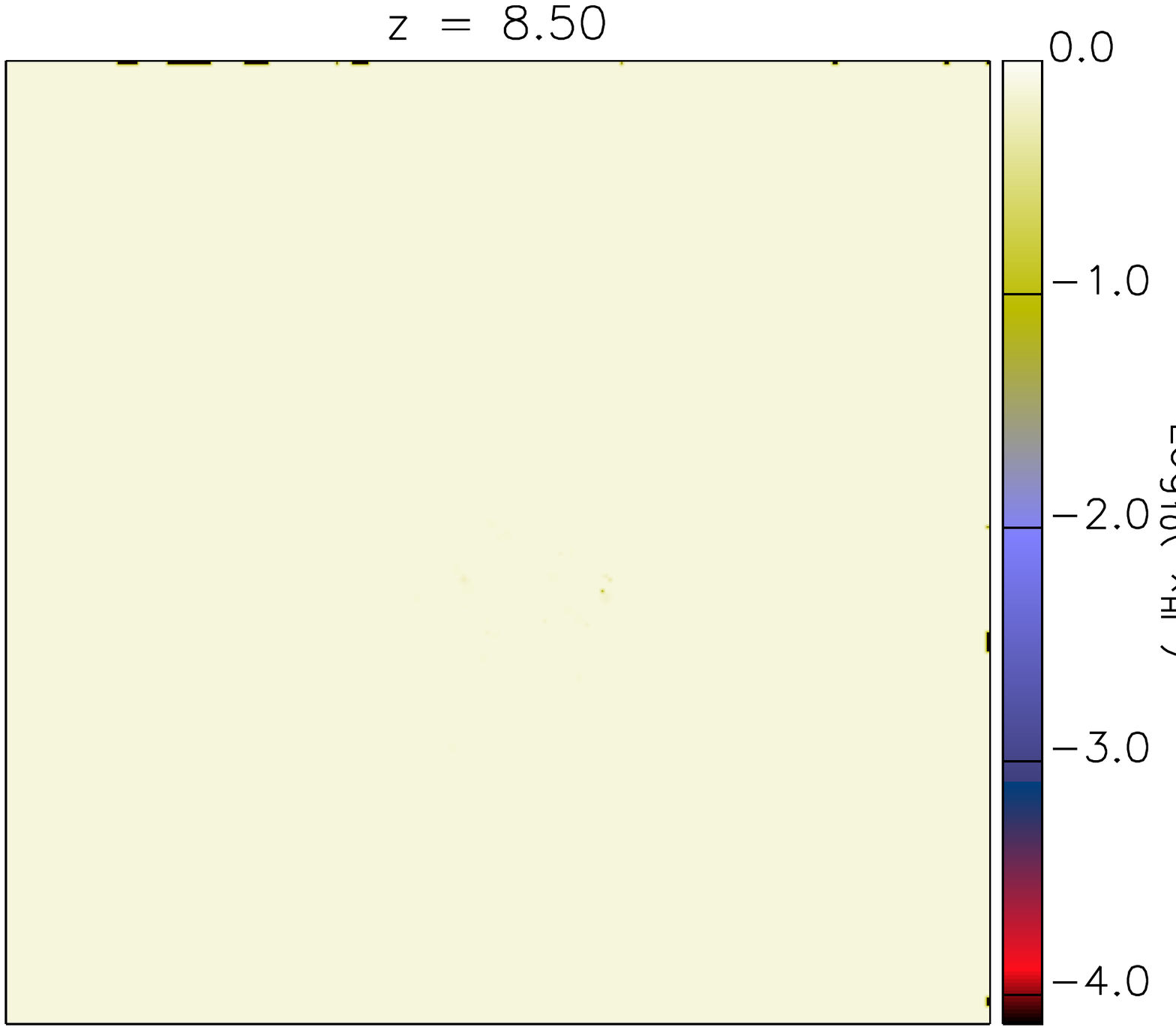}
  \includegraphics[width=0.33\textwidth]{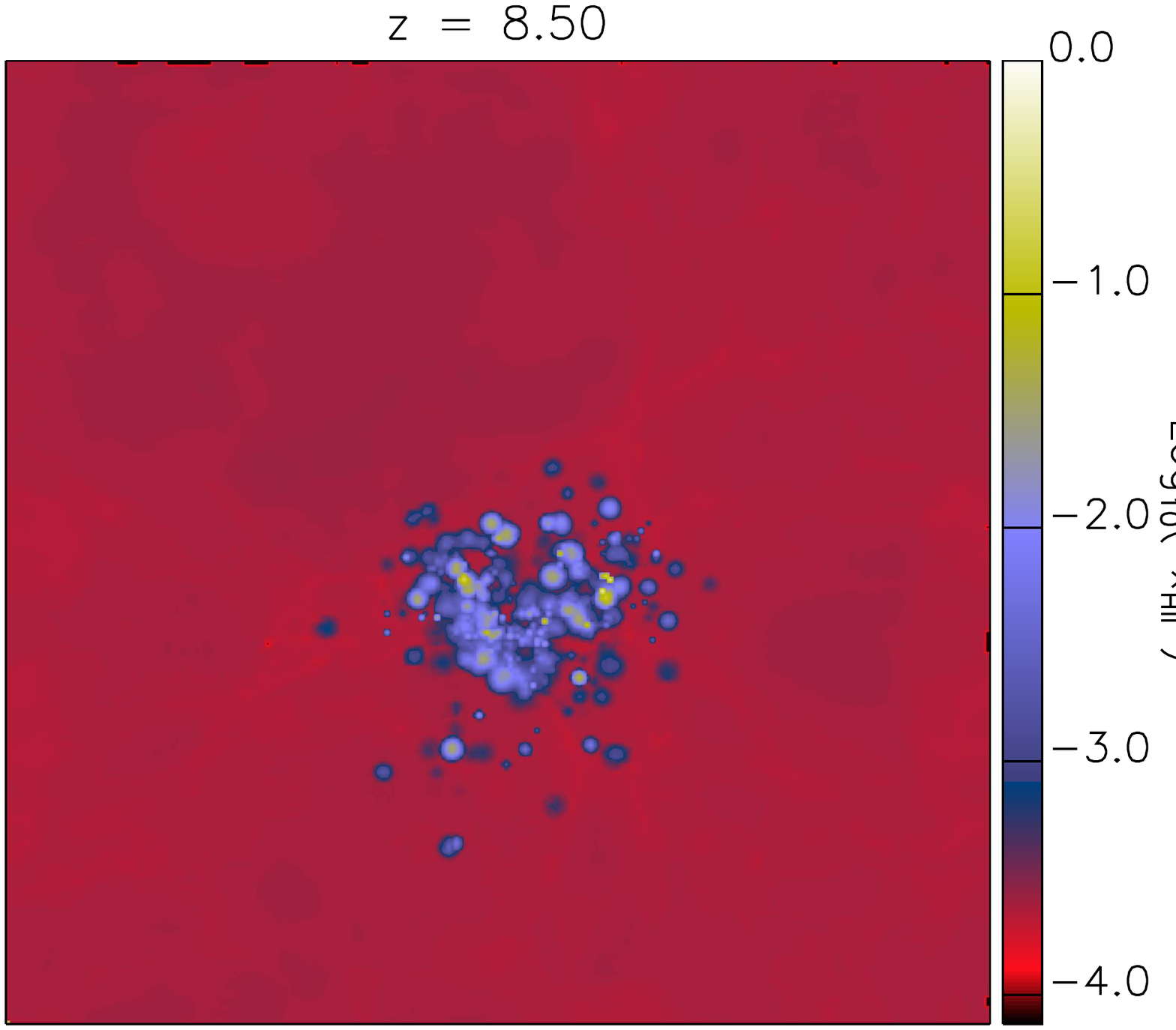}
  \includegraphics[width=0.33\textwidth]{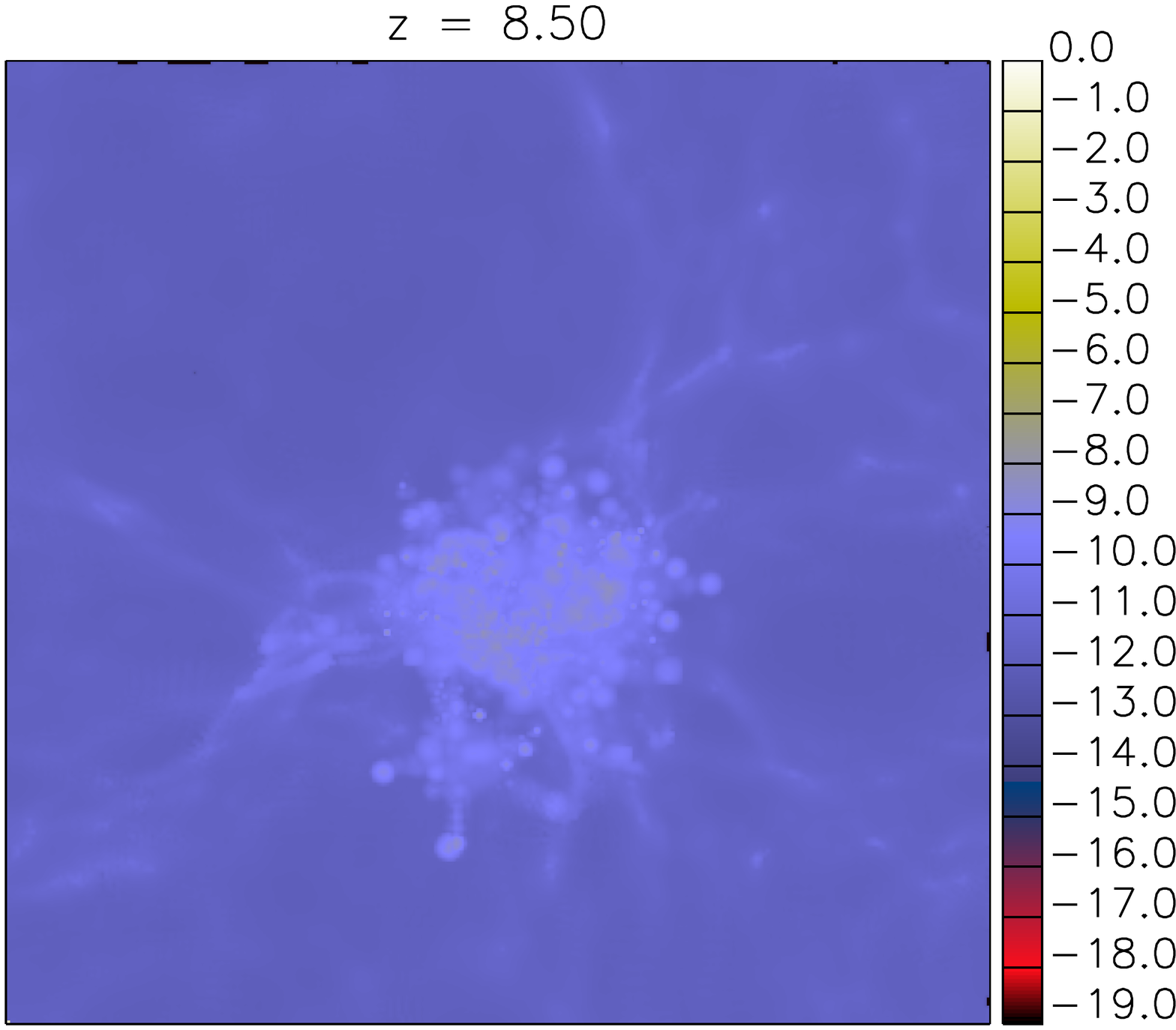}\\
    \vspace{-2cm}
  \includegraphics[width=0.33\textwidth]{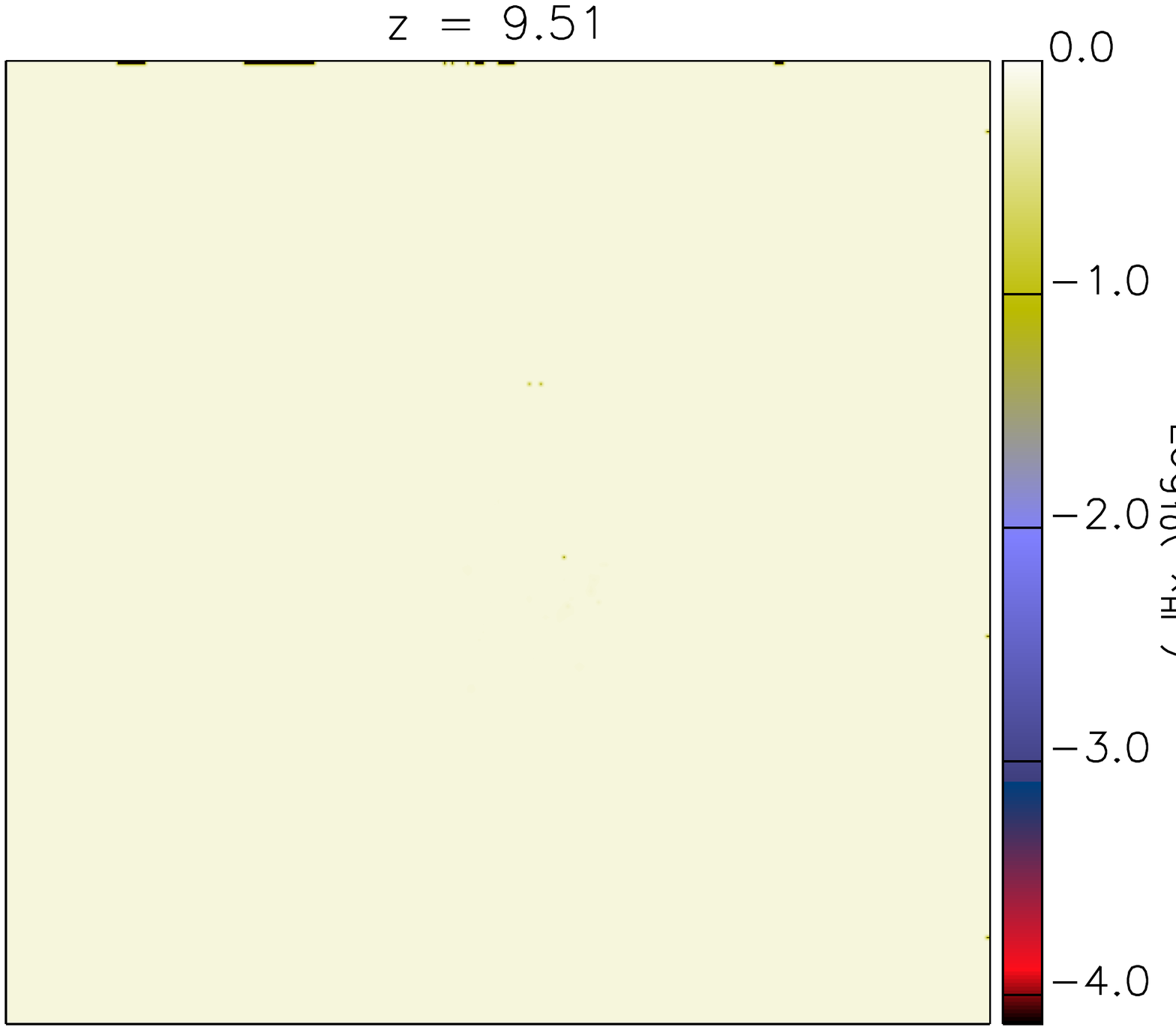}
  \includegraphics[width=0.33\textwidth]{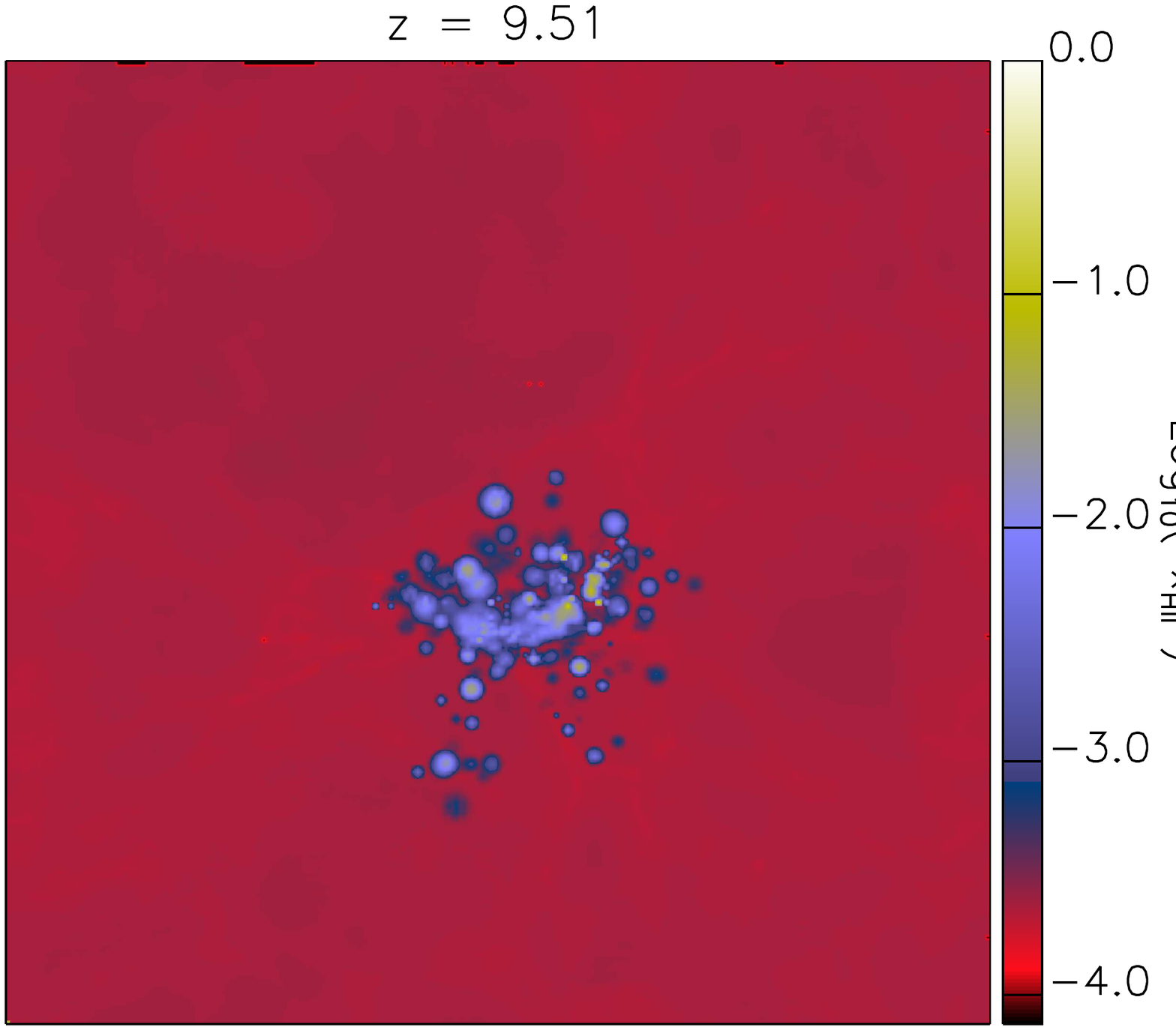}
  \includegraphics[width=0.33\textwidth]{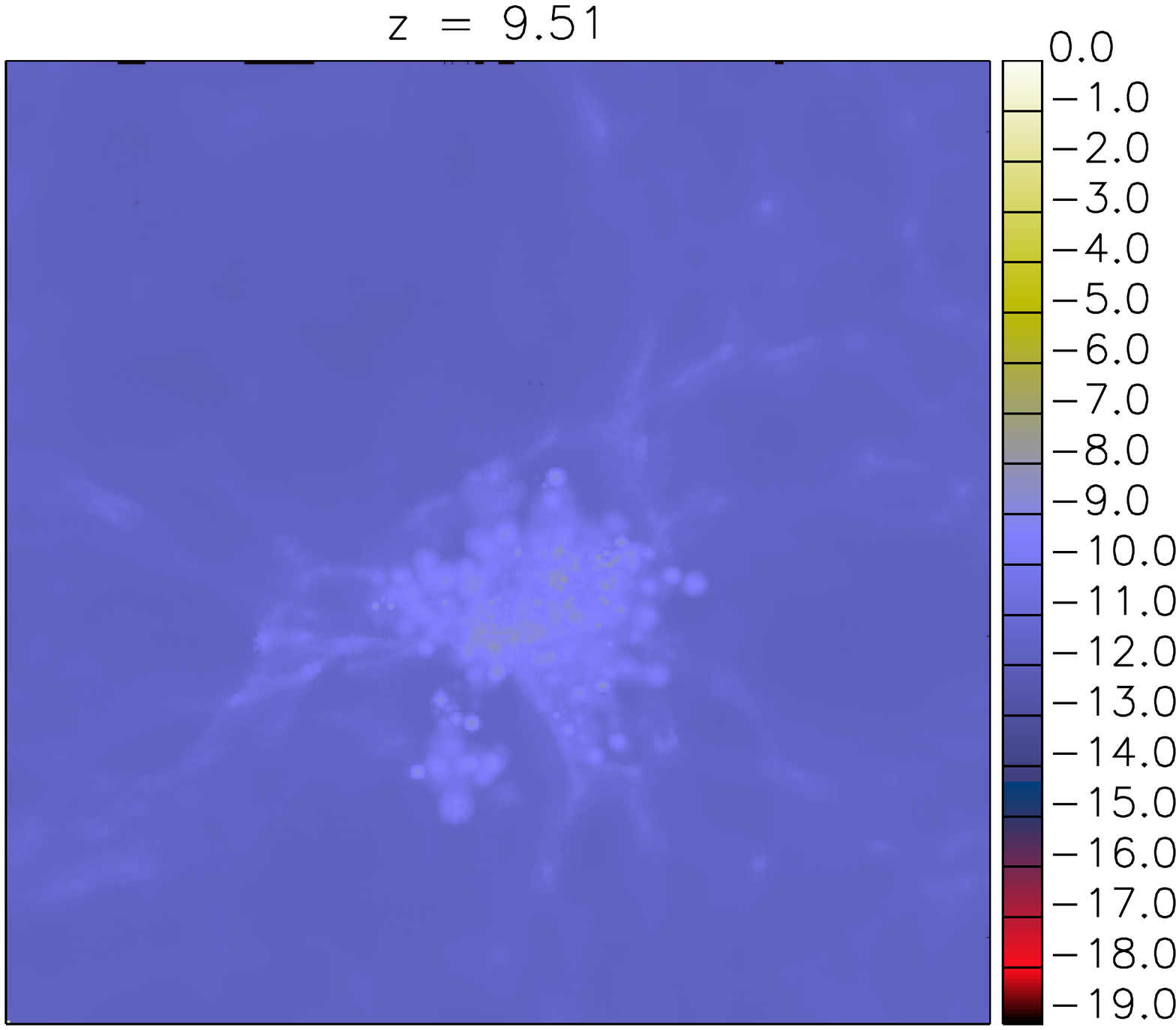}\\
    \vspace{-1cm}
  {\bf HI \hspace{0.3\textwidth} HII \hspace{0.3\textwidth} H$^-$}\\
    \vspace{0.5cm}
  {\underline{\bf with RT} (top-heavy popIII IMF, $\rm T_{\rm eff}=10^5\,K$)}\\
\vspace{-1.5cm}
  \includegraphics[width=0.33\textwidth]{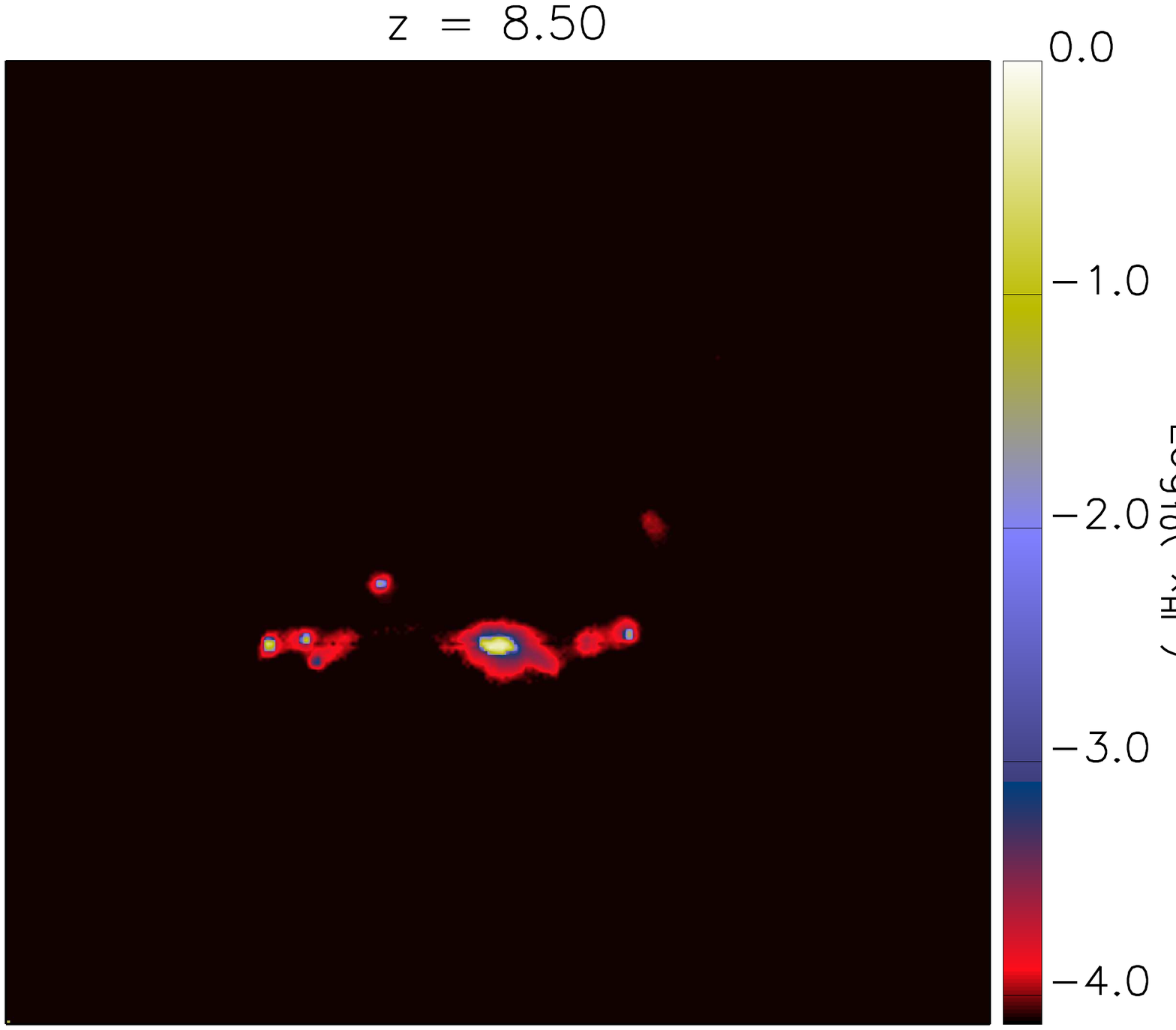}
  \includegraphics[width=0.33\textwidth]{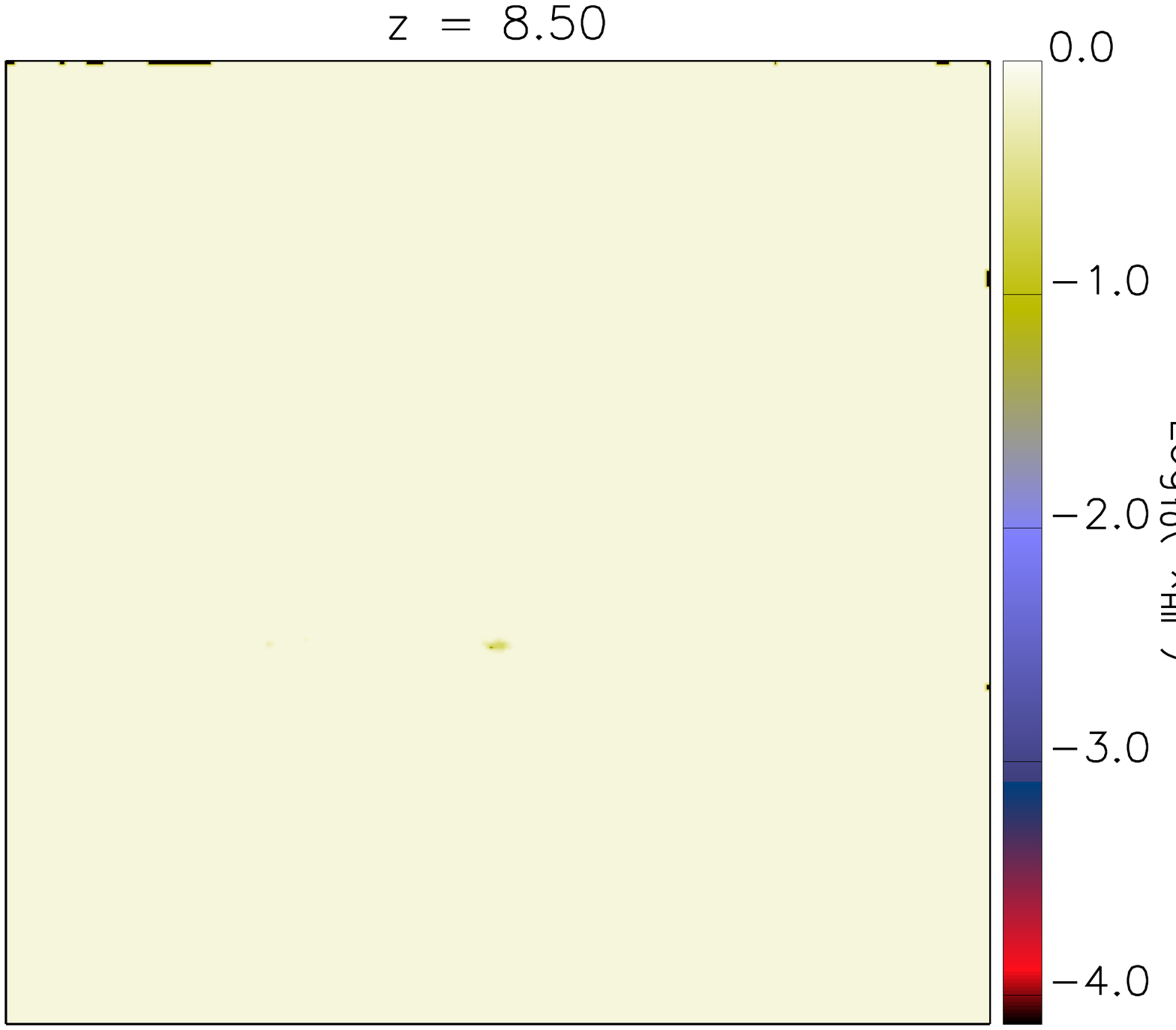}
  \includegraphics[width=0.33\textwidth]{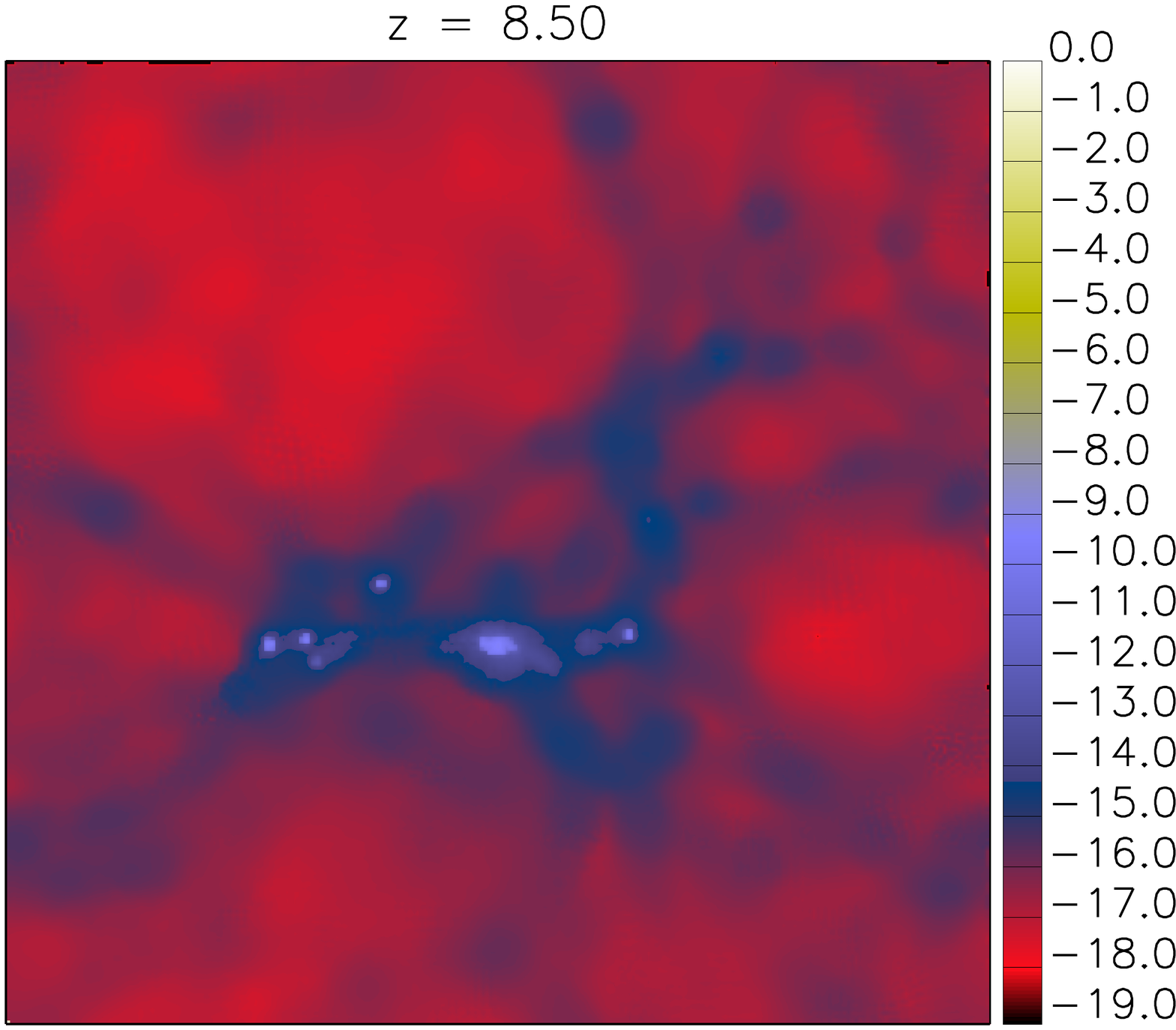}\\
    \vspace{-2cm}
  \includegraphics[width=0.33\textwidth]{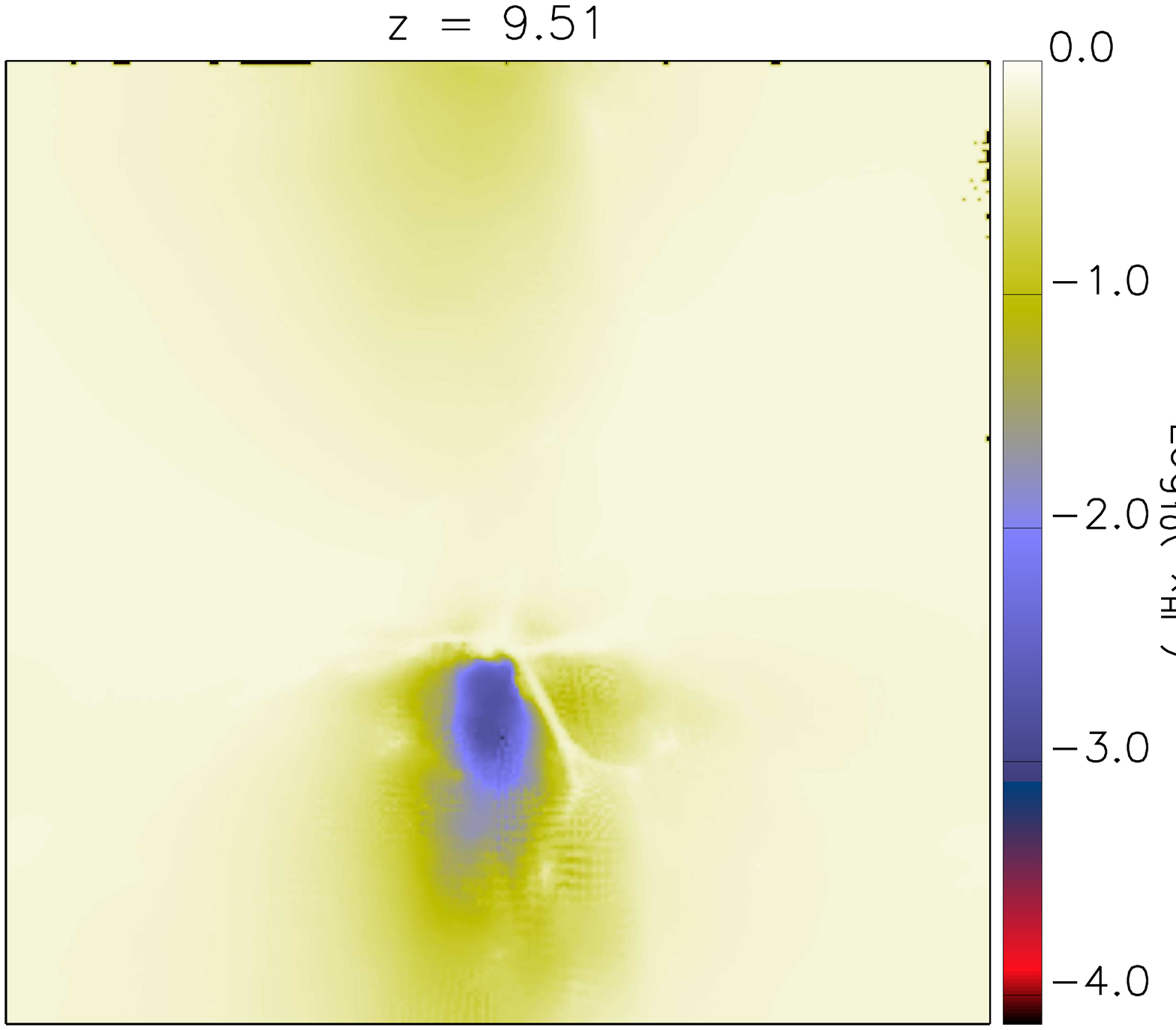}
  \includegraphics[width=0.33\textwidth]{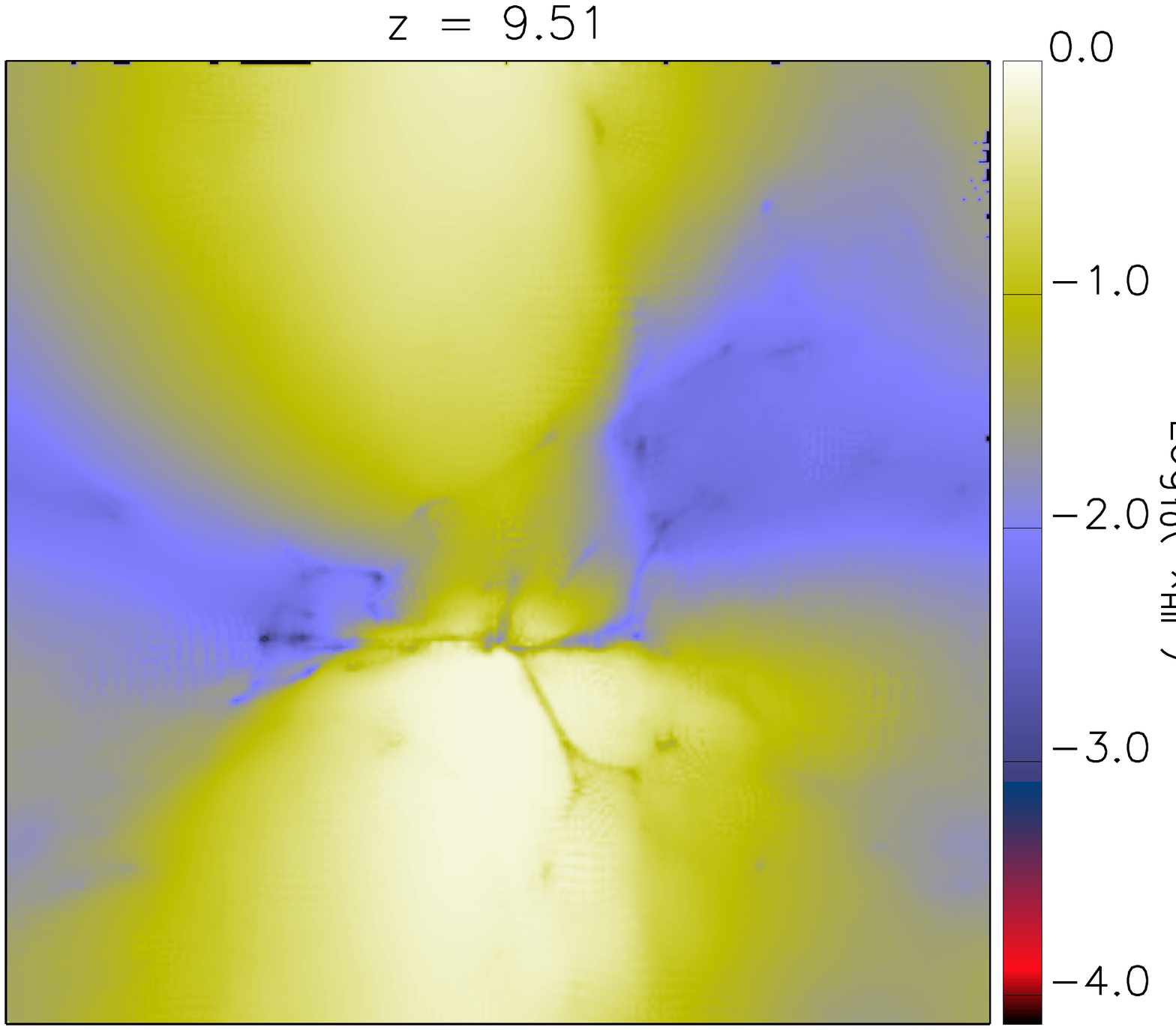}
  \includegraphics[width=0.33\textwidth]{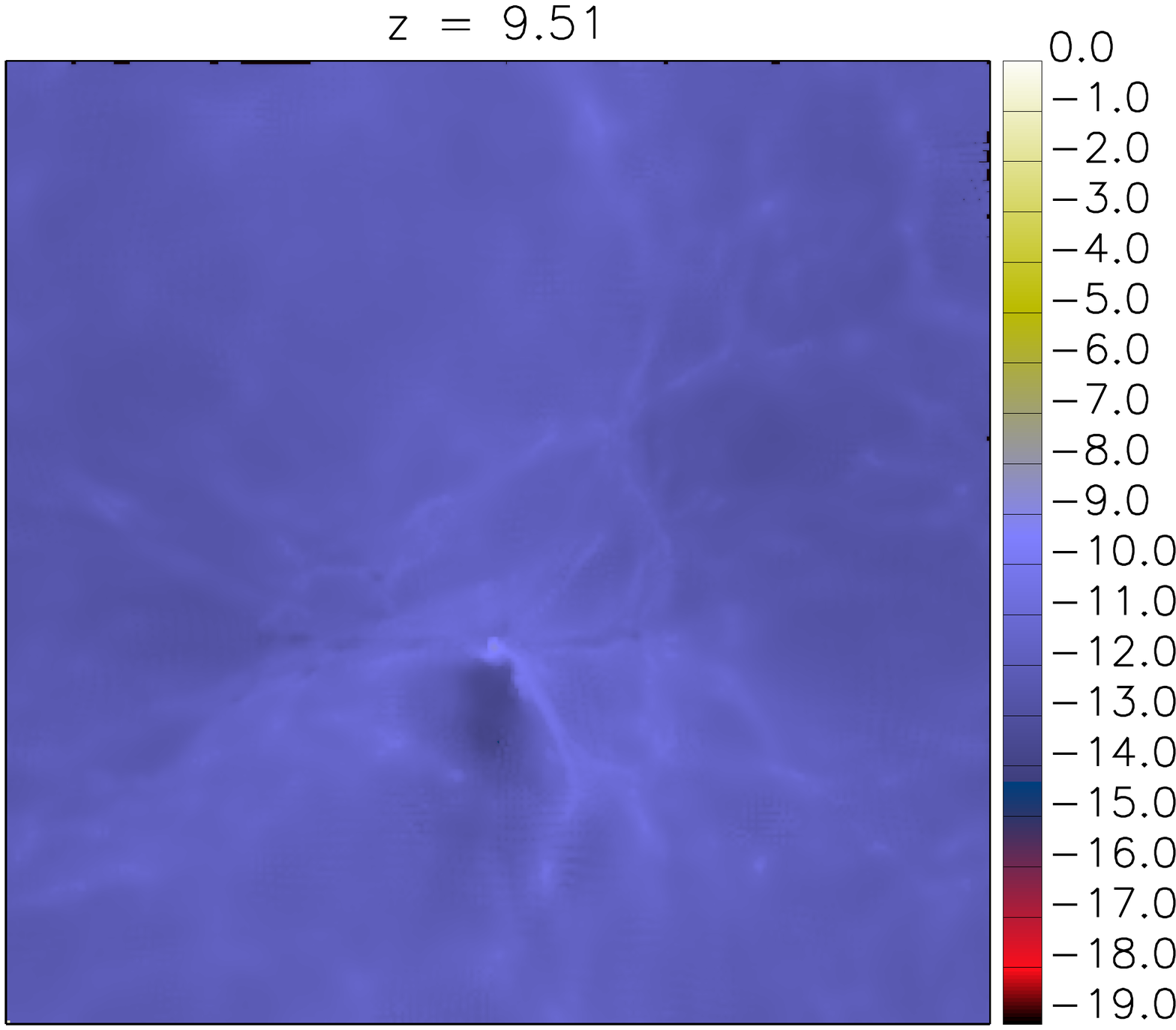}\\
    \vspace{-1cm}
  {\bf HI \hspace{0.3\textwidth} HII \hspace{0.3\textwidth} H$^-$}\\
  \caption[]{\small
    Fractional distributions of neutral H (left), ionized H (center) and H$^-$ (right) at two different redshifts for the no-RT case (upper rows) and the RT case (lower rows) with massive popIII sources.
  }
  \label{fig:Hmaps}
\end{figure*}

\subsubsection{Massive hot stellar sources.}
\begin{figure}
\centering
\includegraphics[width=0.4\textwidth]{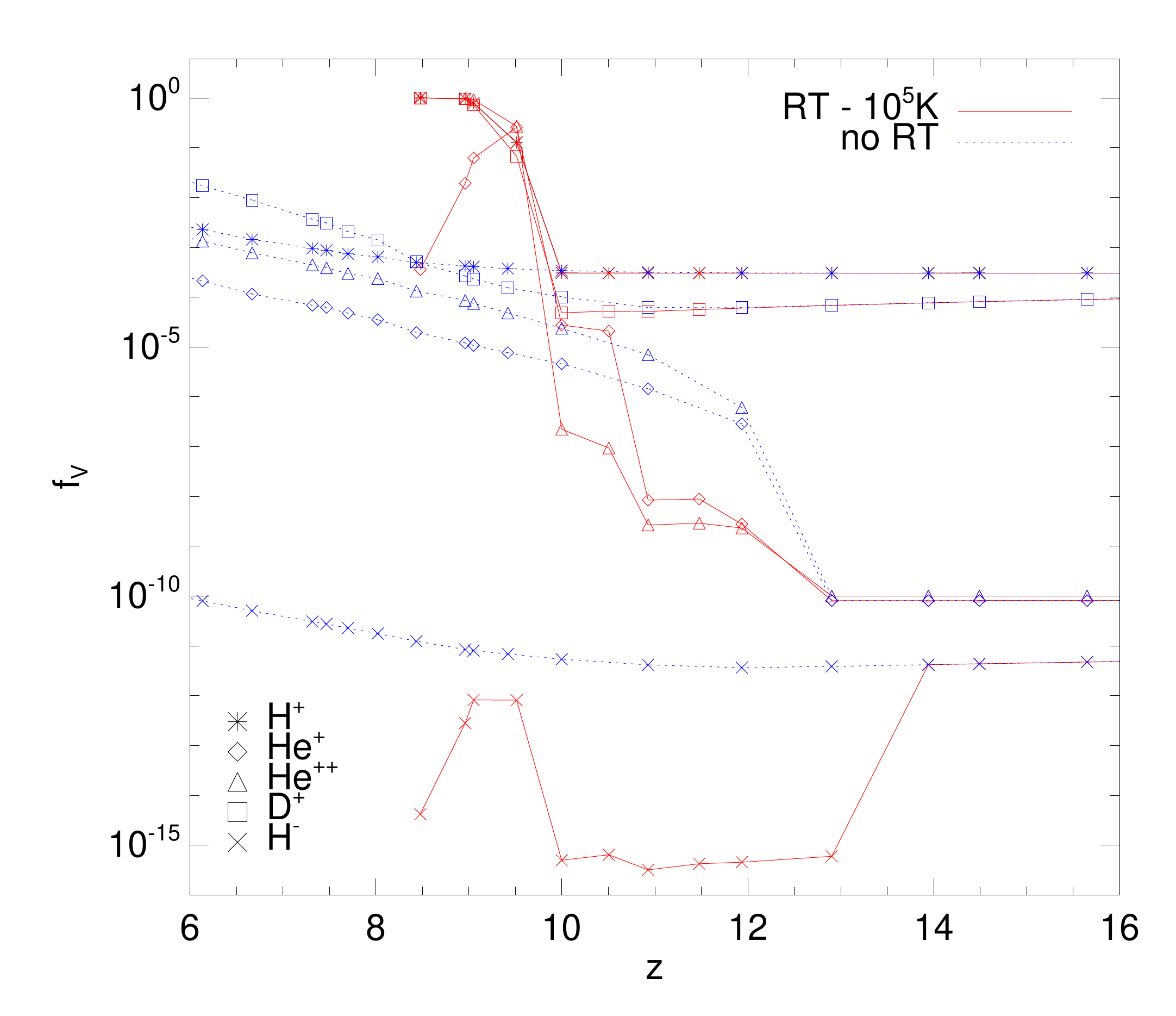}
\caption[]{\small
Filling factors for the two runs having a top-heavy popIII IMF (with and without RT). The different symbols refer to (see legend): HII (asterisks), HeII (rhombs), HeIII (triangles), DII (squares) and H$^-$ (crosses) volume filling factors, for the run including RT with $\rm T_{\rm eff} = 10^5\rm~K$ (solid lines) and the one not including RT (dotted lines).
}
\label{fig:ff}
\end{figure}

We star our discussion by focusing on hydrogen species, that represent $\sim 93\%$ (in number) of all the chemical species in the Universe and that is expected (see Fig.~\ref{fig:cooling_spectra}) to be mostly affected by the first primordial massive popIII stars.
\\
In Fig.~\ref{fig:Hmaps}, we compare the H ionization stages at different redshift for the case with no RT and for the case with RT implementation in the runs with a top-heavy popIII IMF.
A rapid inspection of the maps shows that the behaviours of HI, HII and H$^-$ are quite different.
In the no-RT case, hydrogen remains mostly neutral even at advanced stages of star formation and only a residual fraction of roughly ten per cent at most is found for HII, quite independently from redshift.
The abundances of H$^-$, that is an important catalyzer for molecule formation, do not suffer significant effects from star formation and remain at $\sim 10^{-12}-10^{-9}$ levels.
In the RT case, instead, very relevant impacts from radiation are visible for the different species.
When first sources start shining their radiation penetrates into the IGM and ionizes hydrogen, mostly in the low-density gas (denser innermost gas tends to recombine rapidly).
This is shown for example by the ``hole'' in the distribution of H fraction close to the first star forming regions at $z\sim 9.5$.
At this time, typical neutral H abundances start dropping down from some tens per cent (greenish regions) up to a few per cents (blueish central region within the denser filaments), in correspondence of which, the medium is thinner and can be easily heated up (compare to Fig.~\ref{fig:maps}) and ionized.
In fact, the ionized fraction (center) shows a complementary trend with values going from a few per cent or less (blueish regions), where H is mostly neutral, up to several tens of per cent in the regions reached by ionizing fronts (greenish and yellowish areas).
We note that the background distribution of cosmic gas influences the shape of the ionizing fronts as it allows radiation to propagate more or less easily into low- or high-density environments and filamentary structures.
Radiation is also responsible for dissociation of H$^-$ (right), that drops of a few orders of magnitudes, differently from the no-RT case.
In short time, by $z\sim 8.5$, radiative feedback is basically able to completely ionize almost all the atoms of H in the volume with residual fractions of roughly a few $10^{-4}$.
Also in this case (as in the previous section), the lack of radiative transfer calculations might cause misleading estimates of H ionization stages and distribution.
Indeed, while in the no-RT run HII is only found close to bursty star forming sites, in the full RT run it is also found in (photoionized) thin low-density gas, where recombination is less efficient.
\\
When looking at He, He$^+$, He$^{++}$ species the picture is consistent.
In the no-RT run, there is no substantial evolution of He, with a neutral fractions which always remains close to $\sim 7\%$ (in number).
The ionized fractions reach at most $\sim 10^{-4}-10^{-3}$ values only in very locally limited regions adjacent to star formation events -- similarly to the corresponding ionized-hydrogen distribution in Fig.~\ref{fig:Hmaps}.
\\
In the RT case, instead, the scenario is quite different. Indeed, there is a significant evolution of He species due to the strong emissions from primordial massive stellar sources.
The resulting effect is a continuous decrement of He fractions that in a short time drops below $\sim 10^{-4}$, while He$^+$ and He$^{++}$ increase up to several per cent, with He$^{++}$ finally dominating the whole He content \cite[see also][]{Yoshida_rt_2007}.
In particular, the abundances of He$^+$ and He$^{++}$ are roughly equal when H is almost completely ionised.
\\
In addition, we note that molecules are also expected to be strongly destroyed by RT, since their binding energies are much weaker than H or He atoms.
By a simple comparison it is evident that the effects on molecules are quite strong, because H$_2$ fractions result decreased of several orders of magnitude, with maximum values in the high-density regions that are lowered from almost $\sim 10^{-3}$ in the no-RT case down to $\lesssim 10^{-12}$ in the RT case.
Such strong radiative feedback on H$_2$ \cite[see also e.g.][]{GnedinDraine2014} suppresses significantly star formation in the neighbouring gas.
Simultaneously, H$_2^+$ in the no-RT case reaches abundances of $\sim 10^{-7}$ around the cosmic star forming regions, while in the run with RT typical abundances are always lower than $\sim 10^{-10}$.
Analogous trends are found for HD as well.
\\
The above results can be summarized through the evolution of the volume filling factors, $f_{V}$, of the different species.
\\
In Fig.~\ref{fig:ff} we display the evolution of filling factors for H and He species in the case with no-RT implementation and in the case with RT implementation including massive popIII surces emitting at $\rm T_{\rm eff} = 10^5\rm~K$.
\\
In the former case smoother and more regular trends are found, instead, in the latter case more irregular behaviours are visible, as a consequence of the radiative feedback, that highlights the occurring star formation episodes.
Indeed, Fig.~\ref{fig:ff} clearly shows that in the no-RT run the filling factor of ionized hydrogen (H$^+$) does not evolve very significantly, being around $\sim 10^{-4}$ at early times and increasing only up to $\sim 10^{-3}$ at $z\sim 6$.
This means that most of the hydrogen in the Universe is still neutral by $z\sim 6$.
Similarly, He$^+$ and He$^{++}$ increase slightly only after the first star forming episodes, when SN feedback heats up the local medium and causes (limited) local He ionization.
Never the less, the overall amounts of ionized He are negligible and reach only $f_{V}\lesssim 10^{-3}$.
We note that, together with H and He ionization by SN feedback, also ionized deuterium is produced and increases of a few orders of magnitude, up to $f_V\sim 10^{-2}$ at $z\sim 6$.
The free electrons liberated via the sub-grid SN feedback mechanism enhance H$^-$ production making it reach $f_V\sim 10^{-10}$.
\\
In the RT run, instead, results are much more sensitive to star formation and stellar radiation due to the radiative feedback, that is properly taken into account.
In fact, H$^+$ $f_{V}$ steeply increases, at $z\lesssim 10$, from $\sim 10^{-4}$ up to almost unity and similarly He$^+$ and He$^{++}$.
These are consequence of the powerful hot sources that spread their radiation into the thin low-density medium and are able to ionize both H and He on cosmic scales.
\\
Such scenario of early He ionization seems to be inline with improved measurements of He ionization (around quasars) already at redshift $z\sim 6$ \cite[][]{Bolton2012}.
At later times He$^{++}$ is even more abundant than He$^+$, which drops down of more than two orders of magnitudes, suggesting an almost complete He ionization.
\\
We note that H and He ionization processes take place roughly simultaneously with the mentioned inversion of IGM equation of state, as a consequence of photoheating of the cosmic gas (Sect.~\ref{Sect:IGMimplications}).
In presence of such powerful sources it is also extremely easy to find ionized D, D$^+$, and H$^-$ formation via free-electron attachment.
This latter species is initially destroyed ($z\sim 13$) by the photoheating of first stellar radiative events and, once enough free electrons have been made available (at $z\sim 10$), it gets re-formed catalysing H$_2$ formation at $z\lesssim 9$.
\\
We stress that all these details on chemistry evolution are strongly linked to the nature of primordial sources assumed in this particular case.
In the next section we will see that for the cases of more standard and less powerful stellar sources RT effects are less extreme.

\subsubsection{Standard stellar sources}

\begin{figure}
\centering
\includegraphics[width=0.4\textwidth]{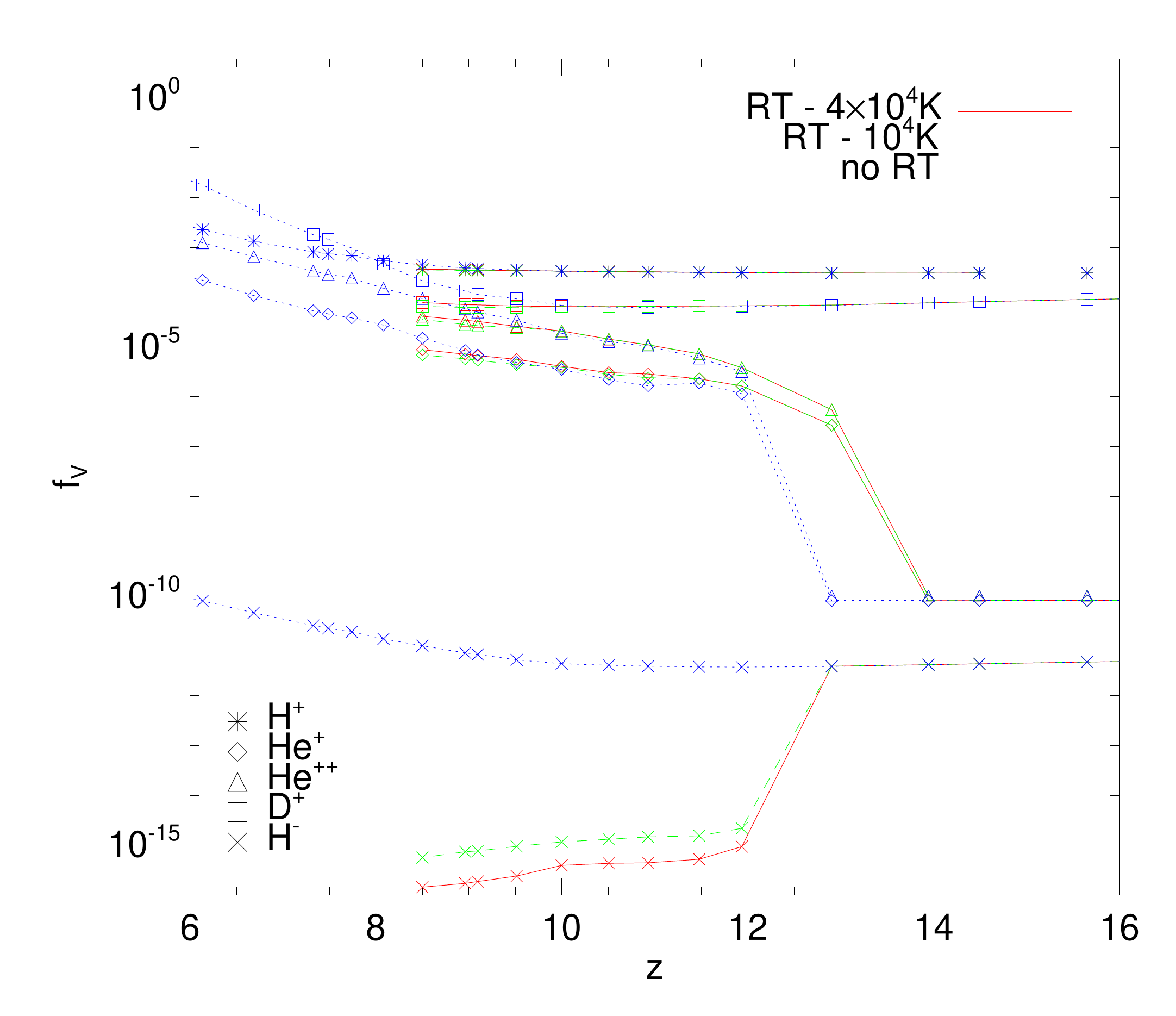}
\caption[]{\small
Filling factors for the three runs having a Salpter-like popIII IMF. The different symbols refer to (see legend): H$^+$ (asteriscs), He$^+$ (rhombs), He$^{++}$ (triangles), D$^+$ (squares) and H$^-$ (crosses) volume filling factors, for the run including RT with  a popIII $\rm T_{\rm eff} = 4\times 10^4\rm~K$ (solid lines), RT with a popIII $\rm T_{\rm eff} = 10^4\rm~K$ (dashed lines) and the one not including RT (dotted lines).
}
\label{fig:ff.lowmass}
\end{figure}
\begin{figure*}
\centering
\includegraphics[width=0.4\textwidth]{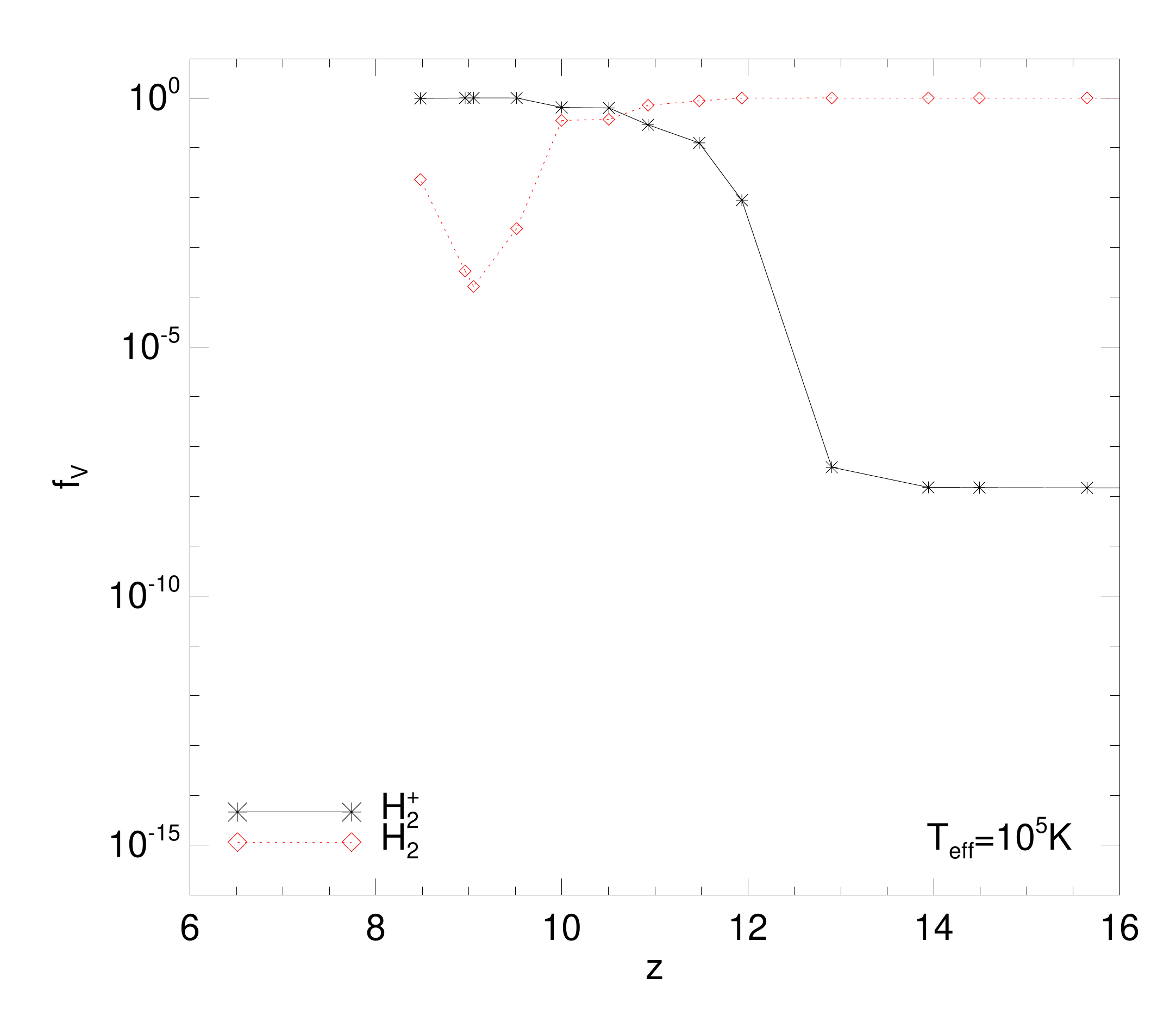}
\includegraphics[width=0.4\textwidth]{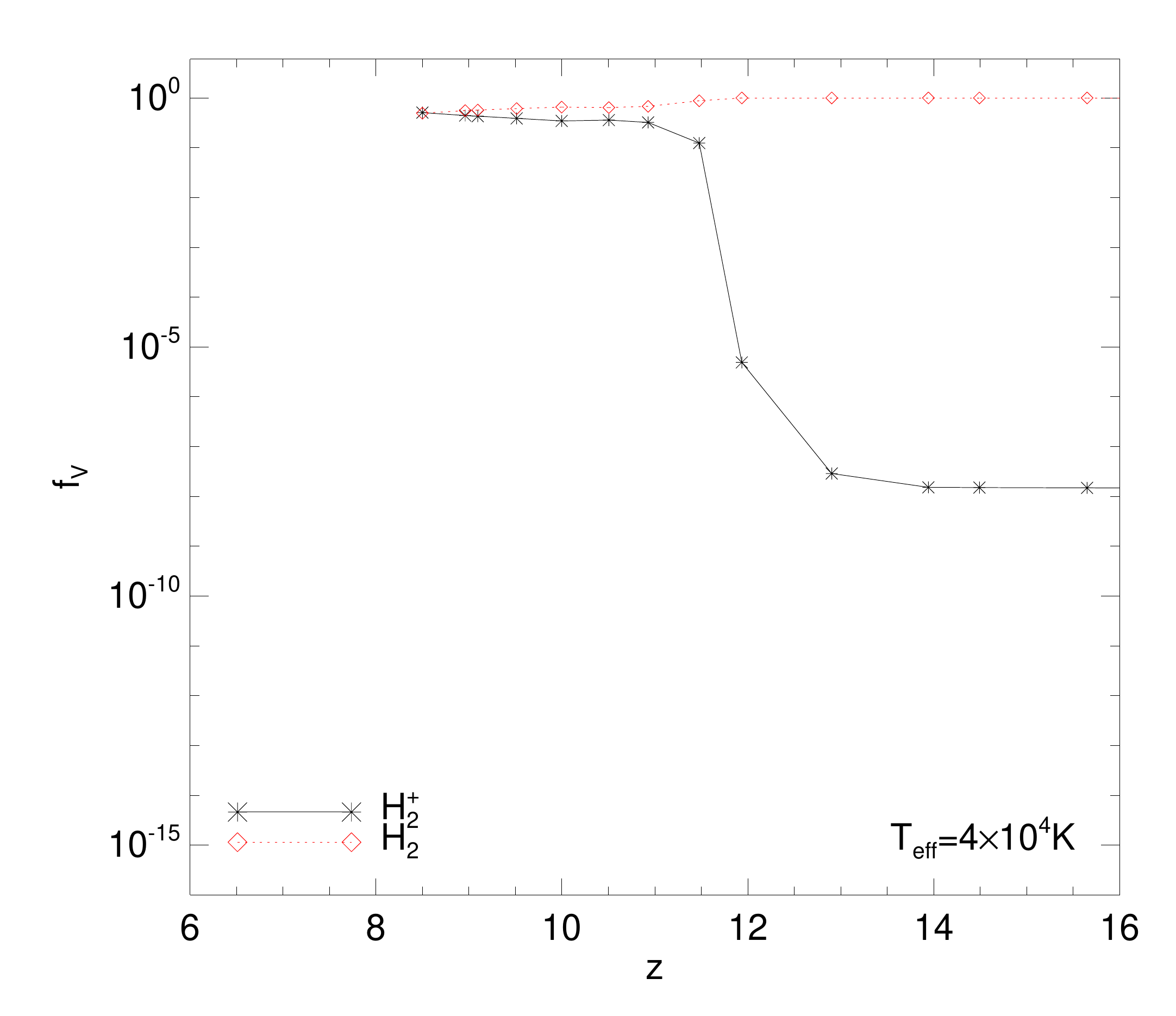}  \\
\includegraphics[width=0.4\textwidth]{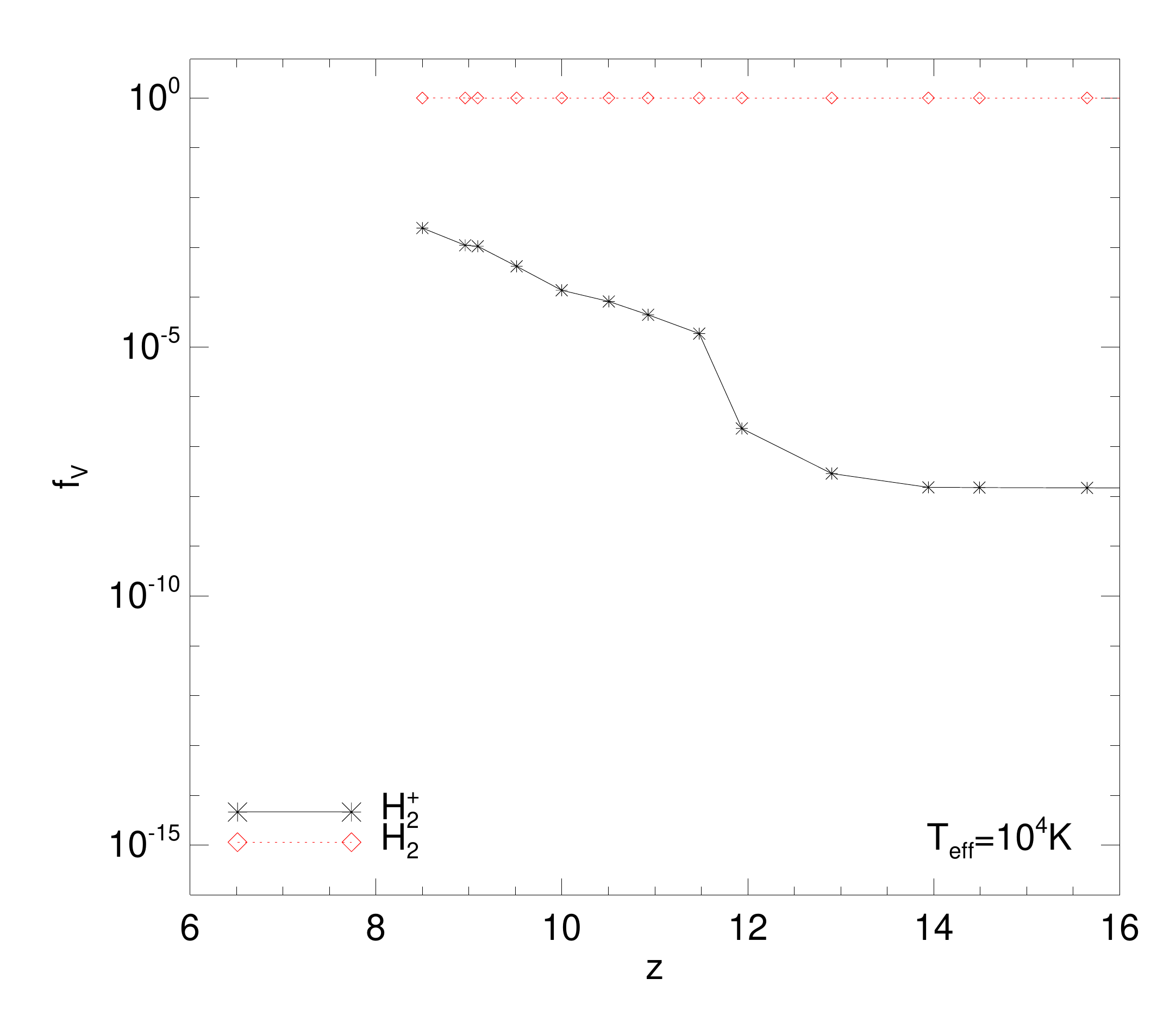}
\includegraphics[width=0.4\textwidth]{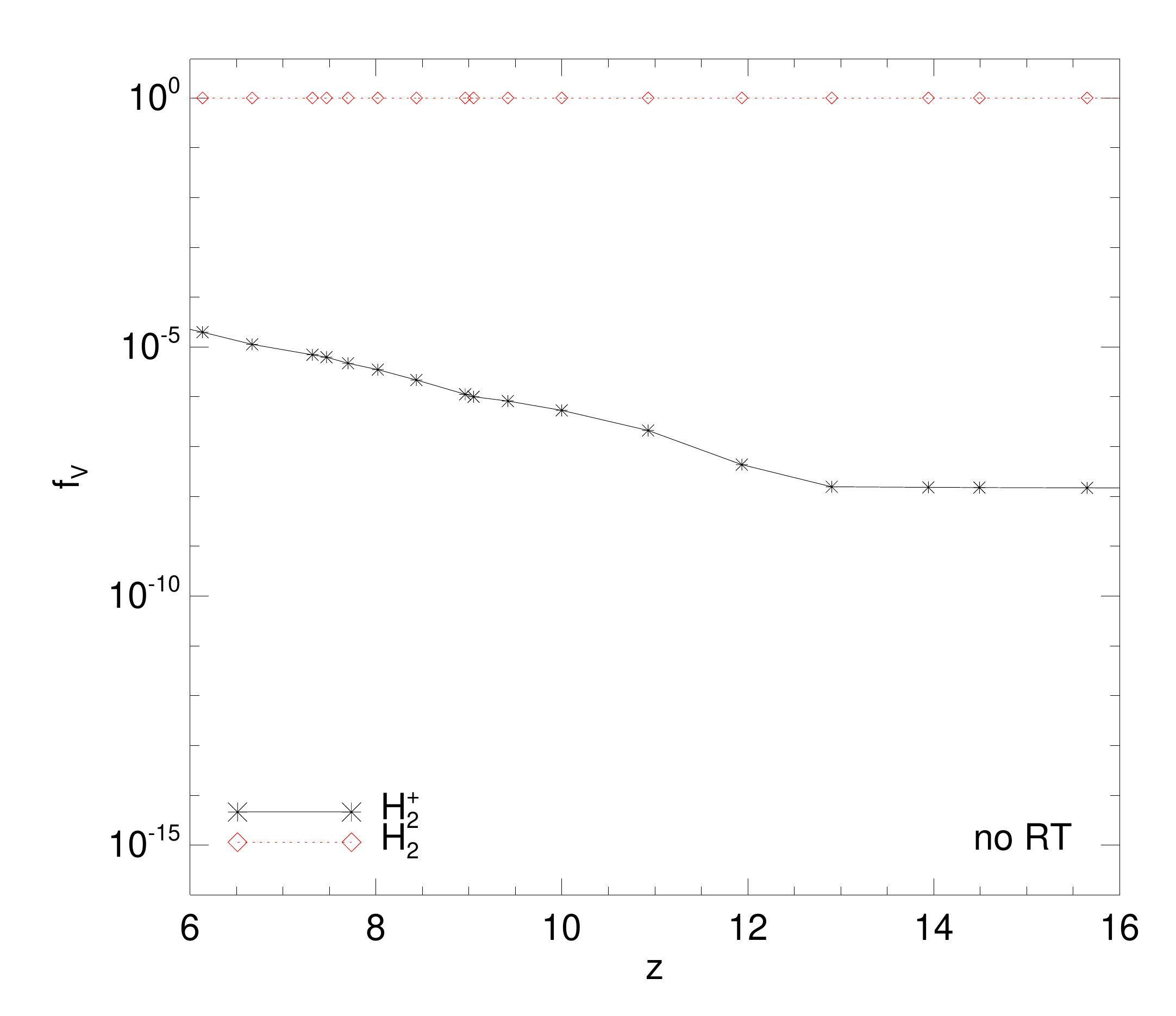}
\caption[]{\small
  Comparison of H$_2$ and H$_2^+$ filling factors. The different symbols refer to the neutral fractions (dotted lines with rhombic points) and firstly ionized fractions (solid lines with asteriscs). The panels refer to the: RT run with top-heavy popIII IMF and $\rm T_{\rm eff} = 10^5\rm~K$ (top left panel), RT run with Salpeter-like popIII IMF and $\rm T_{\rm eff} = 4\times10^4\rm~K$ (top right panel), RT run with Salpeter-like popIII IMF and $\rm T_{\rm eff} = 10^4\rm~K$ (bottom left panel), no RT run with top-heavy popIII IMF (bottom right panel).
}
\label{fig:ff_individual}
\end{figure*}
When a lower-mass IMF is adopted (Fig.~\ref{fig:ff.lowmass}) feedback effects take place later, due to the longer lifetimes of OB stars ($\sim 10^7-10^8\,\rm yr$).
\\
Besides the different timescales, there are no huge differences in the H, H$^+$ and H$^-$ distributions in the no-RT runs and they all show a roughly flat neutral H abundance.
As a consequence of the lack of RT treatment, non-null ionized fractions result only where star formation takes place and ionization is confined just near SN events.
There are no effects on the IGM either.
\\
When we check the effects of stellar radiation by following RT from a black-body spectrum with effective temperature of $4\times 10^4\rm~K$ and $10^4\rm~K$, respectively, the situation is only slightly different.
\\
Given the weaker energies produced by such emissions (whose peak values are below 13.6~eV, the H ionization threshold), no strong variations on cosmic H and H$^+$ are found, however, radiation leaves its imprints on H$^-$.
\\
Indeed, with a ionization threshold of only $\sim\rm 0.7~eV$, H$^-$ is easily destroyed and its abundances drop by a few orders of magnitudes.
On the contrary to the massive popIII case, H$^-$ is not reformed later due to the lack of sufficient free electrons.
Slight differences between H$^-$ and electron fractions in the models with $T_{\rm eff}= 4\times 10^4\rm~K$ and $T_{\rm eff}= 10^4\rm~K$ are due to the different input effective temperatures.
Typical H$^-$ abundances result larger in the former case (where more electrons are available) and smaller in the latter.
\\
For standard stellar sources no appreciable differences in He, He$^+$ and He$^{++}$ are found among the RT and no-RT cases. This is due to the spectral distribution of the emitted photons that almost never reach the He ionization thresholds.
Further discrepancies in molecular fractions for the three runs with a Salpeter-like popIII IMF are modest and the picture emerging from these scenarios with no-RT is quite similar to the previous one.
\\
RT-driven dissociation for H$_2$, H$_2^+$ and HD, according to the power of the stellar spectra adopted, is present, however, changes are less prominent than in the $\rm T_{\rm eff} = 10^5\rm~K$ case.
\\
In Fig.~\ref{fig:ff.lowmass} it is shown the detailed evolution of the filling factors for the pristine atomic species, H$^+$, He$^+$, He$^{++}$, D$^+$ and H$^-$ (as in Fig.~\ref{fig:ff}), for the three runs having a Salpter-like popIII IMF and (i) including RT with a popIII $\rm T_{\rm eff} = 4\times 10^4\rm~K$, (ii) including RT with a popIII $\rm T_{\rm eff} = 10^4\rm~K$ and (iii) not including RT.
As mentioned before, there are no substantial differences between the RT runs and the no-RT run.
This is due to the weaker stellar sources adopted with respect to the more powerful assumption of Fig.~\ref{fig:ff}.
This is particularly true for H and He abundances, while some effects on H$^-$ are more visible.
\\
Fig.~\ref{fig:ff.lowmass} summaries well the high level of similarities among the considered  models and shows a depletion of up to $\sim 5$ orders of magnitude for H$^-$ in the RT runs, although, no drastic differences appear between the two assumptions of $\rm T_{\rm eff}=10^4\rm~K$ and  $\rm T_{\rm eff}=4\times10^4\rm~K$.

\subsubsection{Implications for cosmic molecular content}

To address radiative effects from different stellar populations on cosmic molecule abundances we perform a direct comparison in Fig.~\ref{fig:ff_individual}.
In the panels we show comparisons of H$_2$ and H$_2^+$ filling factors (normalized to unity).
\\
From top left to bottom right, the panels refer to the: 
RT run with top-heavy popIII IMF and $\rm T_{\rm eff} = 10^5\rm~K$ (top left panel); 
RT run with Salpeter-like popIII IMF and $\rm T_{\rm eff} = 4\times10^4\rm~K$ (top right panel);
RT run with Salpeter-like popIII IMF and $\rm T_{\rm eff} = 10^4\rm~K$ (bottom left panel);
no RT run with top-heavy popIII IMF (bottom right panel).
\\
In all the cases, neutral H$_2$ (red dotted lines) is dominant at high redshift, $z>10$, while later on ionization effects become progressively more and more visible with the raise of H$_2^+$ abundance (solid black lines).
\\
The effects of the strong RT from massive popIII stars are evident in the first panel of Fig.~\ref{fig:ff_individual}, where we find an intersection point around $z\sim 10$ in correspondence of which the ionized filling factor overcomes the neutral one.
We note that the large amount of free electrons produced in the meantime boosts the H$^-$ channel and leads to partial H$_2$ reformation at $z\lesssim 10$.
This is consistent with the above discussion about Fig.~\ref{fig:ff}.
\\
The radiative effect is quite impressive when compared to the case without RT implementation (last panel in Fig.~\ref{fig:ff_individual}), in which ionized molecules are underestimated by several order of magnitudes at $z \lesssim 10$.
The bulk of star formation at $z\lesssim 10$ and the consequent strong radiative feedback from the first stars determine molecule dissociation and, additionally, ionization of the surviving H$_2$, reducing the available neutral molecular fraction to $ \sim 10^{-4}$.
\\
In the top right and bottom left panel of Fig.~\ref{fig:ff_individual} we display results for H$_2$ and H$_2^+$ in the RT runs with $\rm T_{\rm eff}=4\times10^4\rm~K$ and $\rm T_{\rm eff}=10^4\rm~K$, respectively.
Here, the radiative feedback is still evident and standard hot sources with $\rm T_{\rm eff}=4\times10^4\rm~K$ can impact H$_2$ and H$_2^+$ in a non-negligible way.
The effects from weaker sources (with $\rm T_{\rm eff}=10^4\rm~K$), instead, are minor.
More quantitatively, in the top right panel H$_2^+$ values rapidly increase to converge with H$_2$ at $z \lesssim 10$, while in the bottom left panel the catch-up is slower, being H$_2^+$ values three orders of magnitude below H$_2$.
\\
The trends in this latter case resemble more closely the ones in the no-RT run, shown in the bottom right panel.
It is worth mentioning, though, that between the H$_2^+$ predictions in the bottom panels there is still a difference of roughly 2 orders of magnitude at redshift $z\lesssim 8$.

%*****************************************************************************

\subsection{Star formation} \label{Sect:sfr}
We continue our discussion by addressing RT effects on star formation and the implications for metal pollution and photoevaporation of primordial haloes.
\\

\subsubsection{Star formation rate density}

\begin{figure}
\centering
\includegraphics[width=0.45\textwidth]{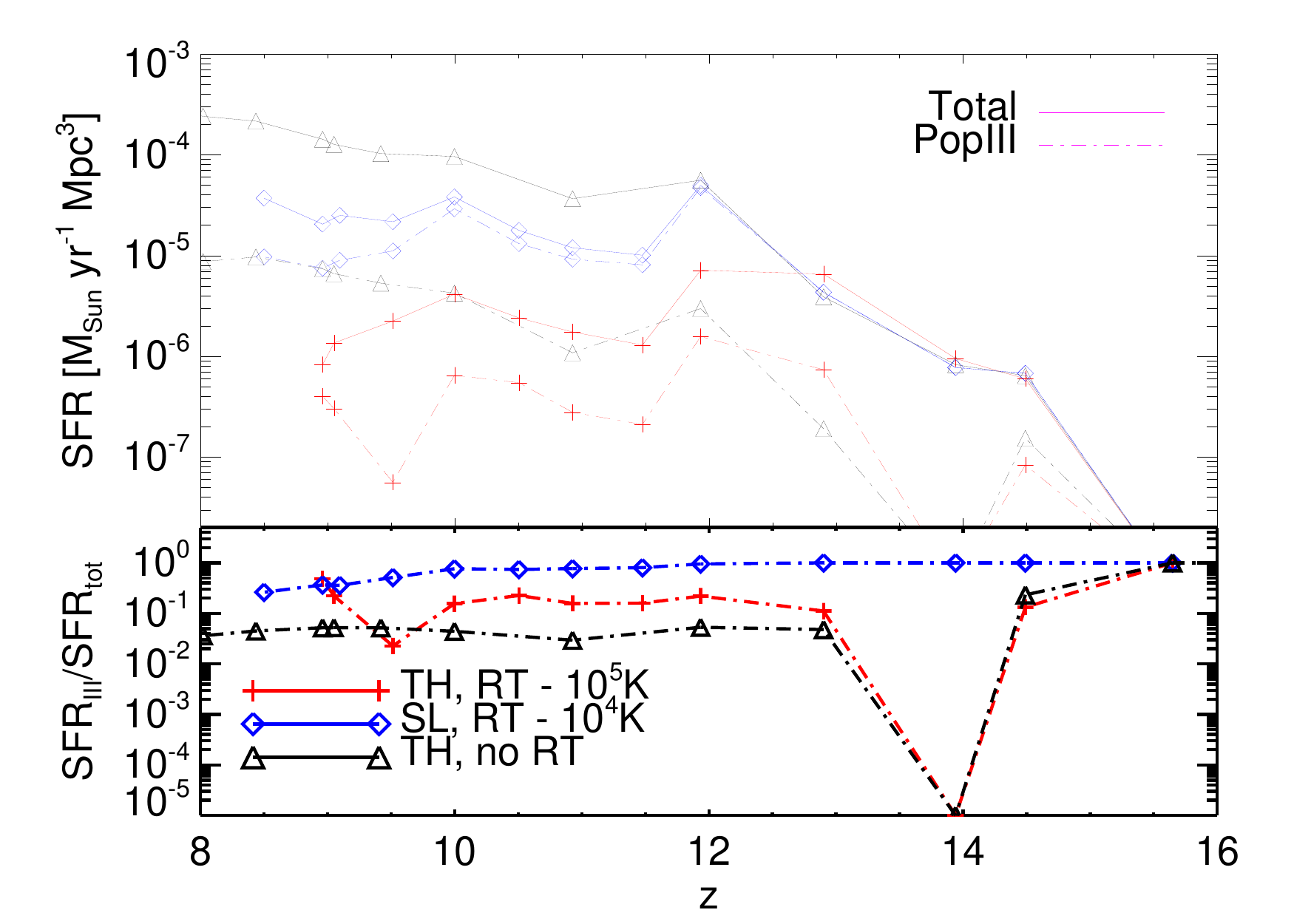}\\
\includegraphics[width=0.45\textwidth]{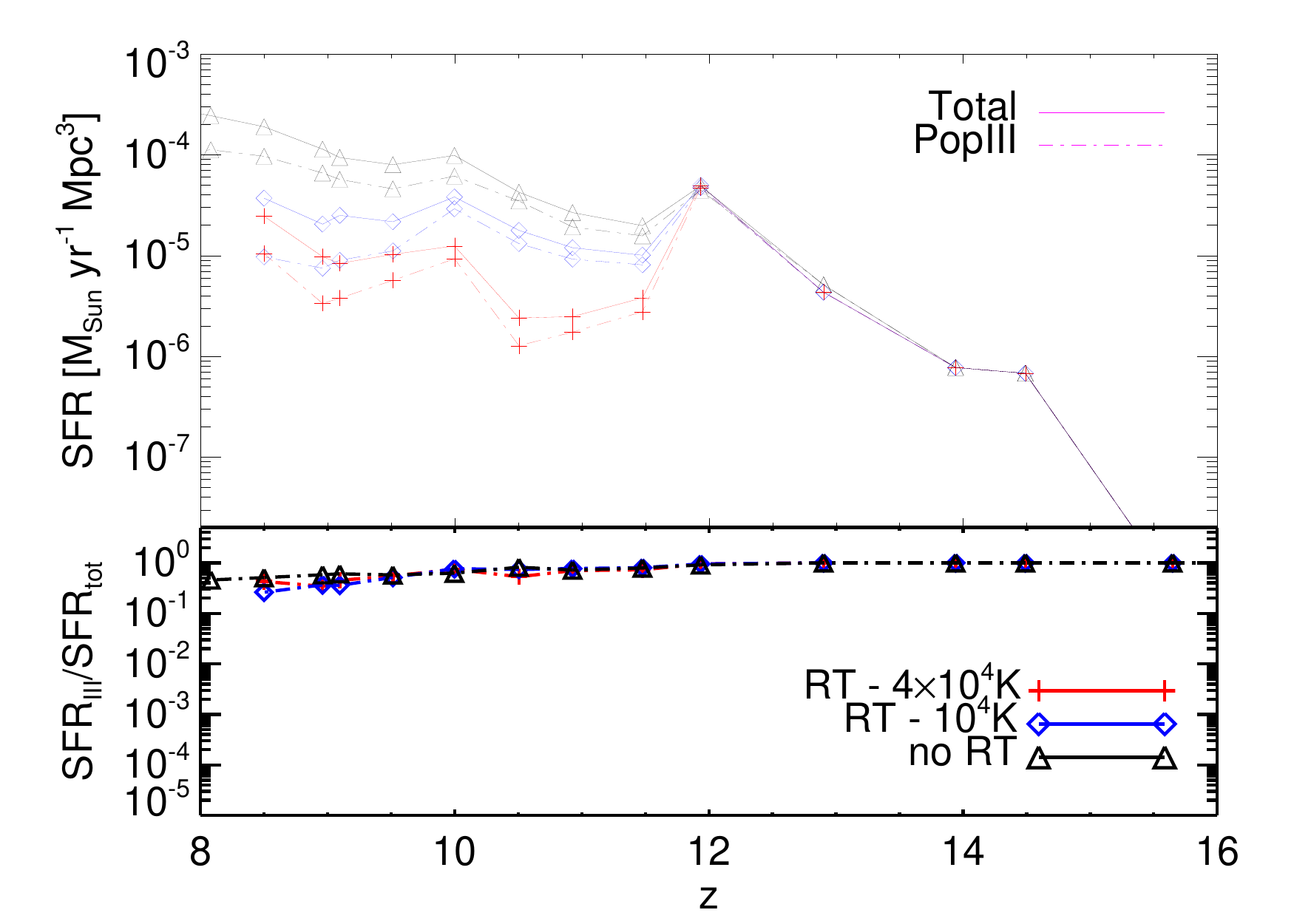}\\
\caption[]{\small
  Cosmic star formation rate densities (top panels) for the models assuming top-heavy (TH) popIII IMFs and Salpeter-like (SL) IMFs in the RT and no-RT scenarios, with corresponding popIII contributions (bottom panels).
The cases shown in the top panel are: 
a model with top-heavy popIII IMF with RT and $ T_{\rm eff} = 10^5\,\rm K $ (black line and crosses);
a model with Salpeter-like popIII IMF with RT and $ T_{\rm eff} = 10^4\,\rm K $ (blue line and diamonds);
a model with top-heavy popIII IMF with no RT (red line and triangles).
The cases shown in the bottom panel refer to SL popIII IMF with:
RT and $ T_{\rm eff} = 4\times 10^4\,\rm K $ (black line and crosses);
RT and $ T_{\rm eff} = 10^4\,\rm K $ (blue line and diamonds);
no RT (red line and triangles).
For a more direct comparison we use the same scales in both panels.
}
\label{fig:sfr}
\end{figure}

Cosmic star formation rate (SFR) densities are displayed in Fig.~\ref{fig:sfr} for the models assuming a top-heavy popIII IMF and Salpeter-like (SL) popIII IMFs.
\\
On the different panels we can retrieve immediately the implications of radiative feedback.
In fact, the total star formation rate density decreases when RT is included and the decrease is larger for more powerful sources.
Broadly speaking, this is understood in terms of the effects of radiation on molecules, which drive gas collapse and star formation. Powerful sources can dissociate more easily H$_2$ and/or HD, while weaker sources have minor impacts.
In detail, when emission by top-heavy popIII stars with $T_{\rm eff} = 10^5\,\rm K$ is considered the drop is up to 1.5 orders of magnitudes with respect to the no-RT case.
If a standard Salpeter-like IMF with sources emitting at $T_{\rm eff} = 10^4\,\rm K$ is considered the drop is milder and the star formation rate is usually quenched by a factor of $\sim 2$.
The resulting popIII contribution (SFR$_{\rm III}$) is affected as well.
After the initial burst that temporarily halts star formation due to the very powerful short-lived stars, the run with top-heavy RT included predicts values for SFR$_{\rm III}$/SFR$_{\rm tot}$ that stay around $0.1-0.5$, while the corresponding top-heavy no-RT run features values that usually are below $0.1$ and approach $\sim 10^{-2}$.
\\
The SL run ($T_{\rm eff}=10^4\,\rm K$) is also plotted for further comparison.
In this case, the delayed evolution of low-mass stars and the less powerful RT emission determine smaller departures from $\rm SFR_{III}/SFR_{tot}\lesssim 1 $ and only at later stages (after a few $10^8\,\rm yr$).
\\
A more complete picture can be drawn by means of the bottom panel, where both the SL cases are reported together with the no-RT run.
As we said, the $T_{\rm eff}=10^4\,\rm K$ case presents SFR deviations of a factor of $\sim 2$ while the $T_{\rm eff}= 4\times 10^4\,\rm K$ gets to total SFR values $\sim 3-10$ times lower.
The corresponding popIII contributions are always comparable, though, and range between 1 (at early times) and $\sim 0.2$ (after some $10^8\,\rm yr$) in all cases.
\\
From an observational point of view it is not trivial to discriminate the models with current determinations, because SFR data at these epochs are few and not tightly constrained.
Thus, it is not easy to infer reliable values for the star formation rate densities.

\subsubsection{Stellar mass and specific star formation rate}\label{Sect:stars}

\begin{figure}
\centering
\includegraphics[width=0.45\textwidth, angle=180]{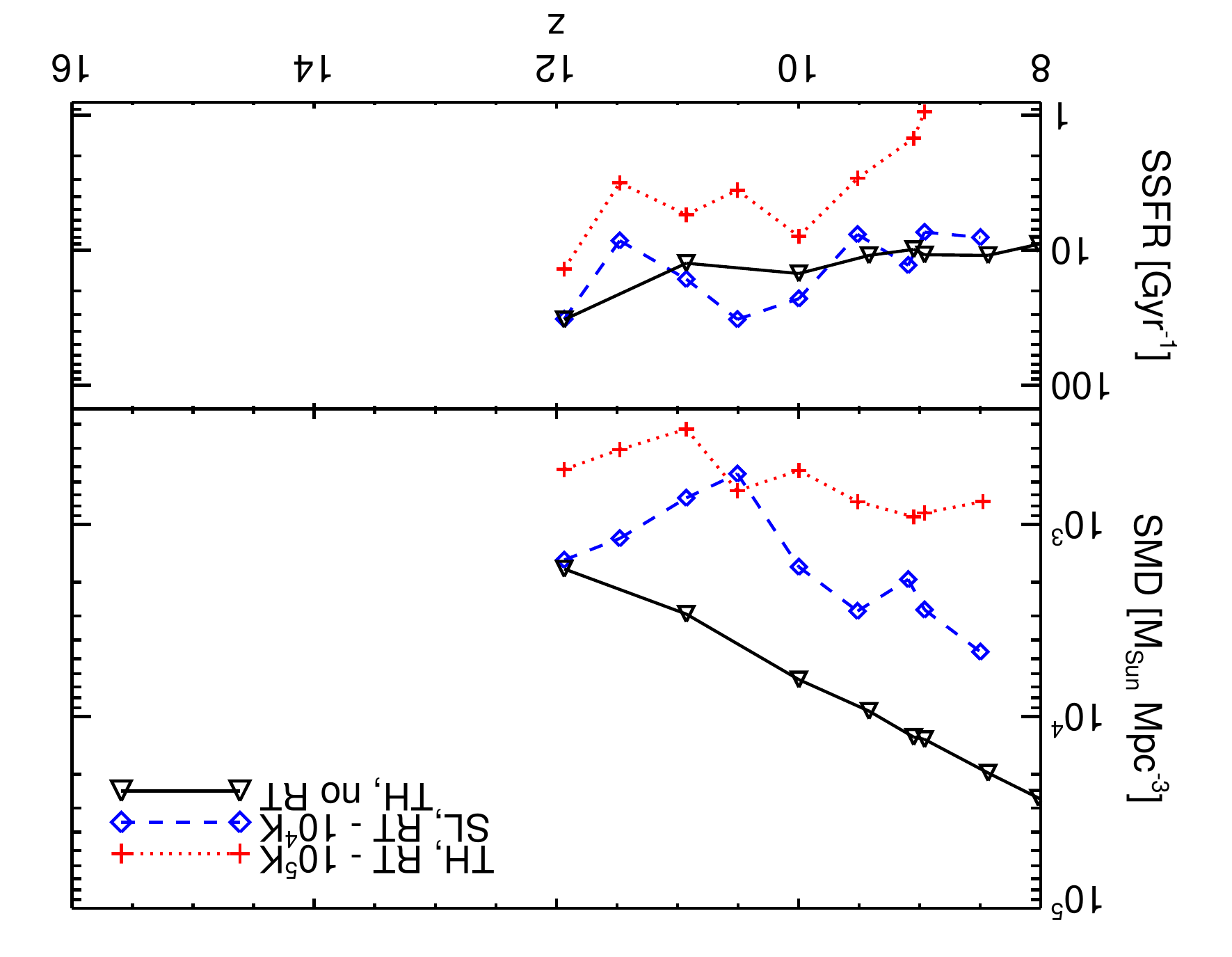}\\
\includegraphics[width=0.45\textwidth]{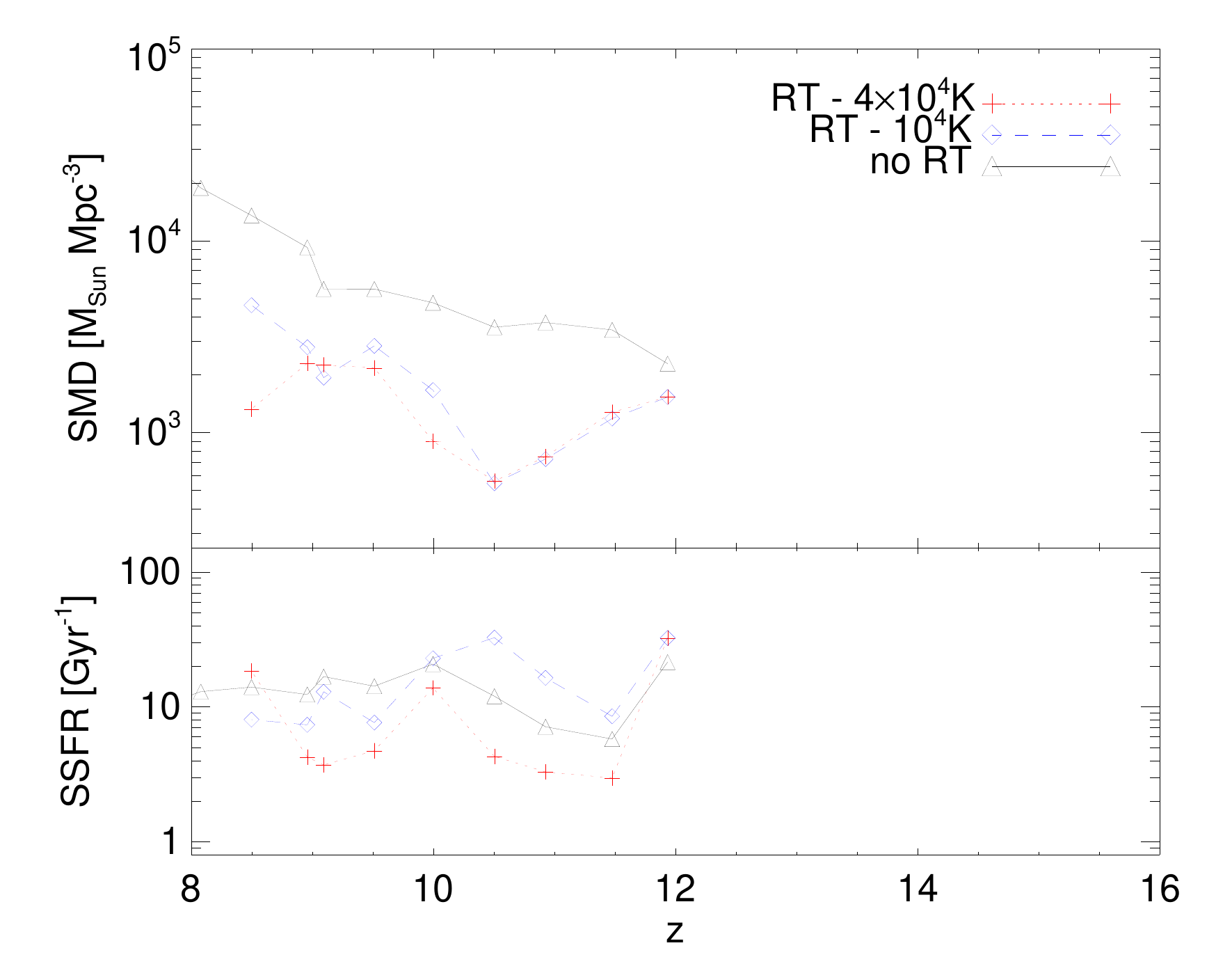}
\caption[]{\small
  Cosmic stellar mass densities, SMD (top), and specific star formation rate, SSFR (bottom), for the models assuming top-heavy (TH) popIII IMFs and Salpeter-like (SL) IMFs in the RT and no-RT scenarios.
The cases shown in the top panel are:
a model with top-heavy popIII IMF with RT and $ T_{\rm eff} = 10^5\,\rm K $ (black line and crosses);
a model with Salpeter-like popIII IMF with RT and $ T_{\rm eff} = 10^4\,\rm K $ (blue line and diamonds);
a model with top-heavy popIII IMF with no RT (red line and triangles).
The cases shown in the bottom panel refer to SL popIII IMF with:
RT and $ T_{\rm eff} = 4\times 10^4\,\rm K $ (black line and crosses);
RT and $ T_{\rm eff} = 10^4\,\rm K $ (blue line and diamonds);
no RT (red line and triangles).
}
\label{fig:smd}
\end{figure}
The implications for star formation processes in primordial epochs can be more clearly understood in terms of stellar mass (M$_\star$) density, SMD, and specific star formation rate, $\rm SSFR = SFR/M_\star$.
\\
These quantities are plotted in Fig.~\ref{fig:smd} and highlight very well the effects of different radiative sources.
While in the no-RT runs SMDs monotonically increase up to a few $10^4\, \rm M_\odot/Mpc^3$, the RT runs present a much less regular behaviour.
\\
In the top panels, the RT case with top-heavy  $ T_{\rm eff} = 10^5\,\rm K $  popIII IMF features SMD always around $10^3\,\rm M_\odot/Mpc^3$, i.e. more then 1~dex less than the no-RT cases.
The RT case with weak  $ T_{\rm eff} = 10^4\,\rm K $ shows a sort of intermediate trend and differ by about 0.5~dex  at $z\sim 9$ from the top-heavy run.
\\
By inspecting the bottom panels it is clear that spectra with $ T_{\rm eff} = 10^4\,\rm K $ and spectra with $ T_{\rm eff} = 4\times 10^4\,\rm K $ behave quite similarly, although the latter case seems a little bit more effective at suppressing SMD values at lower $z$.
These conclusions are in agreement with our previous discussion.
\\
In order to unveil the radiative effects on the efficiency of star formation, we plot the corresponding SSFRs, as well.
Typical values are $\sim 10-30\,\rm Gyr^{-1}$, consistently with existing studies \cite[][]{Salvaterra2013, Dayal2013}, however, in the case of powerful radiative sources (top) the trend is much more irregular and hides local suppression of star formation (see e.g. SFRs at $z\lesssim 10$).
\\
Radiative feedback from sources emitting at $ T_{\rm eff} = 10^4\,\rm K $ produce SSFRs always  oscillating around the no-RT trend, with deviations below a factor of 2 (see both SSFR panels).
\\
For the $ T_{\rm eff} = 4\times 10^4\,\rm K $ spectra (bottom) SSFR experiences a slight drop with respect to the no-RT case, but this is quite limited and SSFR values are still around a few up to $10\,\rm Gyr^{-1}$.
\\
Finally, we note that the doubling time $t_{\rm db} = 1 / \rm SSFR$ is a fraction of the cosmic time at these epochs for all the models but the one with RT from top-heavy popIII sources.
Indeed, in this case $t_{\rm db}$ reaches values larger than cosmic time.
This might seem rather high when compared to typical values inferred from lower-redshift studies \cite[e.g.][]{Salvaterra2013}, but it is a neat and unique feature of the extreme type of sources which determines it.

\subsubsection{Metal enrichment}

\begin{figure}
\centering
\includegraphics[width=0.45\textwidth, height=7.cm]{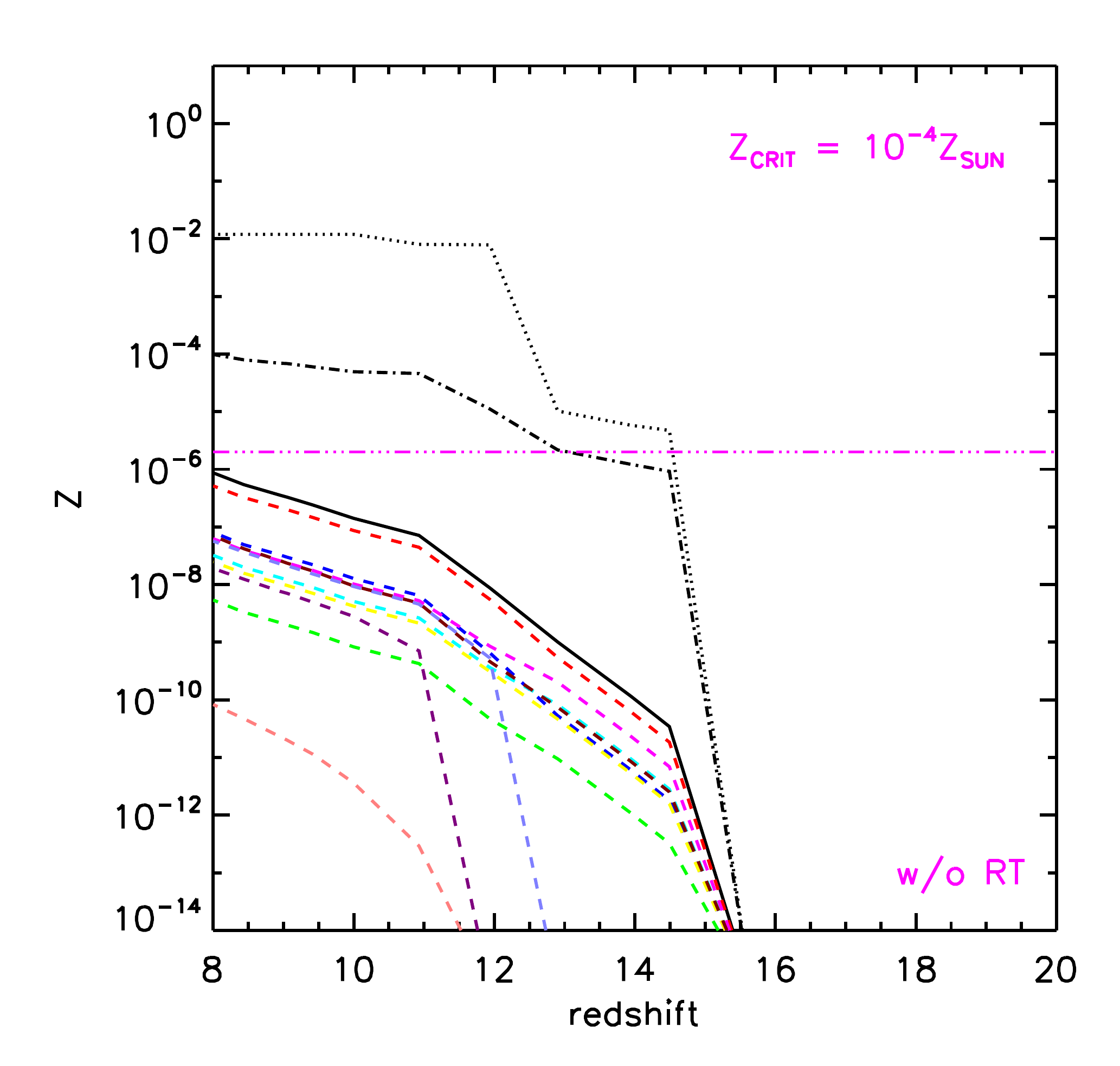}\\
\includegraphics[width=0.45\textwidth, height=7.cm]{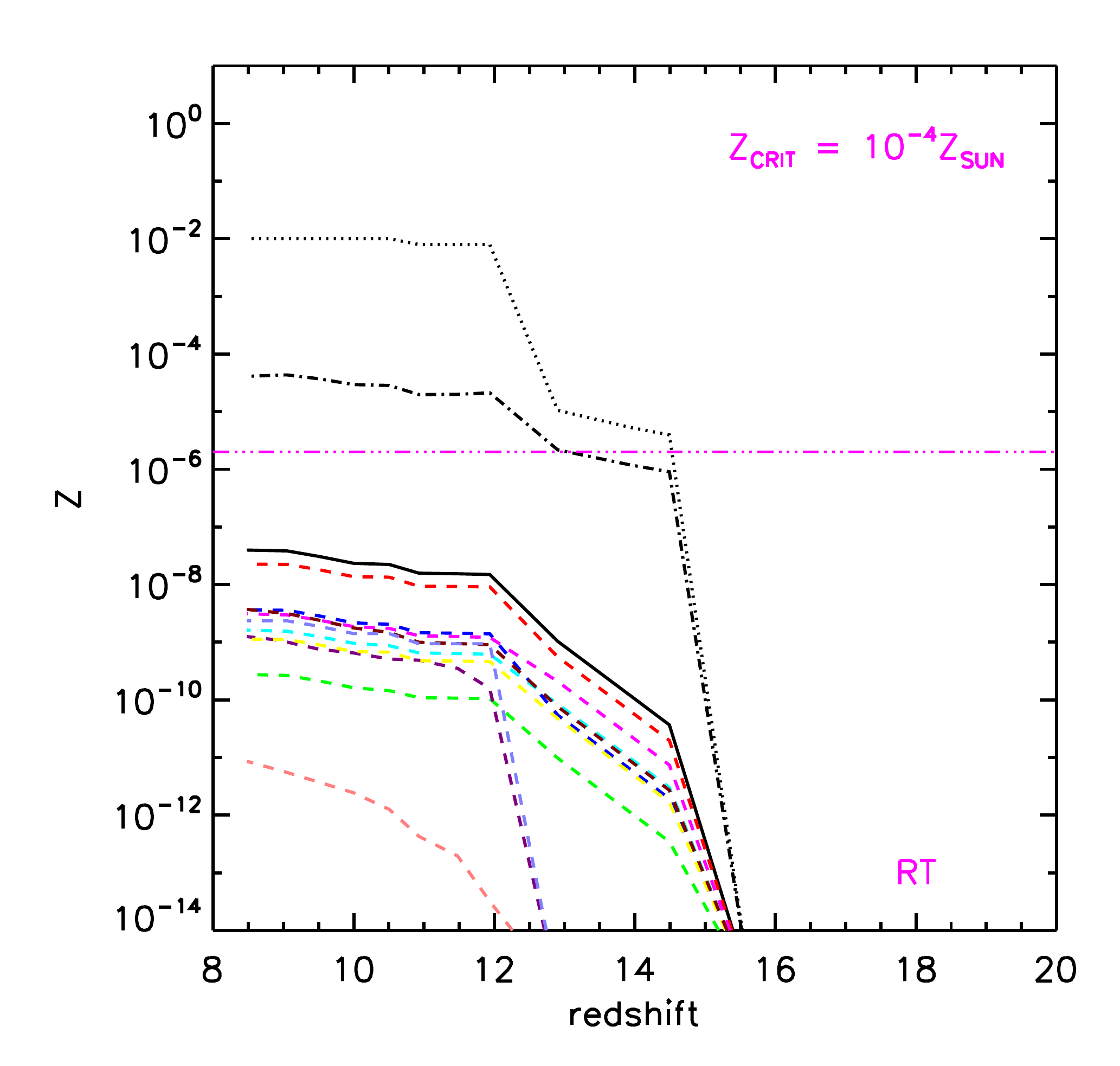}
\caption[]{\small
Evolution of the metal abundances in the Universe. The different lines refer to:
maximum metallicity (dotted), 
average metallicity of spread material (dot-dashed), 
average metallicity of the whole box (solid), 
average abundance for each metal tracked by the simulations (dashed), i.e. C (blue), Ca (green), O (red), N (purple), Ne (pink), Mg (yellow), S (cyan), Si (magenta), Fe (brown), others (light blue).
The horizontal line marks the critical metallicity of $10^{-4}Z_\odot$.
The panels refers to the cases with no-RT and RT from top-heavy popIII IMF, as indicated by the legends.
}
\label{fig:Z}
\end{figure}

A closely related issue is the metal enrichment history of the Universe, which is directly related to cosmic star formation.
In Fig.~\ref{fig:Z} we plot the evolution of the mean metallicity of C, Ca, O, N, Ne, Mg, S, Si, Fe and other elements (dashed lines) spread in the Universe at different epochs and the corresponding mean cosmic metallicity $Z$ (solid line).
The average metallicity (dot-dashed line) of the material enriched by pollution events and the maximum metallicity (dotted line) ejected at different times are overplotted as well.
The horizontal line marks the critical metallicity of $10^{-4}Z_\odot$.
We compare the no-RT case (top) with the corresponding RT one and in both cases a top-heavy popIII IMF is assumed.
\\
It is evident that the overall amounts of metals produced in the RT case is lower by one order of magnitude as a result of inhibited star formation due to radiative feedback (bottom panel).
The amounts of metal fraction ejected by stars (dot-dashed lines) are comparable and the main contributors to metal pollution are oxygen and $\alpha$ elements in both cases.
This is due to the fact that the total amounts of metals ejected and the respective elemental abundances are determined principally by stellar yields and lifetimes, which in these two runs coincide.
\\
Similar trends have been investigated in the SL cases, as well, but we have not found significative differences between the RT and the no-RT runs.
\\
These results further confirm that the main implications of radiative feedback for early structures and cosmic gas come from the more powerful stellar sources.
The impacts of regular sources are, instead, limited to the neighbouring regions only.

%*****************************************************************************

\subsubsection{ Photoevaporation in early haloes } \label{Sect:haloes}

\begin{figure*}
\centering
\includegraphics[width=\textwidth]{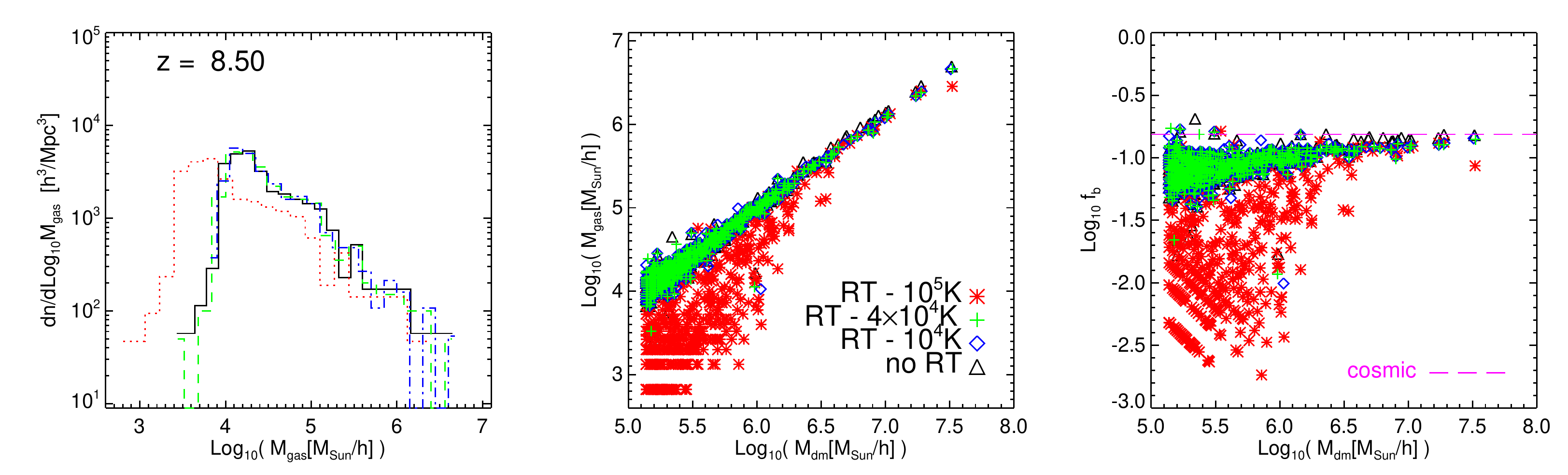}
\caption[]{\small
 Mass distributions for different cases of primordial radiative sources at redshift $z=8.50$. The cases considered are:
RT with $T_{\rm eff} = 10^5\,\rm K$ (red dotted line or asterisks);
RT with $T_{\rm eff} = 4\times10^4\,\rm K$ (green short-dashed line or crosses);
RT with $T_{\rm eff} = 10^4\,\rm K$ (blue dot-dashed line or empty squares);
no RT (black solid line or empty triangles).
The panels show gas cloud distributions (left), gas masses as functions of dark matter (center) and baryon fraction as a function of dark matter (right).
The horizontal long-dashed magenta line in the right panel marks the cosmic baryon fraction for the adopted cosmology.
}
\label{fig:haloes8.5}
\end{figure*}

We finish by discussing the effects on the gas content of early haloes and addressing the main consequence of radiative heating: photoevaporation.
\\
To this purpose, in the panels of Fig.~\ref{fig:haloes8.5} we plot matter distributions for the RT and no-RT runs considered so far at redshift $z=8.5$.
We display the trends of gas mass, M$_{\rm gas}$, and baryon fraction, $f_{\rm b}$, as a function of halo dark mass, M$_{\rm, dm}$, for the models including:
RT from powerful popIII sources with $\rm T_{\rm eff}=10^5\, K$ (red asterisks, dotted line);
RT from sources with $\rm T_{\rm eff}=4\times10^4\, K$ (green crosses, dashed line);
RT from sources with $\rm T_{\rm eff}=10^4\, K$ (blue diamonds, dot-dashed line);
and no RT (black empty triangles, solid line).
\\
The left panel depicts the distributions of gaseous structures as a function of M$_{\rm gas}$, the central panel shows the $\rm M_{\rm gas} - M_{\rm dm}$ trends and the right panel reports the 
$\rm f_{\rm b} - M_{\rm dm}$ relations in the four cases.
We note that dark matter is the main component of cosmic haloes and, as expected, it is little affected by gas evolution within primordial structures, therefore the mass functions \cite[e.g.][]{ShethTormen1999} in the various models practically coincide.
For this reason we do not show it here.
\\
More interestingly, gas distributions are quite sensitive to the ongoing star formation and feedback effects and this is well visible in the left panel.
The RT run with top-heavy popIII IMF (red dotted line) features the main deviations from the no-RT scenario with  a larger abundance of low-mass gas clumps (below $\sim 10^4\,\rm M_\odot$) and a deficit at larger masses (above $10^6\,\rm M_\odot$).
The shift to lower masses is systematic and is a direct consequence of the strong radiative feedback in this model.
In fact, the large power emitted by massive primordial sources heat the gas hosted in the early haloes and increase its thermal energy making it able to escape the structure potential wells.
This photoevaporation process is particularly striking in the RT run with top-heavy popIII IMF, where gas masses get reduced by a factor of $\sim 10$.
In the no-RT case and in the cases with RT from weaker sources, photoevaporation is less important.
\\
The deviations implied by the RT effects are also recognisable in the central panel, where we plot gas mass versus dark mass.
Here, most of the haloes in the no-RT run lie on a roughly straight line with a relatively small scatter.
A similar behaviour is recovered also from the RT models with weak sources.
On the contrary, the trend for the RT run with top-heavy popIII IMF presents a clear decrement for all masses below $\sim 10^7\,\rm M_\odot$.
This means that photoheating from powerful sources is very effective in all primordial haloes and causes relevant gas losses via photoevaporation \cite[see also e.g.][]{Shapiro2004,Jeon2015}.
When compared to less extreme IMFs, we note that only when stellar spectra have $ T_{\rm eff} = 4\times 10^4\,\rm K $ there is some effect on few isolated cases.
\\
Finally, in the right panel we plot the corresponding baryon fraction,  $f_{\rm b}$, within primordial haloes as a function of their dark-matter mass.
This plot neatly highlights radiative feedback effects from the different sources.
Consistently with the aforementioned trends, the large impacts of the emission from massive stars lead to lower $f_{\rm b}$ values by more than 1~dex in most of the haloes, while impacts from weaker sources are minor and not easily distinguishable from the no-RT scenario.

%*****************************************************************************

\section{Discussion}\label{Sect:discussion}

So far, we have shown the main results from our numerical modeling. In the next, we discuss comparisons with data and literature, implications of our findings and possible caveats.

\subsection{Comparisons}

Our coupling of a radiative treatment with a non-equilibrium chemical network makes our results detailed and self-consistent.
The general trends of H ionization are in line with the results from powerful stellar emission in early faint galaxies, as found by e.g. \cite{Yoshida_rt_2007, Wise2012}, and with the temporal evolution inferred by observational data \cite[][]{PlanckXXI2015arXiv}.
\\
Gas fractions are quite sensitive to the adopted cooling and heating sources.
Powerful stellar sources can photoheat haloes quite efficiently, while standard sources are less effective.
As a result, we find that baryon content drops below a few per cent at small masses and recovers values of $\sim 10$~per cent only at masses as large as $\sim 10^7\,\rm M_\odot$.
A similar suppression is found by \cite{Wise2014}, although they recover a fraction of $\sim 10$~per cent only in $\sim 10^8\,\rm M_\odot$ haloes.
We suspect that the reduced amounts of gas fraction in \cite{Wise2014} is due to their lack of fine-structure metal cooling \cite[see][]{Maio2007}.
\\
Metallicities are found to be suppressed by at least one order of magnitude in the most extreme scenario of massive popIII stars, however, in all cases, the typical cosmic values for $Z$ at the epochs of interest here lie below $\sim 10^{-4}~Z_\odot$, compatible with values claimed by observations of damped Lyman-$\alpha$ systems at redshift $z\simeq 7$ \cite[][]{Simcoe2012}.
\\
Our investigation shows that powerful primordial sources might be able to invert the IGM equation of state by the first Gyr with a slope below $-1/5$ and closer to $-1$.
Previous works by \cite{Bolton2008,Bolton2009} suggested a sharper value of about $-1$ to account for the flux distribution in the Lyman-$\alpha$ forest.
Our findings are partly compatible with their results, since a single polytropic relation between gas density and temperature results to be a poor description of the thermodynamical state of baryons in the whole Universe.
This is probably due to the complex interplay among radiative chemical and mechanical feedback.
In fact, the broader range we find for the slope of the IGM equation of state is due to the combination of heating and cooling as result from these feedback effects acting, in a non-trivial way, on gas at different densities and temperatures.
\\
Despite isolated cases, available observations of primordial galaxies in the first Gyr are still poor and it is quite difficult to draw final conclusions from them.
Some hints can be inferred by predicted features of possible primordial star forming objects, such as gamma-ray burst hosts or Lyman-$\alpha$ emitters (LAEs) at redshift $z\sim 6-12$.
Studies in the literature suggest specific SFR around $\sim 10\,\rm Gyr^{-1}$ \cite[][]{Salvaterra2013, Dayal2013}, in line the typical values for the bulk of the galaxy population at such epochs \cite[e.g.][]{BiffiMaio2013, Wise2014}.
Here, we obtain similar orders of magnitude, although we find significant dependencies on the input radiative sources.
Particularly lower SSFRs are found for the case of top-heavy powerful popIII sources.
\\
Several studies in the literature have tried to use LAEs to put constraints on the cosmic reionization near redshift $7$.
Unfortunately, these bear large uncertainties and results from different groups are quite discreprant.
\cite{Taniguchi2005} used 58 LAE candidates to pose lower limits to the neutral hydrogen fraction which was found to be $ > 20-50$ per cent.
This result supported a scenario with a rather delayed cosmic reionization.
On the contrary, \cite{Ouchi2010} detected first signals of LAE clustering from 207 LAEs at the same redshift and put upper limits to the neutral hydrogen fraction of $<20-50$~per cent, concluding that the major reionization process takes place at $z>7$, in agreement with recent Planck data.
When compared to our results, the former case would be compatible with low-mass popIII stars, while the latter case would be compatible with massive popIII stellar generations.
\\
We note that, according to our findings, radiative feedback can be very efficient in ionizing cosmic gas and shutting down star formation, temporarily, only in the case of emission from very powerful stellar sources, such as $\sim 10^2\,\rm M_\odot$ stars and this is consistent with results derived from dynamical expansion of single HII regions by \cite{Hosokawa2011}.
Lately, \cite{Haworth2015} have explored the implications of radiative feedback in star forming regions and HII regions around standard stellar sources. They also found modest effects, mostly when compared to previous results based on simplified radiative treatments, which would likely overestimate the consequences from radiative feedback.
This is in line with our conclusions, as well.

\subsection{Implications}

\begin{figure*}
\vspace{-1cm}
\centering
\includegraphics[width=0.4\textwidth]{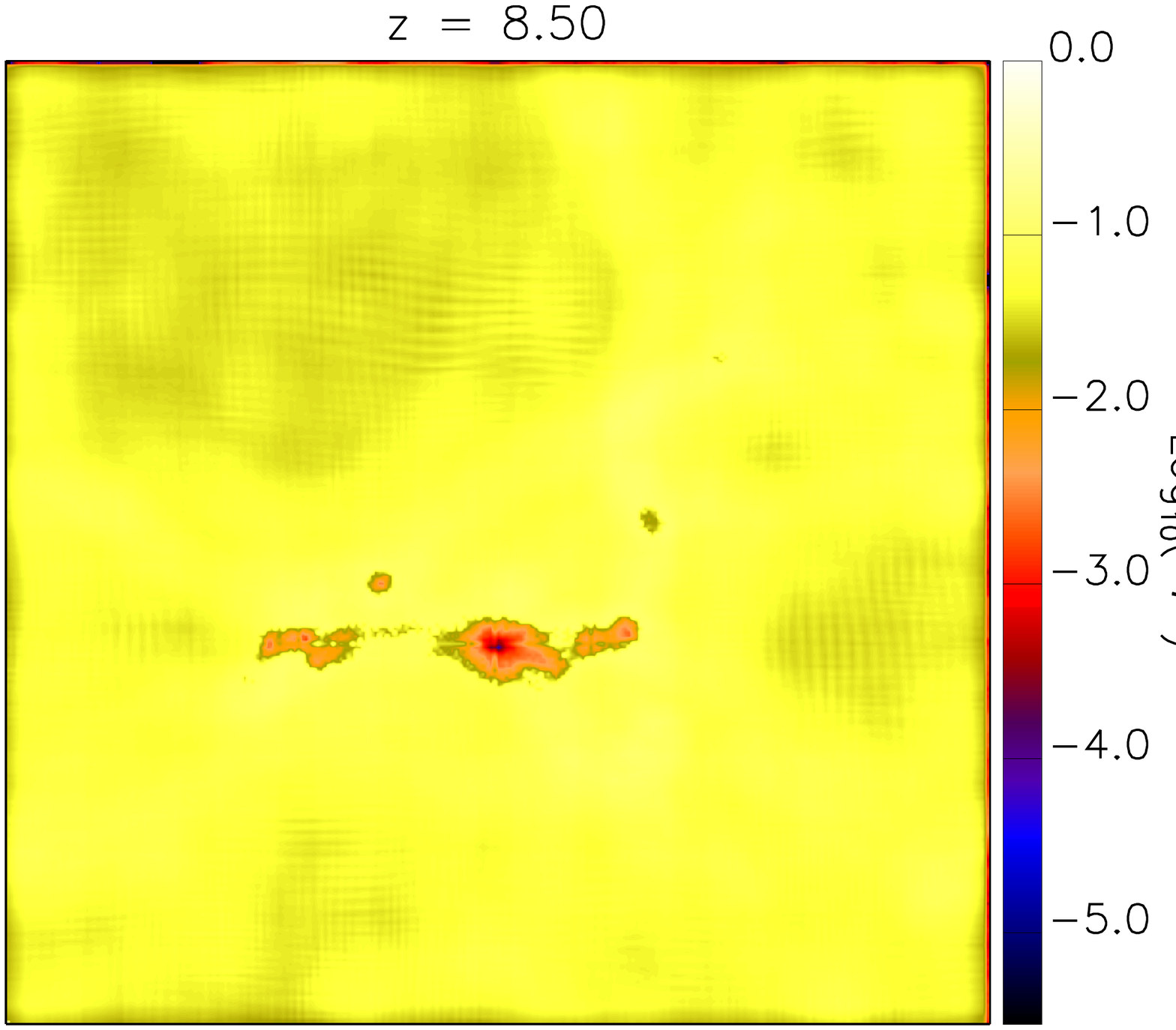}
\includegraphics[width=0.4\textwidth]{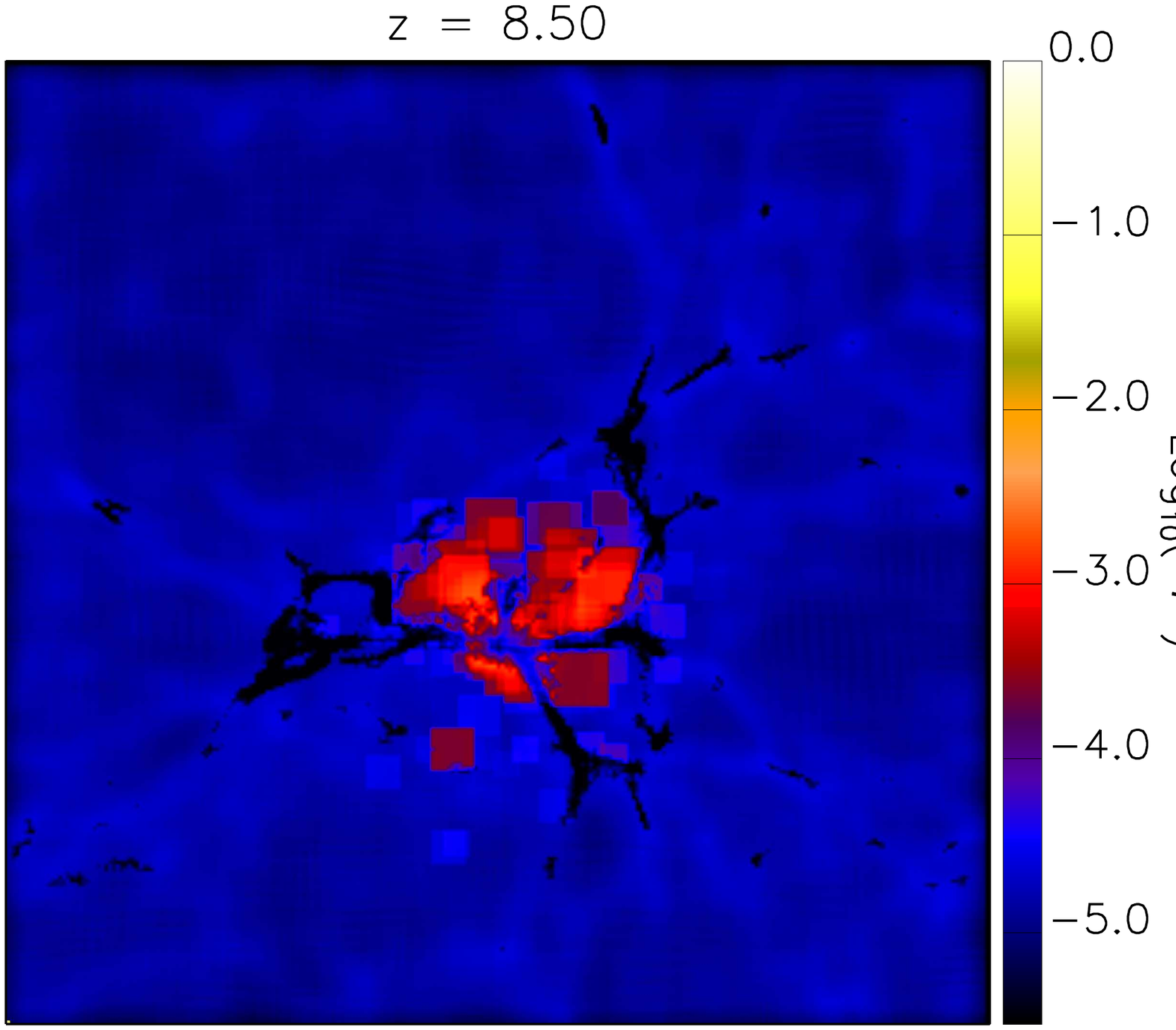}
\vspace{-1cm}
\caption[]{\small
Maps for the Thomson optical depth at redshift $z=8.50$ extracted from the run with RT from powerful sources (left) and with no RT (right).
}
\label{fig:tau}
\end{figure*}

From the results presented in the previous sections we can draw conclusions on the state of the Universe at early times and on the effects from radiative feedback.
\\
As just mentioned, radiative feedback can be effective only in the case of emission from very powerful stellar sources.
The role played by lower-mass stars is minor, though, as their effects are not clearly distinguishable and their influence on the surrounding medium or on the low-density environments is quite limited.
\\
The dependencies on the type of emitting sources are strong for the chemical evolution and ionization state of early gas, as well.
Indeed, very hot sources (in our case, with $\rm T_{\rm eff}\sim10^{5}~K$) have impacts on molecule dissociation, H reionization and the ionization stages of heavier elements, like He, with He$^{++}$ reaching filling fractions of $\sim 50$ per cent already at the beginning of full H reionization, at $z\lesssim 9$.
\\
Interestingly, the powerful sources that can sustain He/He$^+$ reionization, shortly after H reionization, are also those that are able to invert the IGM equation of state, as a consequence of their hard SED.
\\
The reionization history dictated by these powerful sources is consistent with recent Planck data, which report a Thomson optical depth of $\tau \simeq 0.066$, corresponding to a reionization redshift of $z\simeq 8.8$ \cite[][]{PlanckXXI2015arXiv}.
We quantify the expected Thomson optical depth by considering the cosmic history in the different scenarios and by using the physical quantities and chemical abundances retrieved in the simulation snapshots.
Due to the lack of snapshots at late times, we assume full ionization at low redshift.
This is consistent with the trends in Fig.~\ref{fig:ff} at $z\simeq8$ for the case of RT from powerful sources and a quite optimistic assumption for the case of RT from weaker sources in Fig.~\ref{fig:ff.lowmass} (in this latter case, $\tau$ values are to be considered as upper limits).
In this respect, more detailed analyses could be done, however they are beyond the goals of this work.
Our calculations, although rough, should be sufficient to have an idea of the general trends.
Since the optical depth is strongly dependent on the free-electron fraction, RT effects from powerful sources can be rather dramatic, as shown in Fig.~\ref{fig:tau} for the cases with RT from powerful sources and no RT (results for RT from standard sources are similar to this latter case).
Free electrons in the ionized intergalactic medium of the run with RT from powerful sources lead $\tau$ values ranging between $0.01$ and $0.1$, with few dense cold regions that are partially or completely recombined and whose contribution is up to $\tau \lesssim 10^{-2}$.
In the figure the values of $\tau$ are computed over a finite volume while Planck data refer to the whole sky, nevertheless the trend points toward an overall agreement with observations.
The case with no RT is, instead, opposite with most of the gas being cold and largely neutral (accounting for the blue regions where $\tau\lesssim 10^{-4}$) and free electrons present only in the material heated by SN feedback (this accounts for the orange-red regions with $\tau\sim 10^{-2}-10^{-3}$).
\\
This has important implications for the understanding of the formation path of candidate primordial relics, such as low-mass dwarf galaxies.
Indeed, low-mass dwarf galaxies could be explained as fossils of reionization, if this event would have been determined by very powerful sources (i.e. massive popIII stars).
They could have been assembled from gas collapse before reionization and their growth would have been eventually stopped after the explosion of early massive stars \cite[see also][]{Simpson2013}.
Thus, radiative feedback definitely plays a role in suppressing collapse of such low-mass objects, mostly at higher redshift, when typical masses are smaller.
If this is the case, present-day observations should give typical stellar ages in dwarf galaxies around $10\,\rm Gyr$ and rather low metallicities.
Our results seem to be consistent with local dwarf measurements.
For example, \cite{Kirby2011} give [Fe/H] between $-1$ and $-3$ with a tight relation to luminosity, according to which low-luminosity low-mass objects should have lower metal content.
Although a direct comparison is not possible, these observational trends go in the direction of our results.
PopIII events are unlikely to alter significantly the global relation between metallicity and luminosity (mass), however they would leave imprints in the individual abundances present in each object.
Indeed, massive (above $10^2\,\rm M_\odot$) popIII stars dying as PISN produce large amounts of $\alpha$ elements with oversolar patterns of [Si/O] and [S/O] and with $\rm [Mg/O]\gtrsim -0.2$.
The same abundance ratios for more standard popIII stars, instead, are expected to be below solar level for [Si/O] and [S/O] and $\rm -0.5 < [Mg/O] < -0.3 $ \cite[][]{Ma2015}.
\\
Likely, radiative feedback represents a driving mechanism of gas {\it preheating}\footnote{
  This basically means it is a mechanism to delay or halt gas cooling.
}
\cite[][]{LuMo2007, Lu2015} able to halt or delay star formation at early times.
Seen under this point of view, the inversion of the IGM equation of state is a manifestation of the formation of these fossil structures in a medium preheated by the aforementioned radiative effects.
\\
Heating of gas in primordial epochs inhibits gas cooling and fragmentation, thus the existence of massive or super-massive black holes observed at early ($z\sim 7$) times could be in line with a direct-collapse formation scenario\footnote{
	Here we warn the reader about the extensive use in literature of the simple Bondi accretion in modelling such complex process, though.
}.
On the contrary, weak sources induce only limited effects to gas cooling, minor changes to H and He abundances and have no implications for the He ionization.
\\
We note that the scenario with hot sources is able to both induce an inversion of the IGM equation of state and to consistently determine cosmic H reionization (possibly followed by He complete reionization).
\\
This is intriguing as powerful short-lived massive stars could account for both processes, without the need of active galactic nuclei.\\
Obviously, their contribution at redshifts as high as $z\sim6-8$ could be present, but probably it would be minor. \\
From a quantitative point of view, halo masses at these times are mostly distributed below $10^8\,\rm M_\odot$, with a fraction of higher-mass objects which is less than 0.1 per cent the fraction of $\sim 10^6\,\rm M_\odot$ haloes.
This means that, independently from their formation and growth, early AGN occurrence is extremely rare, mostly if we consider duty cycle.
In addition, radiation emitted by AGN, although very powerful, has to travel through usually thick and dusty material that would obscure its emissivity.
As a result, even {\it ad hoc} modelling with parameter selection based on the supposed existence of a hidden population of faint AGNi at $ 4 < z < 6.5$ \cite[][]{Giallongo2015} makes a scenario of reionization led by AGN emission only very marginally compatible with the optical depth measured by the Planck satellite \cite[][]{MadauHaardt2015}.
\\
A good compromise to this issue could be given by the hot popIII sources considered throughout this paper.
Furthermore, radiative feedback seems to facilitate a longer popIII era than the case with no-RT, supporting a larger contribution to the cosmic star formation rate density and  a larger number of ionising photons.
\\
From an observational point of view, diagnostics for typical stellar sources in star forming environments are the Ly$\alpha$ and H$\alpha$ lines, while, in the case of hard spectra from powerful sources, an important feature is represented by the He~II  $\lambda$ 1640 line \cite[][]{Schaerer2002}.
Therefore, He~II flux, $F_{\rm 1640}$, and Ly$\alpha$ flux, $F_{\rm H_\alpha}$, could give indirect hints on primordial stellar populations if detected by future telescopes, such as the {\it JWST}\footnote{
http://www.stsci.edu/jwst
}.
Detailed studies have demonstrated that massive primordial stars (with masses $\sim 10^2\,\rm M_\odot$) are expected to generate an He~II flux $F_{\rm 1640} \lesssim 2 F_{\rm H_\alpha}$, while more massive sources, such as direct-collapse black holes with seed mass above $10^4\, \rm M_\odot$, could determine $F_{\rm 1640} > 2 F_{\rm H_\alpha} $ \cite[][]{JJ2011}.
In the former case, He~II $\lambda$ 1640 emission from a source at redshift $z\sim10$ would not be detectable by the Near-Infrared Spectrograph aboard the {\it JWST} when assuming an observing time of 100 hours and a 3$\sigma$ flux limit.
However, in the latter case, He~II emission would be marginally detectable.
A similar conclusion can be drawn for Ly$\alpha$ line, but not for H$_\alpha$, whose flux is too low for detection.
\\
Average pollution from first stars in early galaxies is sufficient to locally overcome the critical metallicity $Z_{\rm crit}$ after only a couple of massive SN episodes.
However, the larger photoheating effect due to massive popIII stars causes gas photoevaporation and subsequent delay for star formation and cosmic metal enrichment in the simulated volume \cite[][]{Shapiro2004, Jeon2015}.
This results in a factor of 10 lower mean metallicities with respect to the no-RT model and baryon fractions in primordial haloes drop dramatically.
\\
Finally, we mention that delaying star formation at the beginning of galaxy build-up is needed to form extended disks and avoid angular-momentum catastrophe at low redshift \cite[e.g.][]{Stinson2010, TrujilloGomez2015}.
Often, this has been achieved by including powerful thermal SN feedback into sub-grid numerical schemes to halt gas cooling in the early phases of galaxy formation \cite[e.g.][]{Scannapieco2008}.
In the present work, we demonstrated that radiative feedback can be an optimal candidate to face this issue in a consistent way, although further studies are required to assess the relevance of such effects already at $z\sim8$.

\subsection{Caveats}
As always, numerical implementations bear a number of approximations in the sub-grid models and in the treatment of some physical quantities.
\\
Some of these are not crucial for a realistic modelling of early structures.
For example, stellar lifetimes dictate the temporal evolution of cosmic metal spreading and the occurrence of the first heavy elements in the Universe.
Overall, dependencies of stellar lifetimes on metallicity are not very strong \cite[see also][and references therein]{DeLucia2014}.
Lifetimes of high-mass stars are very well determined and do not carry caveats.
Mispredictions among different prescriptions can be present in the case of low-mass stars and this would cause changes in the evolution of e.g. ${}^3$He and ${}^7$Li species.
We highlight, though, that for the epochs we focus in this work this is a very minor problem.
\\
Stellar evolution parameters might play a role for what concerns the heavy-element budget, as individual yields are not always sufficiently known. Nevertheless, metal enrichment is led by mostly oxygen, carbon and $\alpha$ elements.
The yields for C and O are relatively well constrained and uncertainties should not affect gas cooling and star formation. Yields for $\alpha$ elements, instead, present larger uncertainties and could have more severe implications \cite[see e.g.][and references therein for a more extended discussion]{Maio2010,MaioTescari2015}.
\\
SN rates are somewhat affected by model parameters, mostly for SNIa \cite[for a review, see][]{Greggio2005}, with consequent uncertainties in the timing of spreading events and following gas cooling.
In the early epochs we are interested in, these effects should not be dramatic, although dedicated investigations would be needed in order to assess them in detail.
\\
An important limitation is our ignorance of the efficiency of popIII sources. Recent studies \cite[][]{Robertson2015} seem to suggest that at high redshift popII sources should produce at least $10^{53.14}$ ionizing photons per $\rm M_\odot/yr$ of SFR to justify the Thomson optical depth measured by Planck\footnote{
In the work by \cite{Robertson2015} the quoted efficiency can account for $\tau\lesssim 0.035$ at $z \simeq 8.8$ (Planck data suggest $\tau\simeq 0.066 \pm 0.012$ at 1$\sigma$).
}.
Larger values could be requested to keep the cosmic medium ionised at $z\sim7-8$, though.
We have naively assumed for the massive top-heavy scenario a hard black-body spectrum with a $10^5 \,\rm K$ effective temperature which gives luminosities much higher than the popII case, as already visible in Fig.\ref{fig:cooling_spectra} (such luminosities have been assumed to escape completely). Probably describing the emission of an entire stellar population with a single hot black body is an important simplification and better modelling in future works might be performed by adopting more realistic spectral energy distributions.
We note that obtaining a precise empirical value for popIII efficiencies is rather tricky due to the many biases, systematics and calibration issues of observational analysis and stellar population synthesis models \cite[][]{Guidi2015} as well as dust corrections at such early times.
In fact, currently measured SFRs of $z\simeq 10$ galaxies have 1$\sigma$ errors of up to 2~dex, thus large variations in the retrieved efficiencies could be still compatible with data at 1$\sigma$ level.
\\
It is also worth mentioning that when observers try to infer physical properties of primordial galaxies from measured magnitudes, they have to make a number of assumptions on e.g.  underlying IMF, metallicity, age of the stellar population.
The parameters of stellar population synthesis models are usually calibrated on the nearby Universe although they are often blindly applied to galaxies at any $z$. At very high redshift the physical conditions of the medium where galaxies form and evolve are rather different and the recipes used at $z=0$ could dramatically fail.
\\
Besides stellar evolution, possible sources of ionizing photons that have been neglected here are active galactic nuclei (AGNi).
Probably they might give some contribution to the photon budget in the first billion years, but, given their rarity at such early cosmic times, the effects related to AGN activity should not be predominant \cite[e.g.][]{MadauHaardt2015, Keating2015}.
\\
In our implementation we considered self-shielding.
Here we stress that different self-shielding prescriptions are available in literature, however, as as already discussed in \cite{PetkovaMaio2012}, the changes related to varying them should be minor.
\\
An open problem is our complete lack of knowledge of the primordial IMF also due to the quite heterogeneous findings of different studies \cite[][]{Abel2002, Yoshida_et_al_2007, CampbellLattanzio2008, SudaFujimoto2010, Stacy2014, Fraser2015}.
As we have showed, variations of the popIII IMF could result in completely different pictures for structure formation and for the reionization history in the first Gyr.
\\
We have modelled metal spreading according to kernel smoothing, in order to mimic metal diffusion.
This is a simple approximation that can broadly account for such process, however it could lack of the necessary details at very small scales.
In general, this is a very complex issue and, despite some attempts, a fully satisfying treatment is still missing in the literature.
\\
The calculations presented here are done for a rather small box of the Universe.
This type of volumes are often employed in high-redshift studies \cite[a detailed discussion on similar issues can be found in e.g.][]{Maio2010}, although they might miss rare big structures originating from larger modes.
Obviously, the best solution would be to have both large boxes and high resolution, including all the physics described above possibly running down to low redshift.
Current capabilities are still unable to allow us such kind of studies, though.
\\
We warn the reader that cosmic gas collapse and the resulting growth of baryonic structures are also related to the background cosmological model.
In this paper we have adopted a standard $\Lambda$CDM model with a Gaussian input power spectrum, but alternative scenarios are possible.
Indeed, in previous works we checked that different acceleration histories imposed by different early dark-energy models, at redshift $z\gtrsim 10$, can lead to different clumpiness and distribution of primordial gas clouds, with possible effects on luminous matter \cite[][]{Maio2006}.
Similarly, changes in the input power spectrum can have some impact on the cosmic star formation, although only in the case of very large non-Gaussian deviations \cite[][]{MaioIannuzzi2011}.
\\
The intrinsic nature of dark matter is still under debate with a current preference for the cold scenario.
Due to small-scale problems of cold dark matter, an interesting alternative often advocated is the so-called warm dark matter (WDM), i.e. matter composed by particles having a non-negligible thermal velocity.
This would affect the matter power spectrum with a sharp decrease at small scales.
In this latter case, the resulting cosmic star formation density gets delayed and the amount of early objects formed drops by more than one dex.
The implications for reionization are not obvious to infer, since galaxies in a WDM universe, although rarer, are expected to be more bursty \cite[][]{MaioViel2014arXiv}, as a consequence of the initially retarded collapse and the following catch up of primordial WDM structures.
\\
Finally, we mention that primordial supersonic gas bulk flows are predicted to raise at decoupling from non-linear perturbation theory \cite[][]{TH2010}.
They would delay pristine molecule formation and gas collapse due to enhanced gas kinetic energies \cite[][]{Maio2011}, hence they could have repercussions on the onset of star formation at early times as well as on cosmic reionization.

\section{Conclusion}\label{Sect:conclusion}
We have presented results from numerical N-body hydrodynamic simulations including primordial non-equilibrium chemistry (of e$^-$, H, H$^+$, H$^-$, He, He$^+$, He$^{++}$, H$_2$, H$^+$, D, D$^+$, HD and HeH$^+$), high and low-temperature cooling, star formation, stellar evolution (accounting for massive stars, AGB, SNIa), metal spreading for several heavy elements (He, C, Ca, O, N, Ne, Mg, S, Si, Fe, etc.), feedback effects and self-consistently coupled radiative calculations of photon propagation over 150 frequency bins.
We have considered simultaneous evolution of both popIII and popII-I stellar generations according to top-heavy and Salpeter-like IMFs and various possible different assumptions for their input spectral energy distribution, from very powerful massive stars to weaker OB-like emission.
\\
From our analysis, we can conclude that massive popIII stellar sources in the first Gyr are capable of heating the cosmic gas and ionize the hydrogen content in the simulated cosmic volume in a quite short time.
As a consequence of their powerful emission, they affect the equation of state of the intergalactic medium by pushing the corresponding phase-space slope to values below $-1/5$ reaching $-1$ and by enhancing the critical Jeans masses needed for gas collapse.
Furthermore, such type of stellar sources induces subsequent reionization of He and He$^{+}$, fewer amounts of metals spread, lower stellar content in early (proto-)galaxies and a large degree of gas photoheating in smaller haloes.
Radiation from low-mass stars, instead, has less severe effects that result limited to local environments.
\\
Such findings have implications for the thermal state of the Universe at early epochs, the driving mechanism of gas preheating, the build-up and chemical features of low-mass dwarf galaxies, the formation of primordial direct-collapse black holes, the minor role of AGNi during the epoch of reionization and, possibly, for the formation of extended disks and the angular-momentum catastrophe.

%*****************************************************************************

\section*{acknowledgments}
We acknowledge the referee, Nick Gnedin, for his swift and prompt comments which helped improve and clarify the original manuscript.
U.~M.'s research has received funding from the European Union Seventh Framework Programme (FP7/2007-2013) under grant agreement n. 267251.
S.~B. is supported by PRIN-INAF12 grant "The Universe in a Box: Multi-scale Simulations of Cosmic Structures", PRINMIUR 01278X4FL grant Evolution of Cosmic Baryons, and INDARK INFN grant.
The numerical simulations have been performed under the PRACE-2IP project, grant agreement n. RI-283493, resource Abel, NOTUR, Norway, project number NN9903K.
We also acknowledge technical support by T.~R{\"o}blitz, O.~W.~Saastad and M.~Afonso Oliveira.
We made use of the tools offered by NASA ADS for the bibliographic research.

%*****************************************************************************

\bibliographystyle{mn2e}
\bibliography{bibl.bib}

\label{lastpage}
\end{document}